\documentclass[reprint,amsfonts, amssymb, amsmath, showkeys,pra,nofootinbib, superscriptaddress, twocolumn,longbibliography,aps]{revtex4-2}

\usepackage{float}
\makeatletter
\let\newfloat\newfloat@ltx
\makeatother
\usepackage[utf8]{inputenc}
\usepackage[english]{babel}
\usepackage{graphics,overpic}
\usepackage{selinput}
\usepackage[normalem]{ulem}
\usepackage[shortlabels]{enumitem}
\usepackage{braket}
\usepackage{amsthm}
\usepackage{mathtools,soul}
\usepackage[caption=false,font=small]{subfig}
\usepackage{svg}
\svgsetup{inkscapelatex=true}
\usepackage{physics}
\usepackage{booktabs}
\definecolor{darkgreen}{rgb}{0,0.5,0}
\newcommand{\cmark}{\textcolor{darkgreen}{\ding{51}}} 
\newcommand{\xmark}{\textcolor{red}{\ding{55}}}
\usepackage{pifont}
\usepackage{xcolor}
\usepackage{graphicx}
\usepackage{adjustbox}
\usepackage{placeins}
\usepackage[T1]{fontenc}
\usepackage{lipsum}
\usepackage{csquotes}
\usepackage{bm}
\usepackage{bbm}
\usepackage{nameref}
\DeclareMathOperator{\Var}{Var}
\usepackage[linesnumbered,ruled,vlined]{algorithm2e}
\SetKwInput{kwInit}{Init}
\usepackage{changes}
\definechangesauthor[name=Ilya Safro, color=red]{is}
\usepackage[makeroom]{cancel}
\usepackage[toc,page]{appendix}
\usepackage[colorlinks=true,allcolors=blue]{hyperref}
\usepackage{tikz}
\tikzset{every picture/.style=remember picture}

\makeatletter
\newif\ifeqcontrib@this
\newif\ifeqcontrib@any
\eqcontrib@thisfalse
\eqcontrib@anyfalse
\newcommand{\eqcontrib}{\global\eqcontrib@thistrue\global\eqcontrib@anytrue}
\newcommand{\eqcontribmark}{\textsuperscript{\ensuremath{,\dagger\dagger}}}
\newcommand{\eqcontrib@maybe}{%
  \ifeqcontrib@this
    \global\eqcontrib@thisfalse
    \eqcontribmark
  \fi
}

\definecolor{Burgundy}{RGB}{144,0,32}
\usepackage[most]{tcolorbox} 
\newcounter{myresult}
\newtcolorbox{myresult}[1]{
  colback=Burgundy!5!white,    
  colframe=Burgundy!75!black,  
  fonttitle=\bfseries,    
  title={\refstepcounter{myresult}#1},                
  arc=4mm,                 
  outer arc=1mm
}
\newcommand{\printEqContrib}{%
  \ifeqcontrib@any
    \begingroup
      \renewcommand{\thefootnote}{\ensuremath{\dagger\dagger}}%
      \footnotetext{These authors contributed equally to this work.}%
    \endgroup
  \fi
}
\def\doauthor#1#2#3{%
  \ignorespaces#1\unskip\@listcomma
  \begingroup
    #3%
  \@if@empty{#2}{\endgroup{}{}}{\endgroup{\comma@space}{}\frontmatter@footnote{#2}}%
  \eqcontrib@maybe
  \space \@listand
}%
\makeatother

\begin{document}
\preprint{APS/123-QED}
\title{Reaching the Limits of Ground-State Metrology with Many-Body Probes}
\author{Erik L. André\eqcontrib}
\thanks{erik.andre@tuwien.ac.at}
\affiliation{Física Teòrica: Informació i Fenòmens Quàntics, Departament de Física, Universitat Autònoma de Barcelona, 08193 Bellaterra (Barcelona), Spain}
\affiliation{Technische Universität Wien, Atominstitut, Stadionallee 2, 1020 Vienna, Austria}
\affiliation{Institute for Quantum Optics and Quantum Information (IQOQI), Austrian Academy of Sciences, Boltzmanngasse 3, 1090 Vienna, Austria}
\author{Víctor Izquierdo\eqcontrib}
\thanks{victor.izquierdo@uab.cat}
\affiliation{Física Teòrica: Informació i Fenòmens Quàntics, Departament de Física, Universitat Autònoma de Barcelona, 08193 Bellaterra (Barcelona), Spain}
\author{Ricard Puig}
\thanks{ricard.puigivalls@epfl.ch}
\affiliation{Institute of Physics, Ecole Polytechnique F\'{e}d\'{e}rale de Lausanne (EPFL), CH-1015 Lausanne, Switzerland}
\affiliation{Centre for Quantum Science and Engineering, Ecole Polytechnique F\'{e}d\'{e}rale de Lausanne (EPFL), CH-1015 Lausanne, Switzerland}
\author{John Calsamiglia}
\thanks{john.calsamiglia@uab.cat}
\affiliation{Física Teòrica: Informació i Fenòmens Quàntics, Departament de Física, Universitat Autònoma de Barcelona, 08193 Bellaterra (Barcelona), Spain}
\author{Martí Perarnau-Llobet}
\thanks{marti.perarnau@uab.cat}
\affiliation{Física Teòrica: Informació i Fenòmens Quàntics, Departament de Física, Universitat Autònoma de Barcelona, 08193 Bellaterra (Barcelona), Spain}
\date{\today}
\begin{abstract}
We investigate the physical requirements to reach the ultimate limits of ground-state metrology—i.e., the estimation of an unknown parameter $\theta$ of an $N$-body Hamiltonian via measurements on its ground state—using many-body probes featuring either short-range or long-range interactions. We characterize the conditions to saturate two fundamental limits of the quantum Fisher information $\mathcal{F}_\theta$: the static ground-state bound $\mathcal{F}_\theta \lesssim N^2/\Delta^2$, where $\Delta$ is the spectral gap, 
and the dynamical Heisenberg limit $\mathcal{F}_\theta \lesssim N^2 \tau^2$, where $\tau$ is the total protocol duration. To saturate the static limit, we argue that short-range interacting systems require a spectral gap that closes with $N$. At second-order quantum critical points this leads to a universal condition on the critical exponents, $\nu(d+z)=1$, which can be approached arbitrarily closely, as we show for the XXZ chain; cat-like ground states at first-order transitions provide an alternative route to static optimality. In contrast, we find that all-to-all interacting models can achieve optimality even in gapped regimes.  Furthermore, reaching the dynamical bound is tightly related to the ground-state preparation time; we demonstrate that its optimal, inverse-gap profile can be obtained via local adiabatic protocols. By analyzing paradigmatic systems—the transverse-field Ising chain, the XXZ chain, a squeezing Hamiltonian, and the two-mode Bose-Hubbard model—we identify specific many-body probes capable of approaching both limits. Overall, our results provide a theoretical basis for the design of (local) many-body ground-state sensors operating at the ultimate limits of quantum precision.
\end{abstract}
\maketitle
\printEqContrib
\section{Introduction}
\begin{table*}[t!]
\centering
\caption{\textbf{Attainability of the fundamental limits of the QFI in the static and dynamical scenarios for the considered models.} A check mark (\cmark) indicates that Heisenberg scaling (HS) can be attained, whereas crosses (\xmark) denote that it cannot. Here, Heisenberg scaling means $\mathcal{F}_\theta \sim \norm{H_{\rm S}}^2 / \Delta^2$ and $\mathcal{F}_\theta \sim \norm{H_{\rm S}}^2 \tau^2$ in the static and dynamical scenarios, respectively. Namely, the QFI saturates the bounds in Eqs.~\eqref{boundGS} and \eqref{boundGSHeis} in scaling, possibly with a non-unit prefactor. In the dynamical scenario, we consider both the Ground-State Unitary Metrology (GSUM) and Ground-State Adiabatic Metrology (GSAM) frameworks introduced in Sec.~\ref{Sec.frame_dyn}. Attainability in GSUM is considered in the limit of $\tau \gg \tau_{\rm prep}$, and entries for the XXZ model hold as $\gamma_z \to 1^-$.}
\label{tab:resultsintro}
\renewcommand{\arraystretch}{1.3}
\setlength{\tabcolsep}{5pt}
\resizebox{\textwidth}{!}{
\begin{tabular}{lcccc}
\toprule\toprule
        & \multicolumn{2}{c}{Short-range many-body systems (critical)} & \multicolumn{2}{c}{Long-range many-body systems} \\
\cmidrule(lr){2-3} \cmidrule(lr){4-5}
        & Transverse-field 1D Ising & External-field XXZ & Spin-squeezing & Two-mode Bose-Hubbard \\
\midrule
Static HS ($\mathcal{F}_\theta \sim \norm{H_{\rm S}}^2 / \Delta^2$) & \xmark & \cmark & \cmark & \cmark\\
\midrule
GSUM HS ($\mathcal{F}_\theta \sim \norm{H_{\rm S}}^2 \tau^2$)& \xmark & \cmark & \cmark & \cmark \\
\midrule
GSAM HS ($\mathcal{F}_\theta \sim \norm{H_{\rm S}}^2 \tau^2$)& \xmark & \cmark & \cmark & \cmark \\
\bottomrule\bottomrule
\end{tabular}
}
\end{table*}
Quantum metrology deals with the precise estimation of unknown physical quantities by taking advantage of phenomena such as superposition, coherence and entanglement to overcome the capabilities of classical sensors~\cite{Giovannetti2011,RevModPhys.89.035002}. To this end, the unknown parameter $\theta$ is encoded into a quantum probe $\rho_\theta$, which can be realized across diverse platforms, including photonic, atomic, solid-state, and spin-based many-body systems. The corresponding measurement uncertainty $(\Delta \theta)^2$ is bounded by the Cramér-Rao bound (valid for unbiased estimators)~\cite{Rao1945,Cramer1946}
\begin{equation}\label{cramerraobound}
    (\Delta \theta)^2 \geq \dfrac{1}{n \mathcal{F}_\theta},
\end{equation}
for a number $n$ of experiment repetitions. Here, $\mathcal{F}_\theta$ is the quantum Fisher information (QFI), which quantifies how sensitive $\rho_\theta$ is to small variations in the unknown parameter. The explicit form of $\mathcal{F}_\theta$, as well as its fundamental limits, naturally depends on the physical mechanism by which $\theta$ is imprinted onto $\rho_\theta$. This is commonly realized through a dynamical process, $\rho_\theta (t) \coloneqq U_\theta (t) \rho_0 U_\theta^\dagger (t)$,  but it can also arise in static scenarios in which $\rho_\theta$ is a ground state~\cite{Montenegro2025,Braun2018,PhysRevA.78.042106,Rams2018SuperHei,Frrot2018,Garbe2020Critical,Mirkhalaf2020Supersensitive,Montenegro2021Global,Sarkar2025exp,PhysRevLett.130.240803, mihailescu2025criticalquantumsensingtutorial,Sahoo2024}, a thermal~\cite{PhysRevLett.114.220405, Mehboudi_2019,gabbrielli_multipartite_2018,Abiuso2024OptimalThermometersSpinNetworks,Mihailescu2024QuantumSensing,Petropoulos2024,GarcaPintos2024Thermal,Abiuso2025}, or a nonequilibrium steady state~\cite{PhysRevA.90.062130,PhysRevB.96.104402,Ding2022,dicandia2023,Montenegro2023Quantum,Iemini2024Floquet,PRXQuantum.6.030309,Alushi2024,Beaulieu2025Crit,Alushi2025,Khandelwal2025Current,Xiao2026Observation}. 

In ground-state metrology, $\rho_\theta$ is the ground state of a $\theta$-dependent Hamiltonian $H_\theta$, which can be decomposed~as
\begin{equation}\label{generalH}
    H_\theta =  \theta H_\mathrm{S} + H_\mathrm{C},
\end{equation}
where $H_\mathrm{S}$ is the Hamiltonian encoding the signal, while $H_\mathrm{C}$ collects all other $\theta$-independent contributions, ranging from unwanted environmental terms to (possibly time-dependent) control Hamiltonians.

At a quantum phase transition, the ground state $\rho_\theta$ becomes extremely susceptible to small variations of $\theta$, and the QFI $\mathcal{F}_\theta$ can diverge—i.e., it can grow super-extensively with the probe size $N$—thus enabling quantum-enhanced precision. This enhancement, however, comes at the price of diverging preparation times for $\rho_\theta$, which arise from the closing of the spectral gap $\Delta$ with increasing $N$ in the vicinity of the critical point.
The metrological potential of critical ground states, including the trade-off between critical slowing down and enhanced  sensitivity, has been extensively investigated in a variety of settings, ranging from interacting spin systems \cite{Rams2018SuperHei} to quantum-optical platforms \cite{Garbe2020Critical}. 

More generally, the precision of a ground-state probe has been recently bounded as~\cite{Abiuso2025}
\begin{equation}\label{boundGS}
    \mathcal{F}_\theta\leq \dfrac{\norm{H_\mathrm{S}}^2}{\Delta^2},
\end{equation}
where we defined the seminorm $\norm{H_\mathrm{S}} \coloneqq \lambda_M - \lambda_m$, with $\lambda_M (\lambda_m)$ being the maximum (minimum) eigenvalue of~$H_\mathrm{S}$, whereas $\Delta$ is the spectral gap of the total Hamiltonian~$H_\theta$ (i.e., the energy difference between its ground and first excited state). This bound makes explicit the intimate connection between the QFI and the spectral gap~$\Delta$, and hence with the state preparation time---indeed,~$\Delta^{-1}$ is associated with the minimal possible preparation time of ground states \cite{boixo2010necessary,PhysRevA.65.042308,albash2018adiabatic,lin2020near,PlanckianTime}. For a fixed gap $\Delta$, Eq.~\eqref{boundGS} implies that $\mathcal{F}_\theta$ can achieve, at most, the Heisenberg-like scaling $\norm{H_\mathrm{S}}^2 \sim N^2$. Therefore, we note that previous claims of ``super-Heisenberg'' scaling (see various examples in \cite{Montenegro2025}), $\mathcal{F}_\theta \sim N^{2+a}$ with $a>0$, necessarily require a closing of $\Delta$ according to Eq.~\eqref{boundGS}. This highlights the importance of accounting for the ground-state preparation time as a resource, which we address below.

Motivated by these considerations, the main goals of this work are two-fold. First, we characterize the form of many-body ground states capable of reaching the bound in Eq.~\eqref{boundGS}. We show that short-range interacting systems can reach this limit only if their spectral gap closes with $N$—either algebraically at a quantum critical point or exponentially across a first-order transition with cat-like ground states \cite{Sarkar2025exp}. All-to-all interactions can achieve it even in gapped regimes. For critical (second-order) probes, we translate the optimality condition in Eq.~\eqref{boundGS} into a universal relation between critical exponents, Eq.~\eqref{eq:condcriticalexponents}. Furthermore, we identify realistic many-body models—including short-range critical systems and all-to-all interacting models (both critical and non-critical)—that are capable of approaching this bound.

Second, we optimize the corresponding ground-state preparation time through local adiabatic protocols. We therefore analyze quantum sensing within two paradigmatic settings: Ground-State Unitary Metrology (GSUM) and Ground-State Adiabatic Metrology (GSAM), see Fig.~\ref{fig:dynSchemes} below. While GSUM is directly linked to interferometric sensing~\cite{Frrot2018,Hauke2016,Montenegro2025}, in GSAM many-body interactions are present during the sensing process~\cite{Garbe2020Critical,Rams2018SuperHei}.  
Both frameworks are subject to the well known dynamical Heisenberg limit (HL)~\cite{PhysRevLett.98.090401}
\begin{equation}\label{boundGSHeis}
\mathcal{F}_\theta \leq \tau^2 \norm{H_\mathrm{S}}^2,
\end{equation}
where $\tau$ is the total protocol duration. We use units with $\hbar=1$ throughout. Ground-state preparation is itself constrained, under broad conditions, by inverse-gap lower bounds, $\tau_{\rm prep}\gtrsim 1/\Delta$ \cite{boixo2010necessary,PhysRevA.65.042308,albash2018adiabatic,lin2020near,PlanckianTime}. Thus, a ground-state probe saturating the static scaling $\mathcal{F}_\theta\sim \norm{H_\mathrm{S}}^2/\Delta^2$ is not guaranteed to reach the HL: whether an inverse-gap preparation time is additionally required depends on how the parameter is encoded, as we discuss below. In GSUM, attaining the bound will be shown to be directly tied to preparing a probe with sufficiently large $\mathrm{Var} \left[H_\mathrm{S}\right]$, since the parameter is encoded only after preparation. In GSAM, the situation is more constrained: the same adiabatic evolution must both prepare the correlated ground state and imprint the signal. We therefore ask under which physical conditions ground-state probes can attain the Heisenberg limit in each framework, and how these conditions relate to those governing optimality in the static ground-state setting.

We then consider different long- and short-range models, some of them capable of approaching both fundamental limits, as summarized in Table \ref{tab:resultsintro}. For long-range interacting systems, we again find that criticality is not a required resource to reach the ultimate precision with adiabatically-prepared ground states. This aligns with recent insights into finite-component quantum phase transitions—which are closely linked to long-range interacting spin systems—where it has been argued that adiabatic protocols without criticality \cite{Hotter2024Combining,Gietka2022Understanding,Gietka2021adiabaticcritical,Mihailescu2026Mind} can outperform previous criticality-based protocols.  For short-range interacting systems, we find that second-order critical ground states can be optimally prepared using local adiabatic dynamics. Hence, the attainability of the HL for these systems is dictated by the constraint on the critical exponents, Eq.~\eqref{eq:condcriticalexponents}. By developing an explicit protocol for the XXZ model, we show that reaching the HL in short-range ground-state probes is fundamentally possible, both in the GSUM and GSAM frameworks. This contrasts with recent claims~\cite{Shi2024SNL} of a universal shot‑noise limit for local Hamiltonians, which  only apply to time‑independent Hamiltonians or controls. 
Our results instead rely on  time‑dependent Hamiltonians to prepare the appropriate ground state.
 
This article is structured as follows: in Sec.~\ref{sec:framework}, we present the mathematical framework of ground-state metrology and dynamical metrology, along with the key quantities in our study. In Sec.~\ref{sec:limitsgroundstatemetrology}, we derive general conditions for many-body probes to reach the fundamental sensitivity bounds in both static and dynamical scenarios. In Sec.~\ref{sec:gs}, we find Hamiltonians whose ground state achieves the maximum precision in the static case, dictated by Eq.~\eqref{boundGS}. In Sec.~\ref{sec:AdiabaticPreparation}, we consider the dynamical preparation of the ground-state probes via local adiabatic protocols and analyze whether it is optimal in time. These results are then used in Secs.~\ref{sec:dynPTS} and \ref{sec:dynPWS}, where we discuss the reachability of Heisenberg scaling in dynamical metrology by including the preparation time as a resource. We conclude and state future directions of our work in Sec.~\ref{sec:Conclusions}. 

\section{Framework}\label{sec:framework}
We consider the estimation of an unknown parameter $\theta$ whose encoding in the probe state is governed by the Hamiltonian $H_\theta$.  We denote the spectrum of $H_\theta$ as $H_\theta\ket{i(\theta)}=E_i(\theta)\ket{i(\theta)}$. To quantify the precision of the task, we use the quantum Fisher information $\mathcal{F}_\theta$ as a figure of merit. For pure states of the probe $\rho_\theta=\ketbra{i(\theta)}{i(\theta)}$, $\mathcal{F}_\theta$ can be written as 
\begin{equation}\label{qfipurestates}
    \mathcal{F}_\theta = 4 \left( \innerproduct{\partial_\theta i(\theta)} - \abs{\innerproduct{i(\theta)}{\partial_\theta i(\theta)}}^2 \right).
\end{equation}
Let us now express $\mathcal{F}_\theta$ for the different considered encodings of $\theta$ onto $\rho_\theta$.
\subsection{Ground-state metrology}\label{sec:II.A}
In order to compute the QFI associated with a given ground-state probe, we expand the quantum fidelity $F$ between the $\theta$-dependent mixed states $\rho_{\theta}$ and $\rho_{\theta+\delta\theta}$ as follows:
\begin{equation}\label{Fid_fidSus}
    F(\rho_{\theta},\rho_{\theta+\delta\theta})=1-\frac{1}{2}\chi_F(\theta)\delta\theta^2+\mathcal{O}(\delta\theta^3),
\end{equation}
where $\chi_F (\theta)$ is the fidelity susceptibility. In the ground-state scenario, the mixed states reduce to the pure ground states of $H_{\theta}$ and $H_{\theta+\delta\theta}$, i.e., $\ket{0(\theta)}$ and $\ket{0(\theta+\delta\theta)}$. This notation makes the local estimation approach explicit: we sense small variations $\delta\theta$ around the known value of $\theta$. Hence, the fidelity between these two pure states simplifies to $F=\abs{\braket{0(\theta)}{0(\theta+\delta\theta)}}$. For the Hamiltonian in Eq.~\eqref{generalH}, the fidelity susceptibility takes the form\footnote{If an excited level is degenerate, matrix elements to a single eigenstate are understood as projected norms. More generally, $P_\Delta$ denotes the projector onto the lowest excited eigenspace with non-zero coupling to the relevant operator, so that $\abs{\matrixel{1}{H_\mathrm{S}}{0}}^2$ is replaced by $\norm{P_\Delta H_\mathrm{S}\ket{0}}^2$. The same convention is used for the local adiabatic condition in Eq.~\eqref{localadiabaticcondition} below. Throughout, we assume a unique finite-size ground state, or work within a fixed symmetry sector.}\cite{You2007, Zanardi2007}
\begin{equation}\label{fid_sus_def}
\chi_F(\theta)=\sum_{i>0}\frac{\abs{\matrixel{i(\theta)}{H_\mathrm{S}}{0(\theta)}}^2}{(E_i(\theta)-E_0(\theta))^2}.
\end{equation}
From this last expression, the QFI follows immediately as $\mathcal{F}_\theta=4\chi_F(\theta)$.
In the particular case of local estimation around $\theta=0$, the eigenstates and energies reduce to those of the control term of the Hamiltonian\footnote{Since we assume some degree of control over $H_\theta$ through $H_{\rm C}$, we can always shift the parameter so that the estimation is centered around a known value; we choose $\theta=0$.}. Hence, considering the spectrum of $H_\mathrm{C}$ to be $H_\mathrm{C}\ket{i}=E_i\ket{i}$, we get
\begin{equation}\label{gs_QFI}
\mathcal{F}_\theta=4\sum_{i>0}\dfrac{\abs{\matrixel{i}{H_\mathrm{S}}{0}}^2}{(E_i-E_0)^2}.
\end{equation}
\begin{figure*}[t]
    \centering
    \includegraphics[width=\textwidth, height=0.31\textheight]{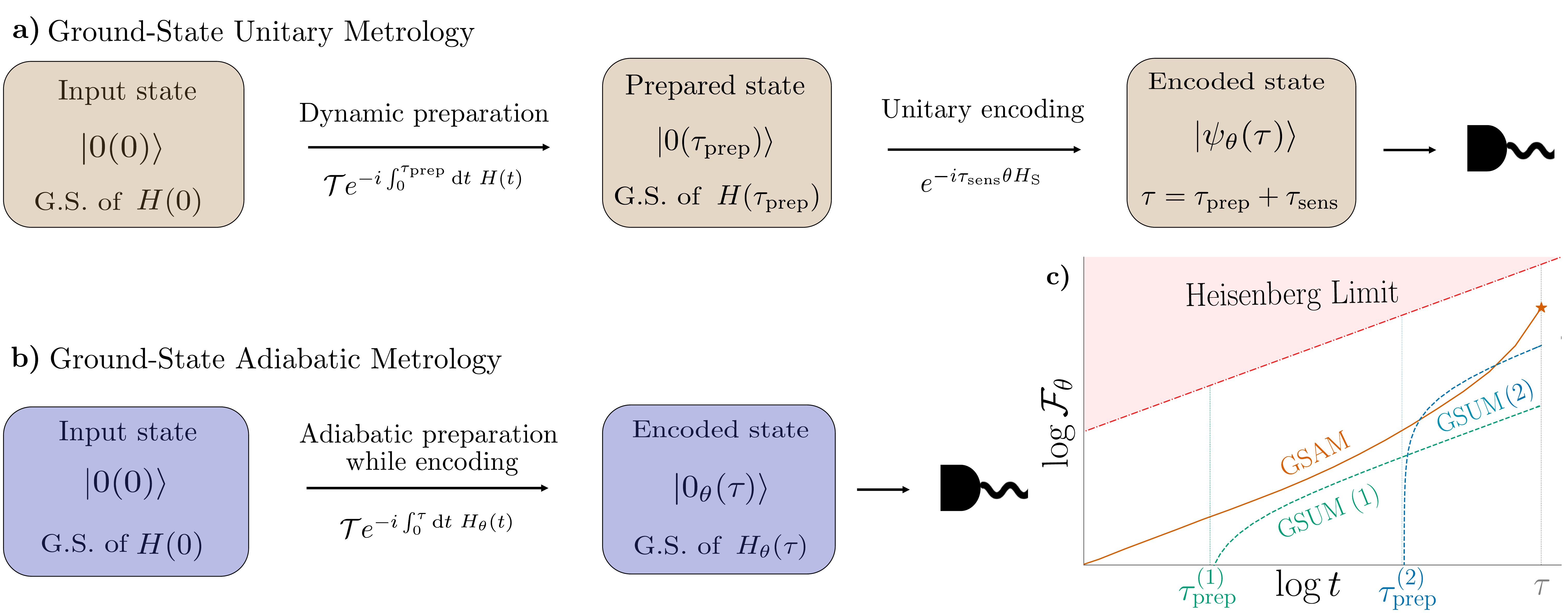}
    \caption{\textbf{Dynamical metrology frameworks considered in this work}. (a) \textit{Ground-State Unitary Metrology} (GSUM): starting from the ground state of a gapped, non-interacting, $\theta$-independent Hamiltonian, the system is evolved under $H (t) \coloneq H_{\theta = 0} (t)$ to the ground state of a target interacting Hamiltonian over a preparation time $\tau_\mathrm{prep}$; the parameter is then encoded unitarily into the probe during a sensing time $\tau_\mathrm{sens}$, with the total time of the protocol being $\tau \coloneqq \tau_{\rm prep} + \tau_{\rm sens}$. The encoded state need not be a ground state. (b) \textit{Ground-State Adiabatic Metrology} (GSAM): the same initial state is adiabatically evolved under $H_\theta(t)=\theta H_\mathrm{S}+H_\mathrm{C}(t)$ over a time $\tau$, so that the probe interacts with the signal $H_{\rm S}$ throughout the evolution. (c) \textit{Schematic QFI dynamical growth in GSUM and GSAM frameworks}: in GSUM, the QFI vanishes during the preparation stage and then exhibits the scaling of the target ground state in $N$ and $\tau$ for sufficiently long protocol times ($\tau\gg\tau_\mathrm{prep}$). If the preparation stage concludes before the target ground state is reached, the optimal scaling with $N$ is lost---see GSUM protocols (1) and (2)---while the asymptotic quadratic scaling in $\tau$ is preserved. In GSAM, the QFI accumulates during preparation, with a protocol-dependent time dependence; at the final time $\tau$ (star) it exhibits the same scaling in $N$ as in the static scenario, while the scaling in $\tau$ will depend on the optimality of the preparation stage. The red dashed curve indicates the Heisenberg bound, Eq.~\eqref{boundGSHeis}.}
    \label{fig:dynSchemes}
\end{figure*}
Following Ref.~\cite{Abiuso2025}, the QFI for ground-state metrology is upper-bounded as follows:
\begin{equation}\label{GS_bound_ineqs}
    \mathcal{F}_\theta \leq \frac{4 \sum_{i>0} \abs{\matrixel{i}{H_\mathrm{S}}{0}}^2}{\Delta^2} = \dfrac{4 \Var_{\ket{0}} \left[ H_\mathrm{S} \right]}{\Delta^2} \leq \frac{\norm{H_\mathrm{S}}^2}{\Delta^2},
\end{equation}
recovering the bound in Eq.~\eqref{boundGS}. In the last expression, we have defined the spectral gap $\Delta \coloneqq E_1 - E_0$ and the seminorm $\norm{H_\mathrm{S}}^2 \coloneqq (\lambda_M-\lambda_m)^2$, with $\lambda_M$ ($\lambda_m$) denoting the maximum (minimum) eigenvalue of $H_\mathrm{S}$. To saturate the first inequality, the state $(H_{\rm S}-\ev{H_{\rm S}})\ket{0}$ must have support only in the lowest excited eigenspace. The second inequality is saturated if and only if $\ket{0}$ has equal weight on the extremal eigenspaces of $H_{\rm S}$, that is, $\ket{0}=\frac{1}{\sqrt{2}}(\ket{\lambda_M}+\mathrm{e}^{i\phi}\ket{\lambda_m})$.
\subsection{Dynamical metrology}\label{Sec.frame_dyn}
Up to now, we only considered the static scenario in which the ground-state QFI is the main quantity of interest. Now, we additionally take into account the preparation time of this ground-state probe, so that it becomes an additional resource. To this end, we consider two different frameworks: \textit{Ground-State Unitary Metrology} (GSUM) and \textit{Ground-State Adiabatic Metrology} (GSAM), which we depict in Fig.~\ref{fig:dynSchemes}.
\subsubsection{Ground-State Unitary Metrology (GSUM)}\label{Sec.PTS_framework}
In this protocol, we first prepare, in a time $\tau_{\rm prep}$, a
$\theta$-independent ground-state probe of a target interacting Hamiltonian. Afterwards, the parameter $\theta$ is imprinted on the probe through an evolution generated by the signal Hamiltonian $H_{\rm S}$ for a sensing time $\tau_{\rm sens}$ (see Fig.~\hyperref[fig:dynSchemes]{\ref*{fig:dynSchemes}a}). The total time of the process is denoted as $\tau \coloneqq \tau_\mathrm{prep}+\tau_\mathrm{sens}$\footnote{Note that we disregard the time needed to prepare the initial ground state $\ket{0(0)}$, as it is negligible compared to the relevant dynamical timescales of the protocol.}. Hence, it is crucial to minimize the preparation time $\tau_\mathrm{prep}$, since no information on $\theta$ is encoded during the preparation stage, as illustrated in Fig.~\hyperref[fig:dynSchemes]{\ref*{fig:dynSchemes}c}. We stress that, in contrast to the preparation time, we regard $\tau_{\rm sens}$ as an unconstrained free parameter of the protocol. In an experimental implementation, $\tau_\mathrm{sens}$ would ultimately be limited by the available coherence time and other sources of noise, which we do not consider here.

For pure states, the dynamical QFI can be directly computed from Eq.~\eqref{qfipurestates} as
\begin{equation}\label{PTSqfi}
    \mathcal{F}_\theta (\tau) = 4 \tau_{\rm sens}^2 \Var_{\ket{0(\tau_\mathrm{prep})}} [H_{\rm S}],
\end{equation}
where the variance is computed with respect to the ground state of the interacting Hamiltonian. Noting that $4\tau_\mathrm{sens}^2\Var_{\ket{0(\tau_\mathrm{prep})}}[H_\mathrm{S}]\leq\tau^2(\lambda_M-\lambda_m)^2=\tau^2\norm{H_\mathrm{S}}^2$, we recover the fundamental bound in Eq.~\eqref{boundGSHeis}.
\subsubsection{Ground-State Adiabatic Metrology (GSAM)}\label{Sec.PWS_framework}
\begin{figure}[H]
    \centering
    \includegraphics[width=1.01\linewidth]{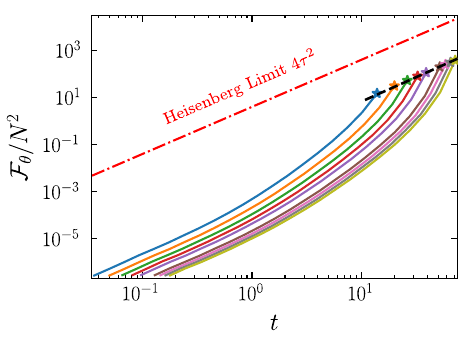}
    \caption{\textbf{Illustration of Heisenberg scaling in the GSAM framework.} Normalized dynamical QFI, $\mathcal{F}_\theta(t)/N^2$, as a function of time along the adiabatic protocol for increasing system sizes $N$ (solid lines, left to right), shown here for the XXZ model with $\gamma_z=0.95$ as a representative critical probe; this model and corresponding protocol are analyzed in detail in Secs.~\ref{Sec_XXZ_GS}, \ref{Sec.PrepTime_xxz} and~\ref{Sec:GSAM_XXZ}. The final attained QFI at the size-dependent preparation time $\tau(N)$ is indicated by a star for each trajectory. Since $\tau(N)$ grows as the inverse critical gap, $\tau\sim N$, and the static QFI grows as a power of $N$ fixed by the critical exponents, the endpoints align along a power-law envelope in $\tau$ (black dashed line; exponent $1.90$ for $\gamma_z=0.95$). As $\gamma_z\to1^-$ this envelope approaches the Heisenberg limit, Eq.~\eqref{boundGSHeis}, $\mathcal{F}_\theta\propto\tau^2N^2$ (red dashed-dotted line).}
    \label{fig:GSAM_multipleN_xxz}
\end{figure}

In contrast to GSUM, in this scenario both the preparation and the encoding happen simultaneously for the duration of the protocol $\tau$, which coincides with the sensing time, as shown in Fig.~\hyperref[fig:dynSchemes]{\ref*{fig:dynSchemes}b}. Hence, information about the unknown parameter is encoded into the probe throughout the whole protocol (see Fig.~\hyperref[fig:dynSchemes]{\ref*{fig:dynSchemes}c}). Although the detailed time dependence of $\mathcal{F}_\theta(t)$ is protocol-dependent, at the end of the evolution the dynamical QFI inherits the scaling in $N$ of the target ground-state QFI. Here, $\tau$ is not a free parameter as in GSUM: it is ultimately fixed by the preparation time of the target ground state and will typically scale with $N$. Consequently, each system size defines a single protocol duration $\tau(N)$, and the scaling of the QFI with $\tau$ can be inferred from the family of protocols at different $N$. In the best case $\mathcal{F}_\theta(\tau)/N^2$ will have Heisenberg-like scaling $\mathcal{O}(\tau^2)$---see Fig.~\ref{fig:GSAM_multipleN_xxz}.
We write the Hamiltonian governing the dynamical evolution as follows:
\begin{equation}\label{H_framework_PAS}
    H_\theta(t)=\theta H_\mathrm{S}+H_\mathrm{C} (t),
\end{equation}
with $t \in [0, \tau]$, such that $H_\theta(\tau) = \theta H_\mathrm{S} + H_\mathrm{C}$ is the target interacting Hamiltonian, where $H_\mathrm{C}\coloneq H_\mathrm{C}(\tau)$.

Throughout the GSAM analysis, the schedule constructed from the $\theta=0$ Hamiltonian is used as a reference and extended to the local-estimation regime. For each finite system size $N$, provided that the instantaneous ground state remains non-degenerate, continuity ensures the existence of a neighborhood \begin{equation}\label{eq:loc_window} 
|\theta|<\theta_{\rm loc}(N) \end{equation} 
in which the $\theta$-dependent gaps and driving matrix elements remain perturbatively close to their values at $\theta=0$. The width $\theta_{\rm loc}(N)$ may shrink as $N$ increases. For each finite $N$, the QFI is defined from the local curvature at $\theta=0$, and only afterwards is its asymptotic scaling with $N$ considered. Thus, the local limit $\delta\theta\to0$ is taken before the thermodynamic limit. If a transition is forbidden exactly at $\theta=0$ but activated by the signal, its leading $\theta$-dependent contribution must instead be evaluated explicitly, as done below in Sec.~\ref{sec:LRpreparation}.

Following Ref.~\cite{PangJordan2017}, the dynamical QFI $\mathcal{F}_\theta (\tau)$ can be computed as follows:
\begin{equation}\label{dynamicalqfieq}
    \mathcal{F}_\theta (\tau) = 4 \Var \left[ h_\theta (\tau) \right]_{\ket{0 (0)}},
\end{equation}
with $h_\theta (\tau) \coloneqq \int_0^\tau \dd{t} U_\theta^\dagger (t,0) H_\mathrm{S} U_\theta (t,0)$. Here, $U_\theta (t,0)$ is the unitary generated by the Hamiltonian $H_\theta (t)$ of Eq.~\eqref{H_framework_PAS}, while $\ket{0(0)}$ is the initial ground state of the protocol. Note that Eq.~\eqref{PTSqfi} can be recovered from Eq.~\eqref{dynamicalqfieq} by restricting to the encoding stage of GSUM, i.e. taking $\ket{0 (\tau_{\rm prep})}$ as initial state and $U_\theta (t,0) = e^{-i t \theta H_{\rm S}}$ for $t \in [0, \tau_{\rm sens}]$.

Although the control Hamiltonian can be time-dependent, the Heisenberg limit still holds, since $H_\mathrm{S}$ is time-independent. Indeed, using that $4 \Var \left[ h_\theta (\tau) \right]_{\ket{0 (0)}} \leq \left[ \lambda_\mathrm{max} (h_\theta (\tau)) - \lambda_\mathrm{min} (h_\theta (\tau)) \right]^2 \leq \tau^2 \norm{H_\mathrm{S}}^2$, we recover the bound in Eq.~\eqref{boundGSHeis}, since unitary conjugation does not change the eigenvalues of $H_\mathrm{S}$.

\section{Limits of ground-state metrology}
\label{sec:limitsgroundstatemetrology}
In this section, we provide general remarks on the saturability of the fundamental bounds of the QFI---namely, Eqs.~\eqref{boundGS} and~\eqref{boundGSHeis} for the static and dynamical scenarios, respectively---using many-body probes. Our analysis applies  to any $H_\mathrm{S}$ that can be decomposed into a sum of local terms:
\begin{equation}\label{localtermsHS}
H_\mathrm{S} = \sum_{i=1}^N o^{(i)},
\end{equation}
where $o^{(i)}$ are local operators acting on site $i$, and we restrict to signals whose seminorm scales extensively, $\norm{H_\mathrm{S}} \sim N$. We further consider control Hamiltonians containing at most two-body interacting terms,
\begin{equation}\label{localtermsHC}
H_\mathrm{C}=\sum_{i} h^{(i)}+\sum_{i<j} h^{(i,j)},
\end{equation}
where $h^{(i)}$ and $h^{(i,j)}$ act on one and two sites, respectively. We will later focus on the paradigmatic task of magnetometry, taking $H_\mathrm{S} = \sum_{i=1}^N \sigma^{(i)}_\mu$. In this case, the control terms are expanded in the Pauli basis as $h^{(i)} = \sum_\mu \alpha_\mu^{(i)} \sigma_\mu^{(i)}$ and $h^{(i,j)} = \sum_{\mu \nu} \gamma_{\mu \nu}^{(i,j)} \sigma_\mu^{(i)} \sigma_\nu^{(j)}$, with $\mu, \nu \in \{ x,y,z \}$ and $\alpha_\mu^{(i)}, \gamma_{\mu \nu}^{(i,j)} \in \mathbb{R}$.
\subsection{Optimal ground-state probes}\label{sec_general_optimal_GS}
Let us discuss which ground states of interacting Hamiltonians $H_{\rm C}$ can potentially saturate the upper bound in Eq.~\eqref{boundGS}. By identifying $\Delta^2\mathcal{F}_\theta$ as the relevant gap-aware figure of merit, let us first restate the attainability condition as follows: 
\begin{myresult}{Ground-State Optimality Condition}
\label{condition:ground-state-optimality}
We consider $N$-body Hamiltonians of the form $H_\theta = \theta H_{\rm S} + H_{\rm C}$ around $\theta = 0$, with $\norm{H_\mathrm{S}} \sim N$, where $N$ denotes the system size. We then say that a ground state is metrologically optimal when $\Delta^2 \mathcal{F}_\theta = \norm{H_{\rm S}}^2$.\\

At the level of scaling with $N$, this optimality condition can be expressed as
\begin{equation}\label{boundGSII}
    \Delta^2 \mathcal{F}_\theta \sim 4 \Var_{\ket{0}}[{H_\mathrm{S}}] \sim \norm{H_\mathrm{S}}^2.
\end{equation}
\end{myresult} 
As we said, this is a gap-aware notion of optimality: it assesses whether the QFI reaches the maximum allowed by the spectral gap $\Delta$ and the signal strength $H_{\rm S}$, rather than ranking probes solely by the growth of $\mathcal{F}_\theta$ with $N$. This notion of optimality will become operationally relevant below, where preparation time is counted as a resource and inverse-gap preparation is needed to attain the dynamical Heisenberg limit.

Note that this condition \textit{cannot} be saturated by ground states of short-range \textit{gapped} Hamiltonians ($\Delta > 0$ as $N\to\infty$). In such systems, the locality of interactions and the existence of a spectral gap imply that correlations decay exponentially with distance, $C(r) \sim \exp(-r/\xi)$, where the correlation length $\xi$ is bounded by the inverse gap as $\xi \lesssim v/\Delta$, with $v$ being the Lieb-Robinson velocity~\cite{Hastings, hastingsSpectralGapExponential2006}. This exponential decay restricts the variance of local observables to extensive scaling. Using $4\text{Var}_{\ket{0}}\left[H_\mathrm{S}\right] = 4 \sum_{i,j} \left( \ev{o^{(i)} o^{(j)}} - \ev{o^{(i)}} \ev{o^{(j)}} \right)$, for each fixed site $i$, the sum over $j$ gives a size-independent contribution, since the correlations $C(r)$ decay exponentially. Hence, $\text{Var}_{\ket{0}}\left[H_\mathrm{S}\right] \sim N$ up to negligible corrections in the thermodynamic limit. 

Given these insights, we identify two main scenarios where the optimal scaling in Eq.~\eqref{boundGSII} can be reached:
\begin{itemize}
\item \textbf{Long-Range Probes:} Ground states of systems featuring long-range interactions, which can exhibit super-extensive sensitivity in both gapped and gapless regimes. 
\item \textbf{Short-Range Probes with a Closing Gap:} Ground states of short-range interacting Hamiltonians whose gap vanishes in the thermodynamic limit, either algebraically at a second-order quantum critical point, $\Delta\sim N^{-z/d}$, which we analyze in detail below, or exponentially across a first-order transition, where the finite-size ground state is a cat-like superposition (see the remark closing Sec. \ref{Sec.TFIM_GS}).
\end{itemize}
In Sec.~\ref{sec:gs} we will construct explicit ground states of two-body interacting Hamiltonians fulfilling the \hyperref[condition:ground-state-optimality]{\textit{Ground-State Optimality Condition}} in both scenarios. Thus, both a closing gap and collective interactions enable the generation of sufficiently strong quantum correlations to achieve optimal ground-state sensitivity. 

For probes near a critical point, where $H_\mathrm{S}$ is the relevant tuning operator, the QFI and the spectral gap exhibit the characteristic fingerprints of criticality (see Appendix \ref{Appx_critical_theo} for a short introduction to the subject). Within the finite-size critical region, they scale as $\mathcal{F}_\theta \sim N^{2/(d \nu)}$ and $\Delta \sim N^{-z/d}$, respectively, where $d$ is the physical dimension and $\nu, z$ are the correlation-length and dynamical critical exponents respectively~\cite{Montenegro2025, PhysRevA.78.042106}. Moreover, in Appendix \ref{Appx_var_crit} we show that $\Var_{\ket{0}}[H_\mathrm{S}]\sim N^{2/(d\nu)-2z/d}\sim \Delta^2\mathcal{F}_\theta$ within this same region provided that $1/\nu-z>d/2$. For critical systems, the \hyperref[condition:ground-state-optimality]{\textit{Ground-State Optimality Condition}} can therefore be translated into a constraint on the critical exponents:\\
\begin{myresult}{Ground-State Optimality Condition for Critical Models}
\label{condition:ground-state-optimality-critical}
For a short-range interacting Hamiltonian undergoing a second-order quantum phase transition described by the finite-size scalings above, the ground state is metrologically optimal, at the scaling level, if and only if
\begin{align}
\label{eq:condcriticalexponents}
\nu (d+z) = 1,
\end{align}
where $d$ is the physical dimension of the system.
\end{myresult} 

Interestingly, Eq.~\eqref{eq:condcriticalexponents} represents a limiting condition rather than one attained at a conventional continuous quantum phase transition. Indeed, the singular part of the ground-state energy density scales as $e_s\sim|\theta-\theta_c|^{2-\alpha}$ with, by hyperscaling arguments, $2-\alpha=\nu(d+z)$~\cite{Continentino2004,Sachdev2011}. Continuity of $\langle H_\mathrm{S}\rangle\propto\partial e_s/\partial\theta$ across the transition requires $\alpha<1$, and hence $\nu(d+z)>1$; exact saturation, $\alpha=1$, corresponds to a jump of $\langle H_\mathrm{S}\rangle$, i.e., to a first-order transition—precisely the cat-state mechanism discussed in Sec.~\ref{sec:GSLR}. Thus, conventional critical probes are generically suboptimal: the transverse-field 1D Ising model at its critical point, for instance, has $\nu(d+z)=2$ (see Sec.~\ref{sec6A}). A family of critical points may nevertheless approach the optimal condition arbitrarily closely, as we show for the XXZ model as $\gamma_z\to1^-$. For non-critical gapped systems, we note that our examples of optimality are based on all-to-all couplings (i.e., the infinite-range limit). It remains an open question whether the gap-aware bound can be saturated by systems with other configurations (e.g., polynomially-decaying interactions).

\subsection{Optimal ground-state preparation via adiabatic dynamics}
\label{Sec:prepTime}
As anticipated in Sec.~\ref{Sec.frame_dyn}, the preparation time of the optimal ground state plays a central role in the dynamical approach to ground-state metrology. In the GSUM framework, the preparation time $\tau_{\rm prep}$ is distinct from the sensing time, while in GSAM it coincides with the total duration of the protocol, $\tau_{\rm prep} \equiv \tau$, since preparation and encoding occur simultaneously. Furthermore, it is known that $\tau_\mathrm{prep}$ satisfies $\tau_\mathrm{prep}\gtrsim 1/\Delta$ under very general conditions \cite{boixo2010necessary,PhysRevA.65.042308,albash2018adiabatic,lin2020near,PlanckianTime}. This naturally motivates the following definition:
\begin{myresult}{Preparation-Time Optimality Condition}
\label{condition:preparation-time}
The time required to prepare a target ground state is said to be optimal if it follows the profile
    \begin{equation}
        \tau_{\rm prep} \sim \dfrac{1}{\Delta},
    \end{equation}
    where $\Delta$ is the smallest relevant spectral gap along the preparation path.
\end{myresult}
In this section, we discuss how to attain this condition with adiabatic protocols. Starting from an easy-to-prepare, $\theta$-independent gapped ground state, we
perform an adiabatic evolution such that, after a time $\tau_{\rm prep}$, the
system reaches the target optimal ground state. The Hamiltonian governing the evolution is given by Eq.~\eqref{H_framework_PAS} in the particular case of $\theta=0$, since we are interested only in ground-state preparation. We consider that the time-dependence enters through a driving parameter $g(t)$, such that
\begin{equation}\label{eq15}
H(t) \coloneqq H_{\theta = 0} (t) = H_\mathrm{C} \left( g(t) \right),
\end{equation}
where the instantaneous gap is denoted by $\Delta (t) \coloneq \Delta (g(t))$. Here, we consider that $g(t)$ is tuned such that $\Delta(t=0)\gg \Delta(t=\tau_{\rm prep})$, with a monotonically decreasing spectral gap. If we denote the instantaneous spectrum as $H(t)\ket{i(t)}=E_i(t)\ket{i(t)}$, the standard adiabatic theorem typically yields a preparation time $\tau_{\rm prep} \propto 1/\Delta_\mathrm{min}^2$, where $\Delta_\mathrm{min}$ is the minimum gap along the time evolution \cite{farhi2000quantumcomputationadiabaticevolution}.
Note that this scaling of the preparation time limits the dynamical QFI to scale linearly with time (see Eq.~\eqref{boundGSHeis}).

Faster schedules can be obtained by adapting the driving speed locally along the path. We therefore consider a local adiabatic schedule in which the instantaneous rate is chosen according to the lowest excited state, or excited subspace, coupled by the driving operator $\partial_g H$
\cite{PhysRevA.65.042308,morita_mathematical_2008}:
\begin{equation}\label{localadiabaticcondition}
    \abs{\dv{g}{t}} \left[ \dfrac{\abs{\matrixel{1 (g)}{\partial_g H}{0 (g)}}}{(E_1 (g) - E_0 (g))^2}  \right] \leq \epsilon,
\end{equation}
with $\epsilon \ll 1$ controlling the local slowness of the driving. Here, $\ket{1(g)}$ denotes the lowest excited state (or eigenspace; see the discussion around Eq.~\eqref{fid_sus_def}) with a non-vanishing coupling to the ground state through $\partial_g H$. Consequently, the relevant spectral gap $E_1(g)-E_0(g)$ is defined with respect to this state, rather than the global first excited state. 

The preparation time $\tau_{\rm prep}$ is then obtained by integrating Eq.~\eqref{localadiabaticcondition} as follows:
\begin{equation}\label{preptimeintegral}
    \tau_{\rm prep} \gtrsim \dfrac{1}{\epsilon} \int_{g(0)}^{g(\tau_{\rm prep})} \abs{\dd{g}} \dfrac{\abs{\matrixel{1 (g)}{\partial_g H}{0 (g)}}}{(E_1 (g) - E_0 (g))^2}. 
\end{equation}
This time constitutes a lower estimate relative to an analogous local criterion that includes all allowed transitions in the full many-body spectrum. Therefore, when this reduced criterion is used, the validity of the resulting schedule should be checked a posteriori, for instance by computing the overlap with the target state numerically along the protocol, as we do for the two-mode Bose-Hubbard and XXZ models in Figs.~\ref{fig:BH_GSUM_prep_time}(b) and~\ref{fig:XXZ_tprep}(b), respectively.

Let us now study ground-state preparation in both long- and short-range interacting probes.
\begin{figure}[H]
    \centering
    \includegraphics[width=\linewidth]{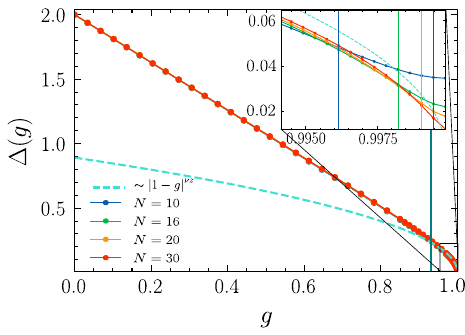}
    \caption{\textbf{Spectral gap of the XXZ model as a function of the adiabatic tuning parameter $g$ for different $N$}. We identify three regions along the driving path. Region (I) lies far from criticality and exhibits an $N$-independent spectral gap, with the crossover to region (II) indicated by a teal solid vertical line. Region (II) spans until finite-size effects emerge---that is, until the critical window $\abs{g-g_c}\lesssim N^{-1/(d\nu)}$ starts---and exhibits a universal $N$-independent spectral gap for sufficiently large values of $N$, i.e., $\Delta\sim\abs{g-g_c}^{\nu z}$. Region (III) coincides with the critical window, where the spectral gap has an explicit dependence on $N$. The critical window depends on $N$ and is indicated by colored solid vertical lines in the inset plot. Numerical data were obtained for the XXZ model with $\gamma_z=0.92$, for which the critical exponents read $z=1$ and $\nu \approx 0.52$. The same qualitative behavior persists as $\gamma_z \to 1^-$.}
    \label{fig:gap_g_xxz}
\end{figure}
\subsubsection{Infinite-range interacting systems}\label{sec:LRpreparation}
We first consider the all-to-all interacting Hamiltonians studied in this work, which we describe as ensembles of $N$ spin-$1/2$ particles in terms of the collective spin operators $J_\alpha = \frac{1}{2} \sum_{i=1}^N \sigma_\alpha^{(i)}$ and $J_\alpha^2 = \frac{1}{4} \sum_{i,j=1}^N \sigma_\alpha^{(i)} \sigma_\alpha^{(j)}$, with $\alpha \in \{ x,y,z \}$. These models can attain the optimal scaling in Eq.~\eqref{boundGSII} even in gapped regimes. In contrast to short-range critical systems, however, general arguments for ground-state and preparation time optimality are more restricted, since the universal finite-size scaling relations used above do not apply. Although ground-state optimality therefore remains generally model-dependent, some general statements about optimal preparation can still be established for a restricted class of collective Hamiltonians, as we show below.

Here, we focus on a restricted class of collective models in which the ground-state QFI is asymptotically dominated by an isolated low-energy doublet $\ket{0(g)}$ and $\ket{1(g)}$, with $\mathcal F_\theta\sim 4|\matrixel{1(g)}{H_{\rm S}}{0(g)}|^2/\Delta(g)^2$ and $|\matrixel{1(g)}{H_{\rm S}}{0(g)}| \sim\norm{H_{\rm S}}\sim N$, where $\Delta(g):=E_1(g)-E_0(g)$. In addition, for the models considered here, the preparation drive does not couple the two states of this doublet at $\theta=0$, so that $\matrixel{1(g)}{\partial_gH}{0(g)}=0$, whereas the signal Hamiltonian does, $\matrixel{1(g)}{H_{\rm S}}{0(g)}\neq0$.

At $\theta=0$, the doublet transition is therefore irrelevant to the ground-state preparation problem considered in GSUM, although higher excited states may still be coupled by $\partial_g H$ (see Sec.~\ref{sec:optimal_prep_LR}). This conclusion cannot be directly extended to GSAM, where the weak but nonzero signal term is present throughout the evolution and modifies both the instantaneous eigenstates and the relevant driving matrix elements. In particular, the signal activates the transition within the doublet already at first order in $\theta$. This signal-activated transition sets the leading scaling of the total GSAM preparation time provided that the remaining coupled transitions contribute only subleading times, as assessed separately for each model below. We now determine the preparation cost of this signal-activated transition within the local-estimation window. 

Within the relevant subspace (i.e., neglecting contributions from states outside $\{ \ket{0 (g)}, \ket{1 (g)} \}$), first-order perturbation theory in $\theta$ gives
\begin{align}
    \ket{0_\theta (g)} &= \ket{0 (g)} - \dfrac{\theta}{\Delta (g)} \matrixel{1 (g)}{H_{\rm S}}{0 (g)} \ket{1 (g)} + \mathcal{O} (\theta^2),  \\
    \ket{1_\theta (g)} &= \ket{1 (g)} + \dfrac{\theta}{\Delta (g)} \matrixel{1 (g)}{H_{\rm S}}{0 (g)}^* \ket{0 (g)} + \mathcal{O} (\theta^2).
\end{align}
Consequently,
\begin{equation}\label{eq:LR_matrix_element}
\abs{\matrixel{1_\theta (g)}{\partial_g H}{0_\theta (g)}} = \dfrac{\abs{\theta} \abs{\matrixel{1 (g)}{H_{\rm S}}{0 (g)}}}{\Delta (g)} \abs{\partial_g \Delta} + \mathcal{O} (\theta^2),
\end{equation}
where we used the Hellmann-Feynman theorem, $\matrixel{n (g)}{\partial_g H}{n (g)} = \partial_g E_n (g)$, so that $\matrixel{0(g)}{\partial_g H}{0(g)} - \matrixel{1(g)}{\partial_g H}{1(g)} = -\partial_g \Delta$.

Before computing $\tau_{\rm prep}$, let us make explicit the local-estimation window of Eq.~\eqref{eq:loc_window}. Since the QFI describes the local distinguishability of neighboring ground states (see Sec.~\ref{sec:II.A}), the perturbative regime is characterized by $\abs{\theta} \sqrt{\mathcal{F}_\theta} \ll 1$. For the class of models considered here, this condition translates to
\begin{equation}\label{localestimationwindow}
    \dfrac{\abs{\theta} \abs{\matrixel{1 (g)}{H_{\rm S}}{0 (g)}}}{\Delta (g)} \ll 1.
\end{equation}
That is, $\theta_{\rm loc} (N) \sim \min_g \Delta(g) / \abs{\matrixel{1(g)}{H_{\rm S}}{0(g)}}$. Here, $\theta$ is fixed throughout a given protocol, and Eq.~\eqref{localestimationwindow} is required to hold for all values of $g$ along the preparation path. Combining Eq.~\eqref{eq:LR_matrix_element} with Eq.~\eqref{localestimationwindow}, the signal-activated matrix element is bounded along the whole path as
\begin{equation}
    \abs{\matrixel{1_\theta (g)}{\partial_g H}{0_\theta (g)}} \leq \abs{\partial_g \Delta}.
\end{equation}
A local schedule controlling this signal-activated transition can therefore be constructed by imposing
\begin{equation}
\abs{\dv{g}{t}} \dfrac{\abs{\partial_g\Delta(g)}}{\Delta(g)^2} \leq \epsilon.
\end{equation}
Choosing the fastest schedule compatible with this condition, i.e., saturating the inequality, we obtain the preparation time:
\begin{equation}
    \tau_{\rm prep} = \dfrac{1}{\epsilon} \int_{g (0)}^{g (\tau_{\rm prep})} \abs{\dd{g}} \dfrac{\abs{\partial_g \Delta}}{\Delta (g)^2} = \dfrac{1}{\epsilon} \left( \dfrac{1}{\Delta (\tau)} - \dfrac{1}{\Delta (0)} \right).
\end{equation}
Again, since we are in the GSAM setting, $\tau_{\rm prep}$ should not be interpreted as a free parameter. Rather, for a fixed system size and target ground state, the final gap $\Delta (\tau)$ is fixed, and so is the corresponding preparation time. For the protocols of interest, where $\Delta (0) \gg \Delta (\tau)$, the preparation time scales as $\tau = \tau_{\rm prep} \sim 1 /  \Delta (\tau)$, hence satisfying the \hyperref[condition:preparation-time]{\textit{Preparation-Time Optimality Condition}}.

\subsubsection{Short-range second-order critical Hamiltonians}\label{Sec:tprep_crit}
In this section, we show that an optimal preparation time profile can be attained (asymptotically in system size) for ground-state preparation of a short-range second-order critical many-body Hamiltonian. To this end, we consider the previously introduced local adiabatic driving.

Before evaluating the preparation time, two preliminary remarks are in order. First, for Hamiltonians of the form in Eq.~\eqref{eq15}, $g$ is the sole relevant tuning parameter for the phase transition. Second, while the local adiabatic condition in Eq.~\eqref{localadiabaticcondition} strictly depends on the gap to the first state with non-vanishing coupling to the ground state through $\partial_g H$---which may differ from the true spectral gap to the absolute lowest excited state---both gaps scale identically in critical systems. This occurs because all low-lying excitations are governed by the same universal energy scale, $\xi^{-z}$, with $\xi$ the correlation length (see Appendix~\ref{Appx_critical_theo}) \cite{Sachdev2011}. Consequently, for the scaling analysis that follows, we can safely evaluate the preparation time scaling using the standard global spectral gap, $\Delta$.

To integrate Eq.~\eqref{preptimeintegral}, we divide the preparation path into three regions, defined in terms of $s \coloneq |g-g_c|$ (see Appendix~\ref{Appx_critical_theo} for more details), which quantifies how close the system is to criticality. These regions are illustrated in Fig.~\ref{fig:gap_g_xxz} for the particular case of the XXZ model, which will be analyzed later in the manuscript. 

In region (I), which corresponds to a gapped phase ($\Delta\sim\mathcal{O}(1)$), we have $s\sim\mathcal{O}(1)$. Using arguments based on the Lieb-Robinson bound, analogous to those employed in Sec.~\ref{sec_general_optimal_GS}, we bound the matrix element as $\abs{\matrixel{1{(g)}}{\partial_g H}{0 {(g)}}}^2\leq\sum_{i>0}\abs{\matrixel{i{(g)}}{\partial_g H}{0{(g)}}}^2=\mathrm{Var}_{\ket{0}}[\partial_g H]\sim N$. Thus, the contribution of region (I) to the ground-state preparation time can be integrated from Eq.~\eqref{preptimeintegral} to yield the worst-case estimate, $\tau_\mathrm{prep}^{\rm (I)}\sim N^{1/2}$.

Region (II) lies immediately before the critical window (defined in Eq.~\eqref{crit_window_def}), with $s$ fulfilling $1\gg s> N^{-1/(d \nu)}$. In this regime, the spectral gap follows the universal scaling $\Delta\sim s^{\nu z}$ for sufficiently large $N$. Additionally, $\abs{\matrixel{1{(g)}}{\partial_g H}{0{(g)}}}^2/\Delta^2\leq \chi_F(g)$, with equality whenever $\partial_g H$ only couples the ground and first excited states (see Eq.~\eqref{fid_sus_def}). Within this region, the susceptibility scales as $\chi_F\sim N s^{d\nu-2}$ \cite{Gu2009, CamposVenuti2007}. Hence, we find the following conservative estimate to $\tau_{\rm prep}$:
\begin{equation}\label{preptime_reg_II}
\begin{split}
    \tau_\mathrm{prep}^{\rm (II)}&\sim N^{1/2}\int_{N^{-1/(d\nu)}}^{\mathcal{O}(1)}\dd s\hspace{0.1cm}s^{\nu(d/2-z)-1}\\\\
    &\sim\begin{cases}
        N^{z/d},\hspace{1.4cm}d<2z\\
        N^{1/2}\log{N},\hspace{0.55cm}d=2z\\
        N^{1/2},\hspace{1.4cm}d>2z
    \end{cases}
\end{split}
\end{equation}
That is, $\tau_\mathrm{prep}^{\rm (II)}\sim N^{z/d}\sim1/\Delta_c$ if $d<2z$.

In region (III), defined by the condition $N^{-1/(d\nu)}\gtrsim s\geq 0$, finite-size effects emerge, since this region corresponds to the critical window. Hence, $\Delta\sim N^{-z/d}= \Delta_c$ and $\abs{\matrixel{1{(g)}}{\partial_g H}{0{(g)}}}^2/\Delta^2\sim \chi_F(g)\sim N^{2/(d\nu)}$ \cite{Gu2009, GU2010FidelityQPT}, using the same argument that all low-lying excitations in the critical window are governed by the same energy scale. Therefore,
\begin{equation}
    \tau_\mathrm{prep}^{\rm (III)}\sim N^{(1/\nu+z)/d} \int_{0}^{N^{-1/(d\nu)}}\dd s=N^{z/d}\sim 1/\Delta_c.
\end{equation}
From the previous computations, it immediately follows that for sufficiently large values of $N$ and provided that $d\leq2z$ (with logarithmic corrections at equality),
\begin{equation}
    \tau_\mathrm{prep}=\tau_\mathrm{prep}^{\rm (I)}+\tau_\mathrm{prep}^{\rm (II)}+\tau_\mathrm{prep}^{\rm (III)}\sim N^{z/d}\sim1/\Delta_c,
\end{equation}
where we neglected the contribution from region (I) since $\tau_{\rm prep}^{(I)} \sim N^{1/2}$ is subleading to the $N^{z/d}$ scaling of regions (II) and (III) as $d \leq 2z$.

This derivation strongly suggests that the \hyperref[condition:preparation-time]{\textit{Preparation-Time Optimality Condition}} is satisfied for critical systems following the local adiabatic protocol. We will later confirm this  for several many-body systems.

Because the Lieb--Robinson bound \cite{Lieb1972} implies $z\ge1$ for short-range critical systems, the condition $2z\ge d$ naturally holds in one and two dimensions, with a logarithmic correction when $d=2z$. A sharper estimate can be obtained for translationally invariant systems in which a symmetry-preserving drive couples the ground state predominantly to pairs of quasiparticles with opposite momenta, as assumed in Appendix~\ref{App:matrels_crit_models}. For $d\nu<2$, inserting the corresponding matrix-element scaling into Eq.~\eqref{preptimeintegral} yields $\tau_{\mathrm{prep}}\sim1/\Delta_c$ within the lowest-transition approximation, including in higher dimensions.

Although this result was derived for the Hamiltonian in Eq.~\eqref{eq15}, it can also be extended to Hamiltonians in the GSAM framework (see Eq.~\eqref{H_framework_PAS}). In that case, the target Hamiltonian sits at a critical point in the two-parameter space $(\theta,g)$, with $g$ driven in time and $\theta$ corresponding to the fixed signal.

First, since these two parameters are perturbations of the same fixed point, they inherit its dynamical exponent: along either direction, the gap closes as $\Delta\sim\xi^{-z}$ with a single value for the dynamical critical exponent $z\coloneq z_\theta=z_g$.

Second, we work in the local estimation paradigm and restrict the driving endpoint to be located within the critical window, where static optimality is found. In this regime, the estimation window fulfills $\abs{w_\theta}\coloneqq\abs{\theta-\theta_c}N^{1/(d\nu_\theta)}\ll1$, i.e., $\theta_{\rm loc} (N) \sim N^{-1/(d \nu_\theta)}$. Within the last two regions, the spectral gap is assumed to obey the standard two-parameter finite-size scaling ansatz $\Delta(\theta,g)=N^{-z/d}f((g-g_c)N^{1/(d\nu_g)}, w_\theta)$~\cite{Fisher1972, Brzin1985, Binder2010}, with $f$ regular and non-zero over the preparation path. To leading order, this can be approximated as $\Delta(\theta,g)= N^{-z/d}[f((g-g_c)N^{1/(d\nu_g)}, 0)+\mathcal{O}(w_\theta)]$. Assuming that the driving matrix elements entering Eq.~\eqref{preptimeintegral} likewise admit a regular expansion in $w_\theta$, the leading size dependence coincides with that obtained at $\theta=\theta_c=0$, and the preceding preparation-time scaling results remain applicable within the local-estimation window.

As a final remark, we note that in general the matrix elements in regions (II) and (III) involve $\nu\coloneqq \nu_g \neq\nu_\theta$, as $\partial_g H$ is independent of $\theta$ and the instantaneous eigenvalues carry a subleading $\mathcal{O}(w_\theta)$ dependence. The equality $\nu_g=\nu_\theta$ holds whenever $H_\mathrm{S}$ and $\partial_gH$ project onto the same relevant direction. We show in Sec.~\ref{sec:AdiabaticPreparationShortRange} that this is the case for the TFIM and the XXZ model.

\subsection{Ground-State Unitary Metrology: Reaching Heisenberg scaling}\label{GSUMframework}
Having discussed optimal ground-state preparation,  we now discuss the attainability of Heisenberg scaling within the GSUM framework introduced in Sec.~\ref{Sec.PTS_framework}. From Eq.~\eqref{PTSqfi} it readily follows that:
\begin{equation}\label{QFI_PTS}
    \mathcal{F}_\theta(\tau)=4(\tau-\tau_\mathrm{prep})^2\mathrm{Var}_{\ket{0}}[H_\mathrm{S}].
\end{equation}
The appearance of $\mathrm{Var}_{\ket{0}}[H_\mathrm{S}]$ enables us to directly connect the dynamical QFI to the static ground-state QFI. In particular, if the probe satisfies the \hyperref[condition:ground-state-optimality]{\textit{Ground-State Optimality Condition}}, so that $4\mathrm{Var}_{\ket{0}}[H_\mathrm{S}] \sim N^2$, then Eq.~\eqref{QFI_PTS} implies $\mathcal{F}_\theta(\tau) \sim \tau^2 N^2$ in the limit $\tau_\mathrm{sens} \gg \tau_\mathrm{prep}$. Therefore, the bound in Eq.~\eqref{boundGSHeis} is asymptotically saturated. The converse is not true in general, since the variance upper bounds the ground-state QFI; see Eq.~\eqref{GS_bound_ineqs}. However, for short-range interacting systems close to a second-order quantum phase transition, we show in Appendix~\ref{Appx_var_crit} that quadratic scaling of $\mathrm{Var}_{\ket{0}}[H_\mathrm{S}]$ requires Eq.~\eqref{eq:condcriticalexponents} to hold. Hence, for such systems, optimal ground-state scaling implies optimal GSUM scaling, and vice versa.

In summary, we identify the following close connections between static and dynamical optimality: 
\begin{myresult}{Ground-State Unitary Metrology Optimality Condition}
\label{condition:GSUM}
Both static and dynamical optimal ground states are intimately connected: 
\begin{itemize}[leftmargin=*]
    \item For a general ground state, the \hyperref[condition:ground-state-optimality]{\textit{Ground-State Optimality Condition}} ($\Delta^2\mathcal{F}_\theta\sim \norm{H_\mathrm{S}}^2 $) implies Heisenberg scaling in dynamical GSUM   ($\mathcal{F}_\theta\sim \tau^2\norm{H_\mathrm{S}}^2 $) asymptotically in $\tau$.
    \item For short-range critical Hamiltonians, the relation goes in both directions: 
     $\nu (d+z) = 1 \iff \mathcal{F}_\theta\sim \tau^2\norm{H_\mathrm{S}}^2$. 
\end{itemize}
\end{myresult}
From these considerations it follows that, regardless of the duration of $\tau_\mathrm{prep}$, the system can always interact with the signal for a sufficiently long time to reach dynamical Heisenberg scaling. However, in practical terms, it is desirable to engineer a preparation time satisfying $\tau_\mathrm{prep}\sim1/\Delta$ (see Sec.~\ref{Sec:prepTime}), thereby making the asymptotic requirement $\tau_\mathrm{sens}\gg\tau_\mathrm{prep}$ less stringent.

In Sec.~\ref{sec:dynPTS}, we consider physical models satisfying this optimality condition for all-to-all and short-range critical interacting systems. 
\subsection{Ground-State Adiabatic Metrology: Reaching Heisenberg scaling}
In the GSAM framework, we consider the Hamiltonian $H_\theta (t)$ in Eq.~\eqref{H_framework_PAS} to be driven by a time-dependent parameter $g(t)$ for $t \in [0, \tau]$, with an instantaneous gap $\Delta(t)$. As discussed in Sec. \ref{Sec:prepTime}, the naive application of the adiabatic theorem yields a total time $\tau\propto1/\Delta(\tau)^2$, where $\Delta(\tau)$ is the minimal gap throughout the protocol. From the relation between Eqs.~\eqref{boundGS} and~\eqref{boundGSHeis}, it immediately follows that $\mathcal{F}_\theta\sim N^2 \tau$, thereby failing to reach the Heisenberg bound $\mathcal{F}_\theta \sim N^2 \tau^2$. Alternatively, the consideration of optimized adiabatic protocols yields the following condition:\\

\begin{myresult}{Ground-State Adiabatic Metrology Optimality Condition}
\label{condition:GSAM}
For a given ground state, if the \hyperref[condition:ground-state-optimality]{\textit{Ground-State Optimality Condition}} ($\Delta^2\mathcal{F}_\theta\sim \norm{H_\mathrm{S}}^2 $) and the \hyperref[condition:preparation-time]{\textit{Preparation-Time Optimality Condition}} ($\tau \sim 1/\Delta$) are both satisfied, then Heisenberg scaling follows: $\mathcal{F}_\theta\sim\tau^2\norm{H_\mathrm{S}}^2$.
\end{myresult}
Note that this condition follows immediately from Eq.~\eqref{boundGS}. This GSAM optimality condition guarantees Heisenberg scaling. The local adiabatic criterion in Eq.~(\ref{localadiabaticcondition}) provides one possible route to constructing a suitable protocol. In Sec.~\ref{sec:dynPWS}, we identify both all-to-all and short-range critical systems satisfying these two conditions.

As a last remark, we note that the stringency of the \hyperref[condition:preparation-time]{\textit{Preparation-Time Optimality Condition}} motivates hybrid protocols consisting of two steps. Beginning from an easy-to-prepare, $\theta$-independent state, a GSAM protocol is run to prepare the optimal ground-state probe. After this stage, the parameter is further encoded into the optimal probe via a unitary operation, as in the GSUM framework. Such protocols potentially combine the strengths of both frameworks: the asymptotic attainability of Heisenberg scaling in GSUM, and the accumulation of a non-zero QFI during ground-state preparation in GSAM. The actual usefulness of such hybrid strategies is left as an open question.
\section{Many-body ground-state probes}\label{sec:gs}
\begin{figure}[t]
    \centering
    \includegraphics[width=\linewidth]{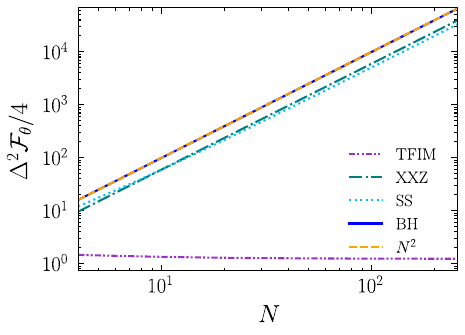}
  \caption{\textbf{Attainability of the ground-state bound in Eq.~\eqref{boundGS}.} Gap-aware figure of merit $\Delta^2\mathcal{F}_\theta/4$, to be compared with the upper bound $\norm{H_\mathrm{S}}^2/4=N^2$ (dashed line), as a function of even $N$ at $\theta=0$ for the four models considered: the transverse-field 1D Ising model, Eq.~\eqref{H_ising_displaced}; the external-field XXZ model, Eq.~\eqref{XXZ_H}, with $\gamma_z=0.99$; the spin-squeezing Hamiltonian, Eq.~\eqref{H_spin_squeez}; and the two-mode Bose-Hubbard model, Eq.~\eqref{H_BH}, with $U=-5$ and $K=10^{-3}$. The Ising model stays of order one and remains suboptimal; the spin-squeezing model reaches $N^2/2$; the XXZ model follows $(c/4)N^{2-\delta}$ with $c\approx2.45$ and $\delta=0.045$ (Table~\ref{StaticTable}); and the Bose-Hubbard model reaches $N^2$ as $\alpha\to0$.}
    \label{fig:gs_plots_all}
\end{figure}
In this section, we analyze the ground-state QFI of various spin probes for magnetometry. That is, we assume the signal and control Hamiltonians in Eqs.~\eqref{localtermsHS} and \eqref{localtermsHC}, respectively. We consider both collective (long-range) interacting systems and short-range ones. Motivated by the fundamental bound in Eq.~\eqref{boundGS}, we use $\Delta^2\mathcal{F}_\theta/4$ as the main figure of merit, as we consider the norm of the signal Hamiltonian to be $\norm{H_{\rm S}}^2 = 4N^2$. Essentially, we search for many-body probes that satisfy the \hyperref[condition:ground-state-optimality]{\textit{Ground-State Optimality Condition}}. Table~\ref{StaticTable} summarizes the obtained results for the different many-body Hamiltonians. 

\subsection{All-to-all interacting systems}\label{sec:GSLR}
We first consider collective spin models. These have been extensively considered in standard quantum metrology \cite{Ma2011,Schulte2020}, also including experiments \cite{Pezz2018, Riedel2010,Li2026}. In particular, we consider a spin-squeezing Hamiltonian, whose gap is independent of $N$, and the two-mode Bose-Hubbard model, whose gap closes super-exponentially with $N$. We will show that both models satisfy the \hyperref[condition:ground-state-optimality]{\textit{Ground-State Optimality Condition}}. 
\subsubsection{Spin-squeezing Hamiltonian}\label{subsec:SS}
As our first long-range interacting model, we consider the following spin-squeezing Hamiltonian:
\begin{equation}\label{H_spin_squeez}
    H_\theta=gJ_z^2-\lambda J^2+2\theta J_x,
\end{equation}
where $g>0$ and $\lambda > 0$. We focus on estimating a small magnetic field along the $x$-axis around $\theta = 0$. The control term encodes all-to-all interactions, and its eigenstates are Dicke states \cite{Dicke1954}, i.e., simultaneous eigenstates of $\{J_z,J^2\}$, with $J^2 \coloneqq J_x^2 + J_y^2 + J_z^2$ being the total angular momentum collective operator.

For even values of $N$, such that ground-state degeneracies are avoided, the spectrum of the control term in Eq.~\eqref{H_spin_squeez} is $E=gm_z^2-\lambda j(j+1)$, so that the spectral gap is given by 
\begin{equation}\label{gap_SpinSqz}
    \Delta=g    
\end{equation}
with ground state $\ket{j=N/2, m_z=0}$ and first excited states $\ket{j=N/2, m_z=\pm 1}$. Using $J_x=\frac{1}{2}(J_++J_-)$, the only non-vanishing matrix elements are $\matrixel{N/2, \pm 1}{H_\mathrm{S}}{N/2, 0}=\sqrt{N/2(N/2+1)}$. Interestingly, the ground state of this model saturates the first inequality in Eq.~\eqref{GS_bound_ineqs}, with the figure of merit being
\begin{equation}\label{gs_qfi_SS}
    \Delta^2\mathcal{F}_\theta=4\mathrm{Var}_{\ket{0}}[H_\mathrm{S}]=4N(N/2+1)\approx 2N^2,
\end{equation}
where the last approximation holds for large values of $N$. Hence, it follows that the fundamental bound $\norm{H_\mathrm{S}}^2=4N^2$ is saturated in scaling with a prefactor of $1/2$. The behavior of $\Delta^2\mathcal{F}_\theta$ is shown in Fig.~\ref{fig:gs_plots_all}.
\subsubsection{Two-mode Bose-Hubbard model}\label{Sec: BH_GS}
The next model we consider is the two-mode Bose-Hubbard Hamiltonian, which describes a system with a fixed number $N>1$ of bosons, with $N_L$ and $N_R$ of them in the $L$ and $R$ mode, respectively. Hence, we fix $N=N_L+N_R$. The Hamiltonian describing this model reads
\begin{equation}\label{H_BH}
    H_\theta=-K(a^\dagger_La_R+a^\dagger_Ra_L)+\frac{U}{2}\sum_{i\in\{L,R\}}n_i(n_i-1)+\theta(n_L-n_R),
\end{equation}
where $K$ is the hopping amplitude between the two modes, $U$ is the boson-boson interaction parameter, and $a_i^\dagger$ and $a_i$ are the bosonic creation and annihilation operators for the $i$-th mode, respectively. We have also defined $n_i\coloneq a_i^\dagger a_i$ as the number operator with eigenvalue $N_i$. For the sensing task, we focus on attractive interactions, $U<0$, with $K>0$.

We define the dimensionless parameters $\Lambda:=K/|U|>0$ and $\alpha:=\Lambda/N$. The condition $\alpha\ll1$ characterizes the strongly interaction-dominated regime. For the perturbative analysis below, we will consider $\Lambda\ll1$ at fixed $N\geq 2$ and $U<0$. To study the scaling with particle number, we also consider $N\gg\Lambda$ at fixed $K$ and $U$. Both limits imply $\alpha\to0$.

At $\theta=0$, the ground and first excited states approach the symmetric and antisymmetric macroscopic superposition of a pair of well-distinguished states, $\ket{0}=\frac{1}{\sqrt{2}}(\ket{\lambda_M}{+}\ket{\lambda_m})$ and $\ket{1}=\frac{1}{\sqrt{2}}(\ket{\lambda_M}{-}\ket{\lambda_m})$ as $\alpha\to0$ \cite{Spekkens1999, PhysRevA.57.1208, JuliDaz2010, Mazzarella2011_cat, DellAnna2012, Mazzarella2013}. This model can be experimentally realized by means of a Bose-Einstein condensate in a double-well potential. However, preparing these eigenstates via cooling techniques is known to be challenging, which has motivated alternative schemes to overcome this issue \cite{Srensen2001, Carr2010, Bychek2018, Vanhaele2021}.

\begin{figure}
  \centering
  \subfloat[]{%
    \begin{tikzpicture}
      \node[inner sep=0] (img) {\includegraphics[width=0.5\linewidth]{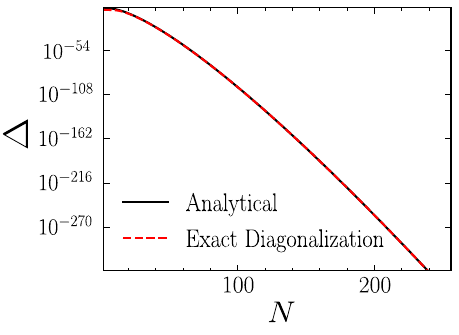}};
      \node[anchor=north west, xshift=0.4cm, yshift=0.5cm] at (img.north west) {\scriptsize\bfseries \textbf{(a)}};
    \end{tikzpicture}%
    \label{fig:BH_gap_anal_QFI}}
  \subfloat[]{%
    \begin{tikzpicture}
      \node[inner sep=0] (img) {\includegraphics[width=0.5\linewidth]{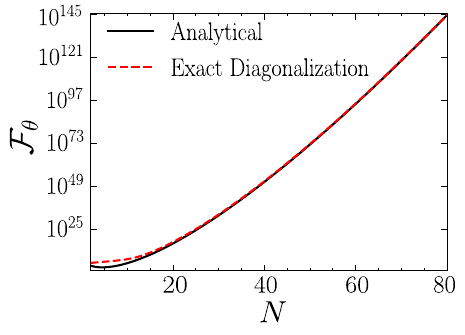}};
      \node[anchor=north west, xshift=0.4cm, yshift=0.5cm] at (img.north west) {\scriptsize\bfseries \textbf{(b)}};
    \end{tikzpicture}%
    \label{fig:BH_QFI_anal_ED}}%
  \vspace{-10pt}
    
  \caption{\textbf{Analytical and numerical comparison for the two-mode Bose-Hubbard model.} Exact diagonalization (ED) results versus the analytical expressions for the two-mode Bose-Hubbard Hamiltonian in Eq.~\eqref{H_BH}: (a) spectral gap in Eq.~\eqref{BHgap}; and (b) ground-state QFI in Eq.~\eqref{BH_QFI}. Numerical data were obtained with parameters $U=-5$ and $K=10^{-3}$, using the Python library \textit{mpmath} \cite{mpmath} for high precision values.}
  \label{fig:BH_anal_ED_plots}
\end{figure}

Using Schwinger's oscillator model of angular momentum \cite{Sakurai_Napolitano_2020}, as shown in Appendix~\ref{App_BH_gs}, the Hamiltonian in Eq.~\eqref{H_BH} can be rewritten in terms of collective spin-$N/2$ operators as follows:
\begin{equation}
H_\theta=-2KJ_x+U(J_z^2+J^2)+2\theta J_z.
\end{equation}
For this model, we also compute the first non-zero contribution to the spectral gap at $N$-th order in perturbation theory with respect to the parameter $\Lambda$---see also \cite{Mazzarella2011_cat}, where the spectral gap is computed using a semi-classical approach\footnote{In general, the results obtained with perturbation theory do not match those from semi-classical methods. However, this is the case for anisotropic spin systems whose Hamiltonian is a linear-quadratic combination of spin operators \cite{Garanin1991, Ulyanov99, Owerre2015, Krizanac2017}.} valid for $|U|N\rightarrow\infty$. At fixed $N\geq2$, the gap has the following expansion as $\Lambda\to0$:
\begin{equation}\label{BHgap}
    \Delta=2\abs{U}\dfrac{N}{(N-1)!}\Lambda^{N}+\mathcal{O}\left(\Lambda^{N+2}\right).
\end{equation}
Similarly, the ground-state QFI can be expressed as an expansion in $\Lambda$,
\begin{equation}\label{BH_QFI}
    \mathcal{F}_\theta=\abs{U}^{-2}\left[(N-1)!\right]^2\Lambda^{-2N}+\mathcal{O}(\Lambda^{-2N+2}).
\end{equation}
Although Eqs.~\eqref{BHgap} and \eqref{BH_QFI} are derived as expansions in $\Lambda$ at fixed $N$, their leading terms also give the large-$N$ asymptotics at fixed $K$ and $U$, with relative corrections vanishing as $N\to\infty$ (see Appendix \ref{App_BH_gs}). Consequently, in the limit $\alpha\to0$ the QFI grows super-exponentially with $N$, while the spectral gap closes with the same scaling. Fig.~\ref{fig:BH_anal_ED_plots} compares these leading expressions with exact diagonalization for increasing $N$, using parameters satisfying $\Lambda\ll1$. This behavior mirrors the exponential scaling of the QFI reported recently in Ref.~\cite{Sarkar2025exp} for Hamiltonians presenting first-order quantum phase transitions, where the gap likewise closes exponentially with system size.

From Eqs.~\eqref{BHgap} and~\eqref{BH_QFI} it follows that, at fixed $N$ the ground-state bound in Eq.~\eqref{boundGS} is asymptotically saturated as $\Lambda\to0$, that is,
\begin{equation}\label{gs_qfi_BH}
\Delta^2\mathcal{F}_\theta=4\mathrm{Var}_{\ket{0}}[H_\mathrm{S}]=4N^2=\norm{H_\mathrm{S}}^2.
\end{equation}
This follows because $H_\mathrm{S}\ket{0}= N\ket{1}$ in this limit, so both saturation conditions in Eq.~\eqref{GS_bound_ineqs} are asymptotically satisfied. The bound is also asymptotically saturated for $N\gg\Lambda$ at fixed $K$ and $U$ as this limit is an equivalent way to achieve $\alpha\to0$ (see also Appendix~\ref{App_BH_gs}). We show in Fig.~\ref{fig:gs_plots_all} given a sufficiently small $\Lambda$ the bound is saturated for any $N$.

\subsection{Short-range interacting systems}\label{sec6A}

Motivated by recent advances in the field of many-body metrology \cite{Montenegro2025, mihailescu2025criticalquantumsensingtutorial}, we analyze two spin-chain critical models: the one-dimensional transverse-field Ising model, which has already been extensively studied (e.g., in Ref.~\cite{mihailescu2025criticalquantumsensingtutorial}), and the XXZ chain in an external field. We will show that, while the transverse-field Ising model does not satisfy the \hyperref[condition:ground-state-optimality-critical]{\textit{Ground-State Optimality Condition for Critical Models}}, the XXZ model approaches this condition arbitrarily closely as $\gamma_z\to1^{-}$.

\subsubsection{Transverse-field 1D Ising model}\label{Sec.TFIM_GS}
We first consider the paradigmatic 1D transverse-field Ising model (TFIM) with periodic boundary conditions and an even number of spins $N$, to avoid exact degeneracies in the ground state \cite{PFEUTY197079, Barouch1970,  Barouch1971, Sachdev2011}. This model corresponds to a one-dimensional chain of $N$ spin-$1/2$ particles with nearest-neighbor interactions embedded in an external transverse field denoted by $\theta$. The Hamiltonian describing this model is expressed as follows:
\begin{equation}\label{H_ising_displaced}
    H_\theta=-\sum_{i=1}^{N}\sigma_x^{(i)}\sigma_x^{(i+1)}-(\theta+1)\sum_{i=1}^N\sigma_z^{(i)},
\end{equation}
where, for simplicity, we fixed the exchange constant to be $J=1$. The ground state of this Hamiltonian undergoes a second-order quantum phase transition at the critical point $\theta=0$, with associated critical exponents $z=1$ and $\nu=1$. Since $\nu (d+z) = 2$, Eq.~\eqref{eq:condcriticalexponents} is not satisfied. Hence, we can already anticipate that this model is suboptimal for ground-state metrology. 

The spectral gap of the control term in Eq.~\eqref{H_ising_displaced} can be found at the critical point by performing a Bogoliubov transformation, yielding $\Delta=4\sin(\pi/2N)\approx2\pi/N$. On the other hand, the ground-state QFI of this model was computed in \cite{Damski2013} as $\mathcal{F}_\theta=(N^2-N)/8$. Hence, it follows that the gap-aware figure of merit at criticality reads
\begin{equation}\label{FOMising}
    \Delta^2\mathcal{F}_\theta=2(N^2-N)\sin^2{\frac{\pi}{2N}}\approx\frac{\pi^2}{2},
\end{equation}
where the approximation is valid for $N\gg1$. Note that this is far from the optimal value $\norm{H_\mathrm{S}}^2=4N^2$. Furthermore, it is of interest to compute $\mathrm{Var}_{\ket{0}}[H_\mathrm{S}]$ at the critical point. As we detail in Appendix~\ref{App_Ising_gs}, one finds
\begin{equation}\label{var_TFIM}
    \mathrm{Var}_{\ket{0}}[H_\mathrm{S}]=N.
\end{equation}
Non-optimality is again evident, as this model does not even saturate the first inequality in Eq.~\eqref{GS_bound_ineqs}. In Fig.~\ref{fig:gs_plots_all}, we show the gap-aware figure of merit for finite values of $N$. Despite the apparent Heisenberg-like scaling of the QFI, $\mathcal{F}_\theta \sim N^2$, we find that the \hyperref[condition:ground-state-optimality]{\textit{Ground-State Optimality Condition}} does not hold for the TFIM.

As a related remark, the cat-state mechanism of the two-mode Bose-Hubbard model in Eq.~\eqref{H_BH} also has a short-range realization. Consider the Ising chain of Eq.~\eqref{H_ising_displaced} at a fixed transverse field, used as a probe of a field applied along the Ising axis, with the control tuned inside the ordered phase rather than at criticality, $H_\theta=\theta\sum_{i=1}^N\sigma_x^{(i)}+H_{\rm TFIM}(h)$, where $H_{\rm TFIM}(h)=-\sum_{i=1}^N\sigma_x^{(i)}\sigma_x^{(i+1)}-h\sum_{i=1}^N\sigma_z^{(i)}$ and $0<h<1$. For finite $N$, the two lowest states are the symmetric and antisymmetric (parity-even and parity-odd) superpositions of the two symmetry-broken states---the hallmark of a $\mathbb{Z}_2$-symmetry-breaking phase \cite{Zeng2019,Beekman2019,rev_SSB,Weinberg1996}---split by a gap $\Delta$ that closes exponentially with $N$ \cite{Damski2015Book,revIsing,Zeng2019}, while the rest of the spectrum lies $\mathcal{O}(1)$ above. Since $\ev{\sigma_x^{(i)}}=0$ in the symmetric ground state while $\ev{\sigma_x^{(i)}\sigma_x^{(j)}}\to m_x^2$ at large separations, with $m_x=(1-h^2)^{1/8}$ the spontaneous magnetization \cite{PFEUTY197079}, the signal variance is $\mathrm{Var}_{\ket{0}}[H_\mathrm{S}]\approx m_x^2N^2$ throughout the ordered phase for $N\gg\xi$ \cite{Frrot2018,Hauke2016}. Moreover, $H_\mathrm{S}\ket{0}$ is dominated by the cat partner, $\abs{\matrixel{1}{H_\mathrm{S}}{0}}\simeq m_xN$, up to $\mathcal{O}(\sqrt{N})$ components above the $\mathcal{O}(1)$ gap, and Eq.~\eqref{gs_QFI} gives
\begin{equation}\label{gs_qfi_TFIM_cat}
\Delta^2\mathcal{F}_\theta\approx 4\,\mathrm{Var}_{\ket{0}}[H_\mathrm{S}]\approx 4m_x^2N^2,
\end{equation}
so that the bound in Eq.~\eqref{boundGS} is saturated in scaling, with prefactor $m_x^2=(1-h^2)^{1/4}$, 
and exactly in the limit $h\to0^+$. This is the short-range instance of the first-order route of Sec.~\ref{sec_general_optimal_GS} 
\cite{Sarkar2025exp}: optimality is reached without second-order criticality, at the price of an exponentially closing gap. Note that 
this probe differs from the critical TFIM of Tables~\ref{tab:resultsintro}--\ref{GSAMresultstable}, where the field is transverse to the 
Ising axis and $h=1$. 

Its dynamical preparation mirrors that of the two-mode Bose-Hubbard model (Sec.~\ref{prep_time_BH}): sweeping from 
the paramagnetic product state across the critical point into the ordered phase, e.g., by extending Eq.~\eqref{H_Ising_prep} to 
$g(\tau)=2/(1+h)>1$, the Hamiltonian commutes with the parity $P=\prod_i\sigma_z^{(i)}$ at all times, so the drive $\partial_gH$ never 
couples the even and odd cat states and the exponentially small splitting is irrelevant for preparation. The lowest coupled gap closes only 
at $g=1$ (i.e., $h=1$), where the local adiabatic schedule of Eq.~\eqref{localadiabaticcondition} gives $\tau_{\rm prep}\sim1/\Delta_c\sim N$ 
(Secs.~\ref{Sec:tprep_crit} and~\ref{Sec.Prep_time_TFIM}), and stays $\mathcal{O}(1)$ inside the ordered phase. This probe therefore behaves 
like the two-mode Bose-Hubbard model in all three settings of Table~\ref{tab:resultsintro}: it is optimal in the static setting, it reaches 
Heisenberg scaling in GSUM after a polynomial (here linear) preparation (Sec.~\ref{sec:dynPTS}), and it reaches it in GSAM with
 $\tau\sim1/\Delta$ once the signal activates the cat transition, since the isolated doublet fulfills the assumptions 
 of Sec.~\ref{sec:LRpreparation} despite the short-range couplings (cf.\ Sec.~\ref{sec:BH_dyn}).

\subsubsection{External-field XXZ model}\label{Sec_XXZ_GS}
This model represents an extension of the transverse-field 1D Ising model considered before. The corresponding Hamiltonian can be written as 
\begin{align}\label{XXZ_H}
    H_\theta=&-\sum_{i=1}^{N-1}(\sigma_x^{(i)}\sigma_x^{(i+1)}+\sigma_y^{(i)}\sigma_y^{(i+1)}+\gamma_z\sigma_z^{(i)}\sigma^{(i+1)}_z) \nonumber \\ 
    &+\theta\sum_{i=1}^N\sigma_x^{(i)},
\end{align}
assuming open boundary conditions and $\gamma_z$ being an anisotropy parameter. Again, we have fixed the exchange constant to unity.

When $|\gamma_z|<1$ the model exhibits a second-order quantum phase transition at the critical point $\theta=0$, with critical exponents $z=1$ and $\nu=\frac{2}{4-\arccos{(\gamma_z)}/\pi}$ \cite{Affleck1999}. Interestingly, in the limit $\gamma_z \rightarrow 1$, where the ferromagnetic point is approached \cite{Giamarchi2003, Alba2012, Franchini2017}, Eq.~\eqref{eq:condcriticalexponents} is satisfied, as $\nu \to 1/2$. For any fixed $\abs{\gamma_z} < 1$ one has $\nu > 1/2$ and hence $\Delta^2 \mathcal{F}_\theta \sim N^{2/\nu - 2}$ with an exponent approaching $2$ from below as $\gamma_z \to 1^-$. Thus, the XXZ model approaches optimality as $\gamma_z\to1^-$; its entries in Tables~\ref{tab:resultsintro}--\ref{GSAMresultstable} are given in terms of $\delta\coloneq\arccos(\gamma_z)/\pi$ (see below). 

As a side remark, we note that $\gamma_z=-1$ corresponds to a Berezinskii--Kosterlitz--Thouless transition, which is an infinite-order quantum phase transition \cite{Zheng2015,Dalmonte2015}. This case, however, is not considered in the present work.

\begin{table*}[t]
\centering
\caption{\textcolor{black}{\textbf{Summary of results in the ground-state scenario for the considered models.} In this setting, Heisenberg scaling is attained whenever $\Delta^2\mathcal{F}_\theta\sim N^2$. Here, $g$ represents the strength of the spin-squeezing in $H_\theta$ as shown in Eq.~\eqref{H_spin_squeez}, while $\Lambda \coloneq K/\abs{U}$, with $K$ and $U$ being the hopping amplitude and boson-boson
interaction parameters, respectively (see Eq.~\eqref{H_BH}). For the XXZ model, $\delta\coloneq\arccos(\gamma_z)/\pi$, while $c$ and $b$ are $N$-independent prefactors; numerically, $c\approx2.45$ and $b\approx 0.04$ at $\gamma_z=0.99$ ($\delta=0.045$), and as $\gamma_z\to1^-$ the numerics (see Appendix~\ref{Appx:XXZ_numeric}) and a heuristic argument indicate $c\to2$, $b\to4(1-\gamma_z)$ and $\delta\to0$, i.e., the values of the spin-squeezing model with $g=4(1-\gamma_z)/N$ (see Sec.~\ref{Sec_XXZ_GS}).}}
\label{StaticTable}
\renewcommand{\arraystretch}{1.3}
\setlength{\tabcolsep}{5pt}

\resizebox{\textwidth}{!}{
\begin{tabular}{lcccc}
\toprule\toprule
        & \multicolumn{2}{c}{Short-range many-body systems (critical)} & \multicolumn{2}{c}{Long-range many-body systems} \\
\cmidrule(lr){2-3} \cmidrule(lr){4-5}
        & Transverse-field 1D Ising & External-field XXZ & Spin-squeezing & Two-mode Bose-Hubbard \\
\midrule
Static HS & \xmark & \cmark & \cmark & \cmark\\
Ground-state QFI & $N^2/8$ & $c\,N^{4-\delta}/b^{2}$ & $2N^2/g^2$ & $\abs{U}^{-2}\left[(N-1)!\right]^2\Lambda^{-2N}$\\
Spectral gap ($\Delta$) & $2 \pi / N$ & $b/N$ & $g$ & $2\abs{U}N\Lambda^{N}/(N-1)!$\\
$\Delta^2\mathcal{F}_\theta/\norm{H_\mathrm{S}}^2$  & $\pi^2/(8N^2)$ & $(c/4)\,N^{-\delta}$ & $1/2$ & $1$ \\
\bottomrule\bottomrule
\end{tabular}}
\end{table*}

Before discussing the optimal regime, let us revisit the super-Heisenberg scaling found in Ref.~\cite{Rams2018SuperHei} for $\gamma_z=0$ in light of the gap-aware optimality criterion of Eq.~\eqref{boundGSII}. In this case, the critical exponents $z=1$ and $\nu=4/7$ imply a ground-state QFI $\mathcal F_\theta\sim N^{\frac{7}{2}}$. Since the critical gap scales as $\Delta\sim N^{-1}$, one obtains $\Delta^2\mathcal F_\theta\sim N^{\frac{3}{2}}$, whereas the gap-dependent bound allows $\norm{H_\mathrm{S}}^2\sim N^2$. Thus, the ground-state probe has a strongly super-extensive QFI, but it does not saturate the optimality criterion of Eq.~\eqref{boundGSII}. Equivalently, its signal variance scales as $\Var_{\ket{0}}[H_\mathrm{S}]\sim N^{\frac{3}{2}}$. 
\begin{figure}[t]
    \centering
    \includegraphics[width=\linewidth]{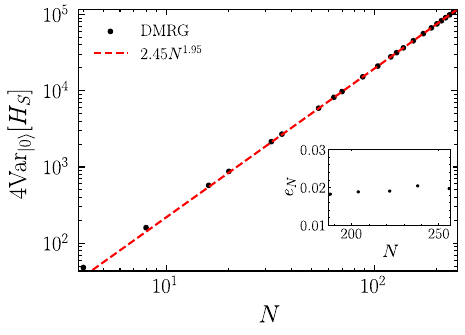}
    \caption{\textbf{Numerical analysis of the XXZ model.} Numerical values of $4\mathrm{Var}_{\ket{0}}[H_\mathrm{S}]$ as a function of $N$. For $N\geq200$, the data are fitted by $ 4\mathrm{Var}_{\ket{0}}[H_\mathrm{S}]\approx 2.45N^{1.95}$. \textit{Inset}: Relative error between $\Delta^2\mathcal{F}_\theta$ and $4\mathrm{Var}_{\ket{0}}[H_\mathrm{S}]$, denoted by $e_N$, as a function of $N$. Numerical data were obtained with $\gamma_z=0.99$ and $\delta\theta=5\times10^{-7}$. Inset data were computed with parameters $\gamma_z=0.9$ and $\delta\theta=5\times10^{-7}$. The same qualitative behavior persists as $\gamma_z \to 1^-$.}
    \label{fig:XXZ_static_numeric}
\end{figure}
To approach the gap-aware optimal regime, one must move towards $\gamma_z\to1^-$. It is convenient to measure the distance to optimality by $\delta\coloneq\arccos(\gamma_z)/\pi$, so that $\nu=2/(4-\delta)$ and, at criticality, $\mathcal{F}_\theta\sim N^{4-\delta}$, $\Delta\sim N^{-1}$ and $\Delta^2\mathcal{F}_\theta\sim N^{2-\delta}$. Hence, the figure of merit falls short of the bound by a factor $N^{-\delta}$, with $\delta\to0$ as $\gamma_z\to1^-$.

We compute the spectral gap, ground-state QFI and variance of $H_\mathrm{S}$ at the critical point with the DMRG algorithm of the \textit{TeNPy} library~\cite{hauschild2018Tenpy}, obtaining the ground states at $\theta=0$ and $\theta=\delta\theta$ and relating their fidelity to the QFI via Eqs.~\eqref{Fid_fidSus} and~\eqref{fid_sus_def}. Fig.~\ref{fig:XXZ_static_numeric} shows $4\mathrm{Var}_{\ket{0}}[H_\mathrm{S}]$ as a function of $N$ for $\gamma_z=0.99$ ($\delta=0.045$). The data follow $4\mathrm{Var}_{\ket{0}}[H_\mathrm{S}]\approx c\,N^{2-\delta}$ with $c\approx2.45$, in agreement with the predicted exponent $2-\delta=1.955$, and the inset shows that $\Delta^2\mathcal{F}_\theta$ and $4\mathrm{Var}_{\ket{0}}[H_\mathrm{S}]$ agree up to a small, $N$-independent relative error. Hence the first inequality in Eq.~\eqref{GS_bound_ineqs} is saturated, while the second is approached as $\Delta^2\mathcal{F}_\theta/\norm{H_\mathrm{S}}^2\approx(c/4)N^{-\delta}$, where $c$ is not a fixed constant but also depends on $\gamma_z$.

The same numerics, as shown in Appendix~\ref{Appx:XXZ_numeric}, reveal how optimality is approached. Repeating the fits for $\gamma_z$ closer to $1$, the exponent keeps following $2-\delta$ ($1.969$ and $1.987$ for $\gamma_z=0.995$ and $0.999$), while the prefactor decreases toward $2$. For $\gamma_z=1-10^{-6}$ and all sizes studied ($N\le192$) we find, to four digits, $\mathrm{Var}_{\ket{0}}[H_\mathrm{S}]=N^2/2+N$ and $\Delta=4(1-\gamma_z)/N$: these are precisely the variance and gap of the spin-squeezing model of Sec.~\ref{subsec:SS} with $g=4(1-\gamma_z)/N$. The XXZ chain thus becomes increasingly similar to the spin-squeezing model as $\gamma_z\to1^-$. Heuristically, this is because at $\gamma_z=1$ the model reduces to the ferromagnetic Heisenberg chain, whose ground states are the fully permutation-symmetric Dicke states $\ket{N/2,m_z}$. To first order, the small anisotropy $(1-\gamma_z)\sum_{i}\sigma_z^{(i)}\sigma_z^{(i+1)}$ only splits these states, and its action within the symmetric subspace is easily computed. Denoting by $P$ the projector onto this subspace, all pairs of sites are equivalent, so the $N-1$ nearest-neighbor bonds can be replaced by the average over all $N(N-1)/2$ pairs,
\begin{equation}
P\sum_{i=1}^{N-1}\sigma_z^{(i)}\sigma_z^{(i+1)}P=\frac{2}{N}\,P\sum_{i<j}\sigma_z^{(i)}\sigma_z^{(j)}P=\frac{4J_z^2-N}{N}\,P.
\end{equation}
 The anisotropy thus acts within the multiplet exactly like $g\,J_z^2$, with $g=4(1-\gamma_z)/N$, up to a constant: this is the spin-squeezing Hamiltonian of Sec.~\ref{subsec:SS}, whose ground state is $\ket{N/2,0}$ and whose gap to the states coupled by $H_\mathrm{S}$, $\ket{N/2,\pm1}$, is $g$.

There is no contradiction with the critical exponent $2-\delta<2$ at fixed $\gamma_z<1$: the deviation from the collective behavior only develops for chains longer than the crossover length $N>(1-\gamma_z)^{-1/2}$, which exceeds all accessible sizes for $\gamma_z=1-10^{-6}$ and diverges as $\gamma_z\to1^-$. The gap seems to be less sensitive to this crossover than the variance: writing $\Delta\simeq b/N$, we find $b\simeq4(1-\gamma_z)$ within $4\%$ already at $\gamma_z=0.99$, which combined with $\Delta^2\mathcal{F}_\theta\approx4\mathrm{Var}_{\ket{0}}[H_\mathrm{S}]$ yields the ground-state QFI entry of Table~\ref{StaticTable}.  In this sense, the XXZ chain appears to interpolate continuously between critical and collective optimality, with $\delta\to0$ and $c\to2$.

\section{Adiabatic preparation of many-body ground-state probes}\label{sec:AdiabaticPreparation}
So far, we have shown that several relevant many-body models can saturate the ultimate precision limits of ground-state metrology. A crucial open question is the preparation time associated with such states. In this section, we address it by focusing on adiabatic protocols. As already presented in Sec.~\ref{Sec:prepTime}, we consider Hamiltonians as in Eq.~\eqref{eq15}, driven from an easy-to-prepare gapped ground state at $t=0$ to the target Hamiltonian at $t = \tau$. Essentially, we check whether the many-body probes introduced in Sec.~\ref{sec:gs} satisfy the \hyperref[condition:preparation-time]{\textit{Preparation-Time Optimality Condition}}.
\begin{figure*}[t]
    \centering
    \begin{overpic}[height=0.3\textwidth]{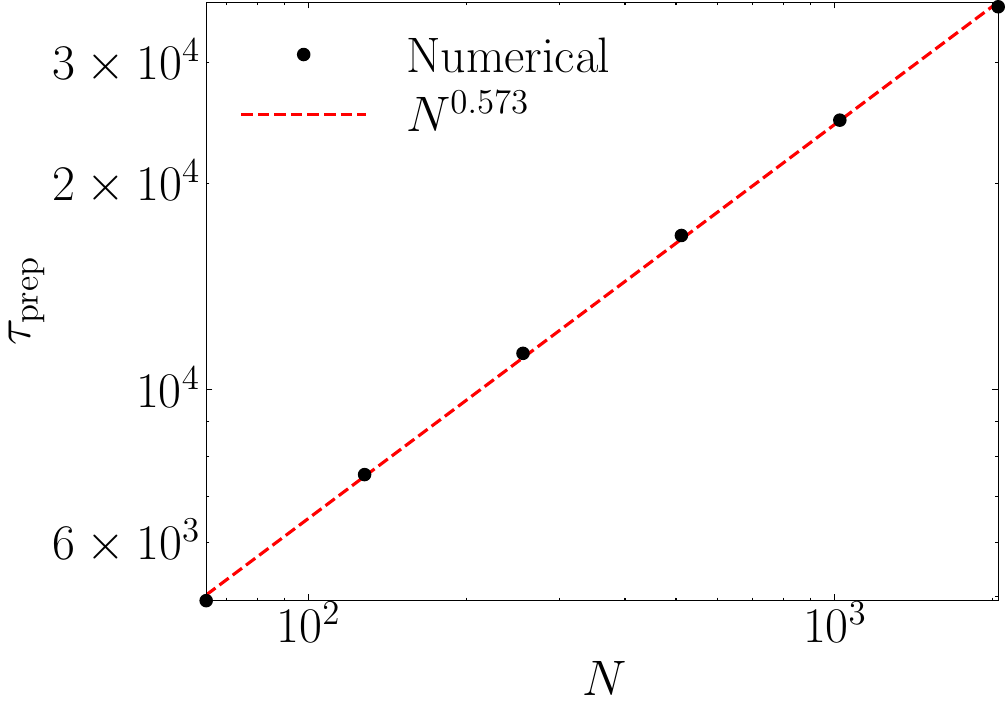}
        \put(1,72){\textbf{(a)}}
    \end{overpic}
    \hspace{1.5cm}
    \begin{overpic}[height=0.3\textwidth]{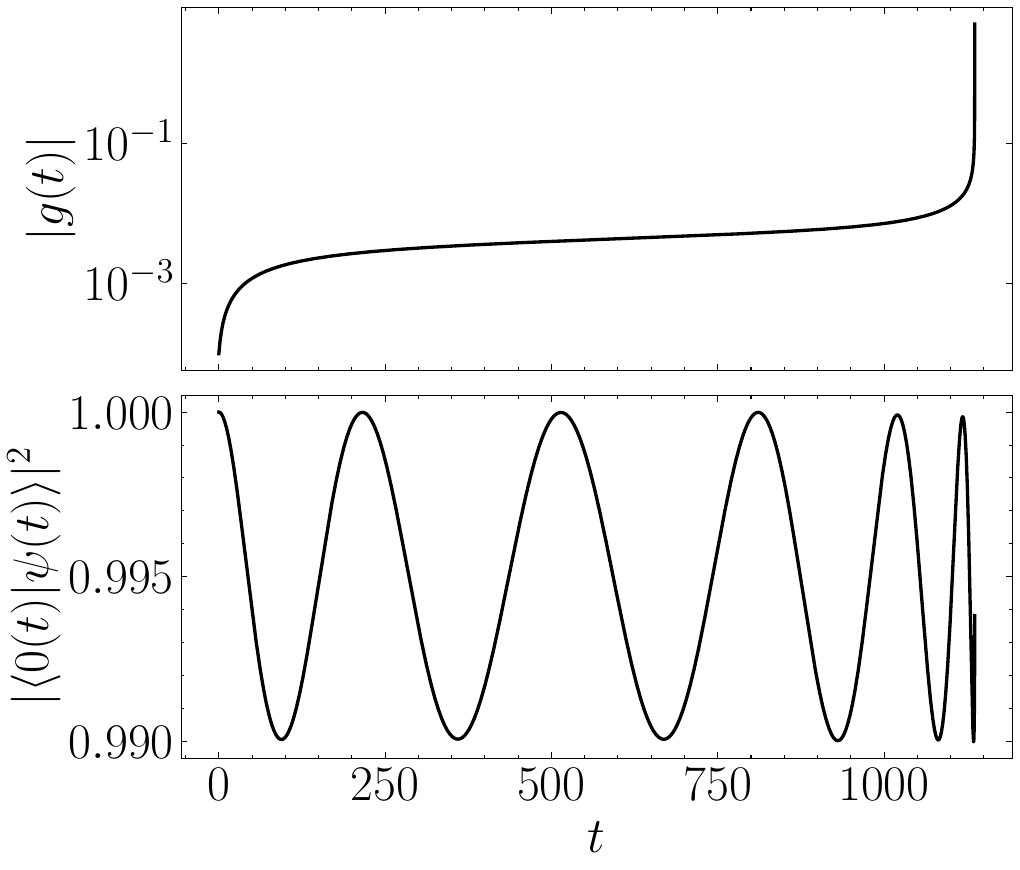}
        \put(-1,87){\textbf{(b)}}
    \end{overpic}
    \caption{\textbf{Ground-state preparation in the two-mode Bose-Hubbard model.} (a) Ground-state preparation time $\tau_{\rm prep}$ as a function of system size $N$, obtained from the local adiabatic protocol with
$K=10^{-2}$, $\epsilon=0.05$ and $\theta=0$. The red dashed line shows a power-law fit,
$\tau_{\rm prep}\sim N^{0.573}$. (b) \textit{Top}: Local adiabatic schedule $\abs{g(t)}$ (in absolute value) as a function of time. \textit{Bottom}: Fidelity between the dynamically evolved state and the instantaneous ground state, $\abs{\langle 0 (t)|\psi(t)\rangle}^2$, along the local
adiabatic preparation. The data were obtained with  $N=16$, and the same values of $K$,
$\epsilon$ and $\theta$ as in (a).}
    \label{fig:BH_GSUM_prep_time}
\end{figure*}
\subsection{All-to-all interacting systems}\label{sec:optimal_prep_LR}
We begin by discussing the ground-state preparation time required for all-to-all interacting models. We show that the spin-squeezing Hamiltonian does not require an adiabatic preparation, since the ground state is uncoupled from all excited states by the driving operator at $\theta=0$. For the two-mode Bose-Hubbard model, we show that the super-exponentially closing gap does not determine the preparation time because the corresponding transition is forbidden by parity. Instead, the preparation is constrained by the lowest coupled excitation, corresponding to the second excited state, whose minimum gap closes only polynomially with $N$. 
\subsubsection{Spin-squeezing Hamiltonian}\label{sec.optimal_prep_SS}
We start by considering a time-dependent spin-squeezing Hamiltonian given by
\begin{equation}\label{H_spin_squeez_time}
    H_\theta (t) = g(t) J_z^2 - \lambda J^2 + 2 \theta J_x. 
\end{equation}
From here, we see that dynamical preparation of the optimal probe is trivial, since the ground state remains unchanged throughout the evolution. This is because, at $\theta = 0$, the driving operator does not couple the ground state to any excited state, namely that $\matrixel{n (g)}{\partial_g H}{0 (g)} = 0$ for all $n \neq 0$. Hence, within our convention of neglecting the time required to prepare the initial ground state, $\tau_{\rm prep} = 0$. If the initialization of the starting ground state is included, for instance through a cooling process, an additional preparation timescale may arise, which under general assumptions is lower-bounded by $\mathcal{O} (1/\Delta)$.
 \subsubsection{Two-mode Bose-Hubbard model}\label{prep_time_BH}
We now consider the time-dependent two-mode Bose-Hubbard Hamiltonian:
\begin{equation}\label{H_BH_dyn}
    H_\theta (t) = -2K J_x + g(t) (J_z^2+J^2) + 2 \theta J_z.
\end{equation}
Throughout the evolution, $g(t)<0$ plays the role of the boson-boson interaction term $U$ in Eq.~\eqref{H_BH}. We start the driving deep in the gapped phase, $|g(t=0)| N \ll K$, and drive the system towards the optimal regime of Sec.~\ref{Sec: BH_GS}, $|g(t=\tau)| N\gg K$.

Since we are interested in the preparation time of the optimal ground state at $\theta = 0$, parity strongly constrains the allowed transitions. In particular, the ground and first excited states are uncoupled by the driving operator, yielding $\matrixel{1 (g)}{\partial_g H}{0 (g)} = 0$. In contrast to the spin-squeezing case, transitions to higher excited states are not forbidden. The lowest coupled excitation is the second excited state, which therefore provides the relevant transition for evaluating the local adiabatic condition. We thus define $\Delta_2 (g) \coloneq E_2 (g) - E_0 (g)$ as the relevant gap, so that the corresponding local adiabatic element is $\abs{\matrixel{2 (g)}{\partial_g H}{0 (g)}} / \Delta_2 (g)^2$. Within a local adiabatic protocol, $\tau_{\rm prep}$ is therefore determined by the integral of this quantity along the preparation path.

In Fig.~\ref{fig:BH_GSUM_prep_time} we numerically evaluate the ground-state preparation obtained from the local adiabatic schedule in Eq.~\eqref{preptimeintegral}, constructed with the lowest-coupled excitation allowed by parity. As shown in Fig.~\ref{fig:BH_GSUM_prep_time}(a), the resulting preparation time scales approximately as $\tau_{\rm prep} \sim N^{0.573}$, which grows only polynomially with the system size $N$. Accordingly, in Table~\ref{GSUMresultstable} we denote the preparation-time scaling simply as $\mathrm{poly}(N)$. The explicit time evolution in Fig.~\ref{fig:BH_GSUM_prep_time}(b) further confirms that the protocol prepares the target cat state with high fidelity (above $0.99$ throughout the protocol). There, the three regimes of the model can be directly identified. At the beginning of the protocol, where $\abs{g (t)} N \ll K$, the ground state is close to the coherent spin state polarized along $x$, and the relevant gap $\Delta_2 \simeq 4 K$ is large and essentially $N$-independent, so that $g$ is swept quickly towards the crossover regime. When $\abs{g(t)} N \sim K$, the ground state starts to split into two macroscopically distinct components. There, $\Delta_2 (g)$ reaches its minimum along the path and the driving speed is lowered accordingly; this slow stage dominates the total preparation time and sets its polynomial scaling with $N$ in Fig.~\ref{fig:BH_GSUM_prep_time}(a). Finally, since the relevant gap reopens in the cat region ($\abs{g(t)} N \gg K$), $g$ is again swept quickly towards the final value $g (\tau)$.
\subsection{Short-range interacting systems}\label{sec:AdiabaticPreparationShortRange}
We now study the ground-state preparation time of short-range critical interacting models. We show that the ground-state preparation time exhibits optimal scaling for both models, analytically for the TFIM and numerically for the XXZ chain.

\begin{figure*}[t]
    \centering
    \begin{overpic}[height=0.3\textwidth]{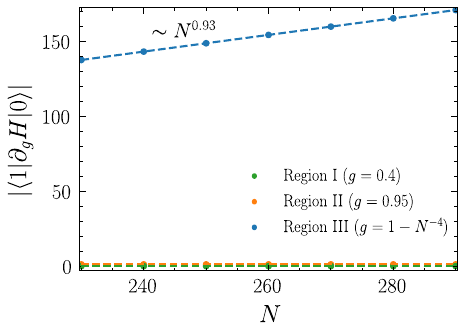}
        \put(1,70){\textbf{(a)}}
    \end{overpic}
    \hspace{1cm}
    \begin{overpic}[height=0.3\textwidth]{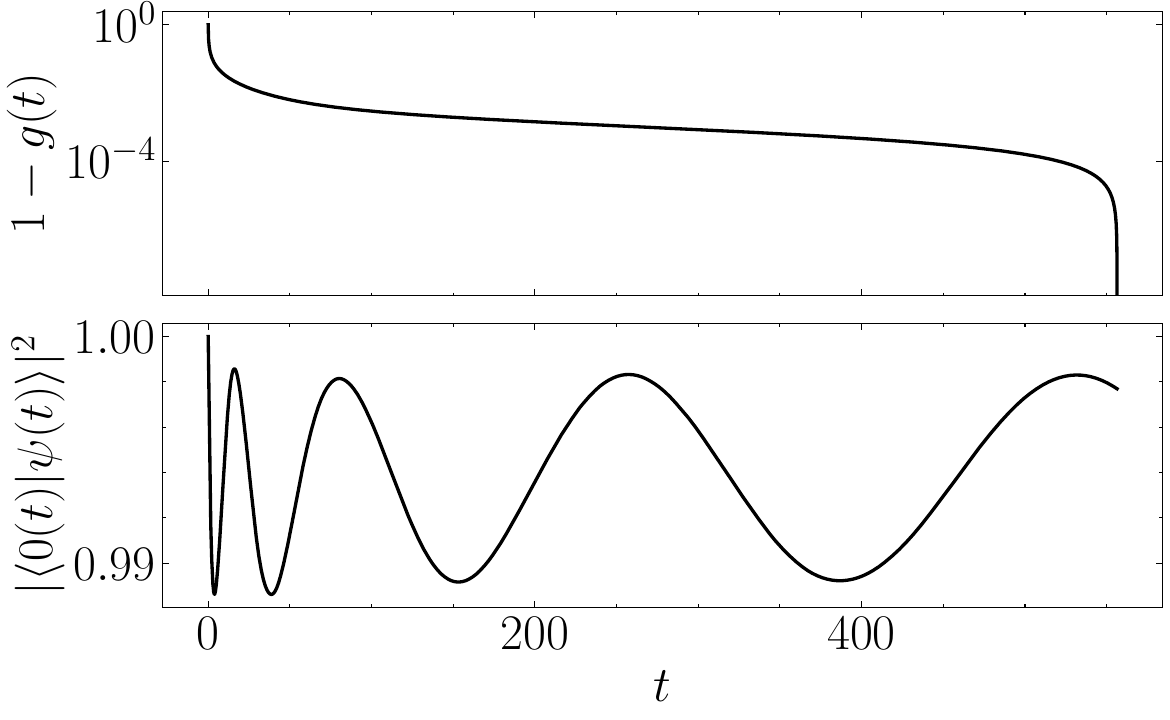}
        \put(-1,60){\textbf{(b)}}
    \end{overpic}
    \caption{\textbf{Ground-state preparation in the XXZ Hamiltonian.} (a) Local adiabatic matrix elements between the ground and the lowest coupled excitation as a function of $N$. Data points are shown for representative values of $g$ across the three regions of interest. Dotted lines indicate power-law fits that illustrate the scaling of the matrix elements with $N$. The numerical data were obtained using $\gamma_z=0.92$. (b) \textit{Top}: Local adiabatic schedule $1-g(t)$ as a function of time. \textit{Bottom}: Fidelity $\abs{\innerproduct{0 (t)}{\psi (t)}}^2$ between the time-evolved state and the instantaneous ground state of the XXZ Hamiltonian in Eq.~\eqref{XXZ_H_time} as a function of time. The local adiabatic schedule is built from Eq.~\eqref{localadiabaticcondition}, considering only the lowest coupled excitation. For the simulation we take $\epsilon = 0.05, \gamma_z = 0.95$ and $N = 10$.}
    \label{fig:XXZ_tprep}
\end{figure*}
\subsubsection{Transverse-field 1D Ising model}\label{Sec.Prep_time_TFIM}
We consider the following Hamiltonian for the time-dependent TFIM:
\begin{equation}\label{H_Ising_prep}
    H_\theta(t)=-g(t)\sum_{i=1}^N\sigma_x^{(i)}\sigma_x^{(i+1)}-\left[2+\theta-g(t)\right]\sum_{i=1}^N\sigma_z^{(i)}.
\end{equation}
This is nothing but a linear interpolation between a gapped Hamiltonian ($g(t=0) = 0$) and the gapless Hamiltonian in Eq.~\eqref{H_ising_displaced} ($g (t=\tau) = 1$). As mentioned, we do not consider the signal term for the purpose of ground-state preparation; that is, we fix $\theta = 0$.

We now analytically confirm the expected scalings for the local adiabatic matrix element and the spectral gap presented in Sec.~\ref{Sec:tprep_crit} for the TFIM. To this end, we first compute the spectral gap of the Hamiltonian. Here $\Delta$ denotes the global gap, which need not coincide with the gap to the lowest excited state in the ground-state parity sector; at criticality, however, both scale identically as $N^{-z}$ \cite{Sachdev2011}. Mapping $J\mapsto g(t)$ and $\theta\mapsto 2-g(t)$ in Eq.~\eqref{spectrum_isign1d}, we obtain
\begin{equation}\label{Gap_ising_dyn}
    \Delta(g(t))\approx2\sqrt{4(g-1)^2-\frac{\pi^2g(g-2)}{N^2}},
\end{equation}
where we only keep terms up to $\mathcal{O}(N^{-2})$. From this expression, we see that $\Delta(g(0))=4$ and $\Delta(g(\tau))=2\pi/N$ correspond to the spectral gaps of the non-interacting and target Hamiltonians, respectively.

Secondly, as we show in Appendix~\ref{Appx:matrels_TFIM}, the local adiabatic matrix element is given by
\begin{equation}\label{main_matrels_TFIM_LA}
    \abs{\matrixel{1}{\partial_g H}{0}}=\frac{4\pi}{N\Delta(g)},
\end{equation}
for sufficiently large values of $N$ and taking $\ket{1}$ as the first excited state within the same symmetry sector as the ground state.

By evaluating the spectral gap from Eq.~\eqref{Gap_ising_dyn} and the matrix element from Eq.~\eqref{main_matrels_TFIM_LA}, we can characterize the system in the three distinct regimes. In region (I), defined by $s\coloneq|1-g|\sim\mathcal{O}(1)$, we find that $\Delta\sim s\sim\mathcal{O}(1)$ and $\abs{\matrixel{1}{\partial_g H}{0}}\sim N^{-1}$. Moving to region (II), where $1\gg s >N^{-1}$, the gap and matrix element scale as $\Delta\sim s$ and $\abs{\matrixel{1}{\partial_g H}{0}}\sim (s N)^{-1}$, respectively. Finally, in region (III), characterized by $N^{-1}\geq s\geq 0$, these quantities scale as $\Delta\sim N^{-1}$ and $\abs{\matrixel{1}{\partial_g H}{0}}\sim\mathcal{O}(1)$.

Therefore, we show that these scaling relations fully agree with the arguments presented in Sec.~\ref{Sec:tprep_crit} for the critical exponents of this model, $d=z=\nu=1$. As a result, for sufficiently large $N$, the preparation time integrates to the optimal profile $\tau_\mathrm{prep}\sim 1/\Delta_c\sim N$. Ultimately, this provides an analytical verification of the \hyperref[condition:preparation-time]{\textit{Preparation-Time Optimality Condition}} for the transverse-field Ising model within the lowest-excitation criterion of Eq.~\eqref{localadiabaticcondition}.

\subsubsection{External-field XXZ model}\label{Sec.PrepTime_xxz}
We now consider the XXZ Hamiltonian with a driving $g(t)$ such that the system ends at criticality. In particular, 
\begin{align}\label{XXZ_H_time}
    H_\theta (t) &= - g(t) \sum_{i=1}^{N-1} \left( \sigma_x^{(i)} \sigma_x^{(i+1)} + \sigma_y^{(i)} \sigma_y^{(i+1)} + \gamma_z \sigma_z^{(i)} \sigma_z^{(i+1)} \right) \notag\\
    &+ \left[ 1 + \theta - g(t) \right] \sum_{i=1}^N \sigma_x^{(i)}. 
\end{align}

Fig.~\ref{fig:gap_g_xxz} illustrates the three regions along the path for the XXZ model, and Fig.~\ref{fig:XXZ_tprep}(a) shows the scaling of the local adiabatic matrix element with $N$ at a representative point $g$ within each region. In regions (I) and (II), the matrix element is independent of $N$, whereas in region (III) the data are consistent with an $N^{0.93}$ scaling. These numerical results support the universal scaling law $\abs{\matrixel{1}{\partial_g H}{0}}\sim N^{(1/\nu-z)/d}$ presented in Sec.~\ref{Sec:tprep_crit}: for $\gamma_z = 0.92$ one has $\nu\approx 0.517$, which for $d=z=1$ predicts $N^{0.934}$. Therefore, the preparation time integrates to the optimal profile $\tau_\mathrm{prep}\sim1/\Delta_c\sim N$ for large enough $N$. Hence, the \hyperref[condition:preparation-time]{\textit{Preparation-Time Optimality Condition}} is satisfied for the XXZ model.

The three regions are also directly visible in the schedule $g(t)$ obtained by integrating Eq.~\eqref{localadiabaticcondition}, shown in Fig.~\ref{fig:XXZ_tprep}(b) (top) for $N=10$, $\gamma_z=0.95$ and $\epsilon=0.05$. In region (I) both the gap and the matrix element are of order one, so $g$ is swept quickly away from its initial value. In region (II), $\Delta\sim(1-g)^{\nu z}$ and the $N$-independent matrix element scales as $(1-g)^{\nu z-1}$, so that Eq.~\eqref{localadiabaticcondition} gives $\dot g\propto(1-g)^{\nu z+1}$ and the critical point is approached algebraically, $1-g(t)\propto(t-t_0)^{-1/(\nu z)}$; this slow stage is where most of the preparation time is spent, reaching the critical window $1-g\sim N^{-1/(d\nu)}$ after a time $\propto N^{z/d}$, in agreement with Eq.~\eqref{preptime_reg_II}. Finally, in region (III) the gap and the matrix element saturate to their finite-size values and $g$ is swept linearly across the critical window, which appears as the sharp final drop on the logarithmic scale. As an a posteriori check, Fig.~\ref{fig:XXZ_tprep}(b) (bottom) shows the fidelity between the instantaneous ground state and the time-evolved state along the same protocol: it remains above $0.99$ throughout, with oscillations whose period grows as the instantaneous gap closes and whose amplitude is set by $\epsilon$, confirming that the schedule built from the lowest coupled excitation alone remains adiabatic.

\section{Dynamical metrology in the Ground-State Unitary Metrology framework}
\label{sec:dynPTS}
\begin{table*}[t]
\centering
\caption{\textbf{Summary of results in the GSUM framework for the considered models.} In this setting, HS is asymptotically ($\tau\gg \tau_\mathrm{prep}$) attained whenever $\mathrm{Var}_{\ket{0}} [H_\mathrm{S}] \sim N^2$, i.e., the \hyperref[condition:GSUM]{\textit{Ground-State Unitary Metrology Optimality Condition}} is met. For the XXZ model, $\delta\coloneq\arccos(\gamma_z)/\pi$ and $c\approx2.45$ at $\gamma_z=0.99$, with $c\to2$ and $\delta\to0$ as $\gamma_z\to1^-$ (see Table~\ref{StaticTable} and Sec.~\ref{Sec_XXZ_GS}).}
\label{GSUMresultstable}
\renewcommand{\arraystretch}{1.3}
\setlength{\tabcolsep}{5pt}

\resizebox{\textwidth}{!}{
\begin{tabular}{lcccc}
\toprule\toprule
        & \multicolumn{2}{c}{Short-range many-body systems (critical)} & \multicolumn{2}{c}{Long-range many-body systems} \\
\cmidrule(lr){2-3} \cmidrule(lr){4-5}
        & Transverse-field 1D Ising & External-field XXZ & Spin-squeezing & Two-mode Bose-Hubbard \\
\midrule
GSUM HS & \xmark & \cmark & \cmark & \cmark \\
$\Var_{\ket{0}} \left[ H_{\rm S} \right]$ & $N$ & $(c/4)\,N^{2-\delta}$ & $N^2/2$ & $N^2$ \\
Preparation time ($\tau_{\rm prep}$) & $N$ & $N$ & $0$ & $\mathrm{poly} (N)$\\
\bottomrule\bottomrule
\end{tabular}}
\end{table*}
In this section, we investigate the attainability of Heisenberg scaling within the GSUM framework introduced in Sec.~\ref{Sec.PTS_framework}. Specifically, we examine whether the \hyperref[condition:GSUM]{\textit{Ground-State Unitary Metrology Optimality Condition}} is satisfied by the ground-state probes introduced in Sec.~\ref{sec:gs}. 

Table \ref{GSUMresultstable} summarizes the obtained results for the different many-body Hamiltonians. In the limit of $\tau \gg \tau_{\rm prep}$, Heisenberg scaling is achieved for Hamiltonians presenting $\Var_{\ket{0}} \left[ H_{\rm S} \right] \sim N^2$, as we anticipated in Sec.~\ref{GSUMframework} (see Eq.~\eqref{QFI_PTS}). This is the case for the XXZ Hamiltonian, as well as the two infinite-range interacting systems.

We additionally revisit the preparation times for the ground states of the different models (see Sec.~\ref{sec:AdiabaticPreparation} for the explicit derivations). While the GSUM framework asymptotically attains Heisenberg scaling for any ground-state optimal probe, the time required to enter this regime depends crucially on the scaling of $\tau_{\rm prep}$. For short-range critical Hamiltonians, we argued in Sec.~\ref{Sec:tprep_crit} that, in general, the optimal preparation time profile $1/\Delta_c$ can be attained with local adiabatic dynamics, with $\Delta_c$ being the spectral gap at the critical point. We further considered infinite-range interacting probes and analyzed the preparation of their ground states. In the GSUM scenario, where $\theta = 0$, we showed that the ground state of the spin-squeezing Hamiltonian requires no dynamical preparation, while the two-mode Bose-Hubbard model only incurs a polynomial overhead in the preparation.

\section{Dynamical metrology in the Ground-State Adiabatic Metrology Framework}\label{sec:dynPWS}
Let us now apply the tools we have developed to check whether Heisenberg scaling remains attainable within the GSAM framework introduced in Sec.~\ref{Sec.PWS_framework}. Specifically, we examine if the \hyperref[condition:GSAM]{\textit{Ground-State Adiabatic Metrology Optimality Condition}} is satisfied for the ground-state probes introduced in Sec.~\ref{sec:gs}. We recall that this condition requires both the \hyperref[condition:ground-state-optimality]{\textit{Ground-State Optimality Condition}} and the \hyperref[condition:preparation-time]{\textit{Preparation-Time Optimality Condition}} to hold. In Table \ref{GSAMresultstable} we summarize our results for the different many-body Hamiltonians.

\subsection{All-to-all interacting systems}
We first consider all-to-all interacting Hamiltonians and show that both models can achieve Heisenberg scaling, since they both satisfy the \hyperref[condition:ground-state-optimality]{\textit{Ground-State Optimality Condition}} (see Sec.~\ref{sec:GSLR}) and the \hyperref[condition:preparation-time]{\textit{Preparation-Time Optimality Condition}} (see Sec.~\ref{sec:LRpreparation}).
\subsubsection{Spin-squeezing Hamiltonian}
Now, consider the spin-squeezing Hamiltonian in Eq.~\eqref{H_spin_squeez_time}. As discussed in Sec.~\ref{sec.optimal_prep_SS}, at $\theta = 0$ the driving operator does not couple the ground state to any excited state, so that no dynamical preparation time is required. In the GSAM framework, however, a small but finite $\theta$ mixes the relevant eigenstates, leading to a non-vanishing adiabatic coupling. Since this Hamiltonian fulfills the assumptions presented in Sec.~\ref{sec:LRpreparation}, within the local-estimation window the preparation time scales as $\tau = \tau_{\rm prep} \sim 1 / \Delta (\tau)$. Hence, together with the optimal ground-state scaling obtained in Sec.~\ref{subsec:SS}, the \hyperref[condition:GSAM]{\textit{Ground-State Adiabatic Metrology Optimality Condition}} is satisfied and Heisenberg scaling is attained.

For this model, the local adiabatic dynamics can be solved explicitly. In Appendix \ref{SSappendix} we derive the spectral gap,
\begin{equation}\label{gap_SS_PWS}
    \Delta(\tau)=g(\tau)=\frac{g(0)}{1+\epsilon g(0) \tau},
\end{equation}
and the dynamical QFI,
\begin{equation}\label{SStimedepQFI}
    \mathcal{F}_\theta (\tau) = \dfrac{2 N^2 \tau^2}{\epsilon^{-2} + 1} + \mathcal{O} (\tau^2 N) + \mathcal{O} (\tau N^2),
\end{equation}
where $\epsilon$ is, again, the parameter that controls the adiabaticity. Eq.~\eqref{gap_SS_PWS} explicitly gives
\begin{equation}
    \tau = \dfrac{1}{\epsilon} \left( \dfrac{1}{\Delta (\tau)} - \dfrac{1}{\Delta (0)} \right),
\end{equation}
and therefore recovers the scaling derived from the general long-range argument. Moreover, Eq.~\eqref{SStimedepQFI} shows that, for any fixed $\epsilon$, the QFI scales quadratically in both $N$ and $\tau$. In the adiabatic regime, this explicitly confirms the attainability of Heisenberg scaling within the GSAM framework. 
\begin{table*}[t]
\centering
\caption{\textbf{Summary of results in the GSAM framework for the considered models.} In this setting, HS is attained whenever the static ground-state probe is optimal and $\tau_\mathrm{prep}\sim 1/\Delta$, i.e., whenever the \hyperref[condition:GSAM]{\textit{Ground-State Adiabatic Metrology Optimality Condition}} is met. Entries for the XXZ model hold as $\gamma_z \to 1^-$.}
\label{GSAMresultstable}
\renewcommand{\arraystretch}{1.3}
\setlength{\tabcolsep}{5pt}

\resizebox{\textwidth}{!}{
\begin{tabular}{lcccc}
\toprule\toprule
        & \multicolumn{2}{c}{Short-range many-body systems (critical)} & \multicolumn{2}{c}{Long-range many-body systems} \\
\cmidrule(lr){2-3} \cmidrule(lr){4-5}
        & Transverse-field 1D Ising & External-field XXZ & Spin-squeezing & Two-mode Bose-Hubbard \\
\midrule
GSAM HS & \xmark & \cmark & \cmark & \cmark \\
Sensing time ($\tau$) & $1/\Delta$ & $1/\Delta$ & $1/\Delta$ & $1/\Delta$\\
\bottomrule\bottomrule
\end{tabular}}
\end{table*}

Interestingly, Eq.~\eqref{SStimedepQFI} also reveals a trade-off between adiabaticity and the prefactor of the QFI. Increasing $\epsilon$, and hence relaxing adiabaticity, enhances the QFI. In the limit of $\epsilon \to \infty$, one obtains $\mathcal{F}_\theta (\tau) \approx \frac{1}{2} \tau^2 \norm{H_\mathrm{S}}^2$, which coincides with the QFI obtained within the GSUM framework in Table~\ref{GSUMresultstable}. Thus, the two protocols coincide at the level of both scaling and prefactor in this non-adiabatic limit. Conversely, when adiabaticity is ensured ($\epsilon \to 0$), Eq.~\eqref{SStimedepQFI} gives $\mathcal{F}_\theta (\tau) \approx 2 N^2 \tau^2 \epsilon^2$, which coincides with the instantaneous ground-state QFI, since $\Delta (\tau)^2 \approx 1 / (\epsilon^2 \tau^2)$ in the same limit.

\subsubsection{Two-mode Bose-Hubbard model}\label{sec:BH_dyn}
We now consider the two-mode Bose-Hubbard Hamiltonian in Eq.~\eqref{H_BH_dyn}. As shown in Sec.~\ref{prep_time_BH}, in the regime where the target ground state approaches the cat-state limit, this model satisfies the assumptions of the preparation argument for infinite-range interacting models presented in Sec.~\ref{sec:LRpreparation}. Therefore, within the local-estimation window, the preparation time in the cat-state region scales as $\tau^{\rm cat} \sim 1/\Delta (\tau)$, where $\Delta (\tau)$ is the gap of the target state in Eq.~\eqref{BHgap}. The parity-allowed transitions governing the pre-cat region contribute only the polynomial preparation overhead found in Sec.~\ref{prep_time_BH}. By contrast, the signal-activated cat-doublet transition introduces the inverse-gap timescale. Since $\Delta(\tau)^{-1}$ grows super-exponentially with $N$, it dominates the polynomial pre-cat contribution asymptotically, yielding $\tau=\tau_{\rm prep}\sim1/\Delta(\tau)$.

Moreover, since the ground-state QFI saturates the bound in Eq.~\eqref{boundGS}, the two-mode Bose-Hubbard model satisfies the \hyperref[condition:GSAM]{\textit{Ground-State Adiabatic Metrology Optimality Condition}} and attains Heisenberg scaling within the GSAM framework.

\subsection{Short-range interacting systems}
Now, we consider the dynamics of short-range second-order critical Hamiltonians. We will see that the TFIM is suboptimal because it does not satisfy the \hyperref[condition:ground-state-optimality]{\textit{Ground-State Optimality Condition}}, whereas the XXZ model approaches Heisenberg scaling arbitrarily closely by satisfying the \hyperref[condition:GSAM]{\textit{Ground-State Adiabatic Metrology Optimality Condition}}. As discussed in Sec.~\ref{Sec:tprep_crit}, the preparation-time scaling at $\theta=0$ remains valid for weak nonzero signals within the local-estimation window.
\subsubsection{Transverse-field 1D Ising model}
Here, the Hamiltonian is given by Eq.~\eqref{H_Ising_prep}. In the GSAM framework, encoding and preparation happen simultaneously, so $\tau = \tau_{\rm prep} \sim 1/\Delta_c\sim N$ from Sec.~\ref{Sec.Prep_time_TFIM}. By construction of the protocol, the QFI at time $\tau$ corresponds to the ground-state QFI in the static scenario, $\mathcal{F}_\theta (\tau) \sim N^2$, far below the Heisenberg scaling $\tau^2 \norm{H_{\rm S}}^2 \sim N^4$. This shows that, although the preparation protocol is optimal, the target ground state does not satisfy the \hyperref[condition:ground-state-optimality]{\textit{Ground-State Optimality Condition}}. Consequently, Heisenberg scaling is not achieved.

\subsubsection{External-field XXZ model}\label{Sec:GSAM_XXZ}
We finally consider the XXZ Hamiltonian in Eq.~\eqref{XXZ_H_time}. From the analysis in Sec.~\ref{Sec.PrepTime_xxz}, we have $\tau = \tau_{\rm prep} \sim 1/\Delta_c \sim N$, while the final QFI is the static ground-state one, $\mathcal{F}^\mathrm{GS}_\theta \sim N^{2/\nu}$ where $\nu=\frac{2}{4-\arccos(\gamma_z)/\pi}$ for this model. Hence,
\begin{equation}
    \mathcal{F}_\theta(\tau)\sim N^2\tau^{2/\nu-2}=N^2\tau^{2-\arccos(\gamma_z)/\pi}= N^2 \tau^{2-\delta}.
\end{equation}
Thus, for any fixed \(|\gamma_z|<1\), the time scaling is sub-Heisenberg. However, as \(\gamma_z\to1^{-}\), one has \(\nu\to1/2\), and therefore (see also Fig.~\ref{fig:GSAM_multipleN_xxz})
\begin{equation}
    F_\theta(\tau)\sim N^2\tau^2.
\end{equation}
Hence, the XXZ family can approach the dynamical Heisenberg limit arbitrarily closely, with exact Heisenberg scaling obtained asymptotically in the ferromagnetic limit \(\gamma_z\to1^{-}\). 
\section{Conclusions and outlook}
\label{sec:Conclusions}

In this work, we have provided a characterization of the physical requirements to reach the ultimate limits of ground-state metrology, identifying several many-body probes capable of saturating them. Our analysis is based on the interplay between two fundamental bounds on the QFI: the static limit of $\Delta$-gapped ground states, $\mathcal{F}_{\theta} \le \norm{H_{\rm S}}^2 / \Delta^2$, and the dynamical Heisenberg limit, $\mathcal{F}_\theta \leq \tau^2 \norm{H_\mathrm{S}}^2$, which accounts for the total protocol duration $\tau$. 

Regarding the static regime, the reachability of the ground-state bound depends fundamentally on the range of interactions. For short-range interacting systems, optimality requires a spectral gap that closes with $N$, either at a second-order quantum critical point or at a first-order transition with cat-like ground states (Secs.~\ref{Sec: BH_GS} and~\ref{Sec.TFIM_GS}). For the former, this led us to a universal condition on the critical exponents, $\nu(d+z)=1$, which can only be approached by a continuous transition, since its exact saturation would correspond to a jump of $\langle H_\mathrm{S}\rangle$, i.e., to the first-order mechanism. 
The XXZ chain approaches it arbitrarily closely as $\gamma_z\to1^-$; our numerics, supported by a heuristic argument, indicate that in this limit its exponents, prefactors and gap tend to those of the spin-squeezing model, suggesting a continuous interpolation between critical and collective optimality. Other paradigmatic models, such as the critical 1D transverse-field Ising model, remain suboptimal. The first-order route, realized by the ordered Ising chain probing a longitudinal field, saturates the static bound exactly and behaves like the two-mode Bose-Hubbard model in the dynamical settings as well. In contrast, all-to-all interacting models, such as the spin-squeezing and two-mode Bose-Hubbard Hamiltonians, can saturate the ground-state bound even in the absence of criticality.

We then accounted for the ground-state preparation time and studied the attainability of the dynamical bound in both the Ground-State Unitary Metrology (GSUM) and Ground-State Adiabatic Metrology (GSAM) frameworks (see Fig.~\ref{fig:dynSchemes} and Table~\ref{tab:resultsintro} for a summary of the results). In the Ground-State Unitary Metrology (GSUM) framework, we have shown that reaching static ground-state optimality directly translates into asymptotic Heisenberg scaling. For short-range interacting systems undergoing a second-order quantum phase transition, this implication holds in both directions: a probe achieves Heisenberg-limited precision if and only if it satisfies (or approaches) the \hyperref[condition:ground-state-optimality-critical]{\textit{Ground-State Optimality Condition for Critical Models}}. This is illustrated by the contrast between the XXZ model, which approaches both static and dynamical optimality, and the 1D transverse-field Ising model (TFIM), which fails the ground-state optimality criterion.

The requirements become more stringent in the Ground-State Adiabatic Metrology (GSAM) framework, where preparation and encoding occur simultaneously. In this case, achieving the Heisenberg limit requires satisfying both the \hyperref[condition:ground-state-optimality]{\textit{Ground-State Optimality Condition}} and the \hyperref[condition:preparation-time]{\textit{Preparation-Time Optimality Condition}}—i.e., the ground-state preparation time should scale optimally as $\tau_{\rm prep} \sim 1/\Delta$~\cite{boixo2010necessary,PhysRevA.65.042308,albash2018adiabatic,lin2020near,PlanckianTime}. For long-range interacting systems, we found examples of protocols satisfying both conditions. The fact that criticality is not required for optimal ground-state performance is consistent with previous results for quantum-optical systems~\cite{Hotter2024Combining,Gietka2022Understanding,Mihailescu2026Mind}, which can be mapped onto long-range interacting systems. For short-range critical systems, our analysis reveals that local adiabatic protocols generally achieve an optimal ground-state preparation time, $\tau_{\rm prep} \sim 1/\Delta_c$. However, optimal preparation alone is insufficient to reach Heisenberg scaling in the GSAM framework; the target ground state must also be statically optimal, which is dictated by the constraint on the critical exponents, Eq.~\eqref{eq:condcriticalexponents}. This dual requirement explains why the 1D transverse-field Ising model---which achieves an optimal preparation time but lacks static optimality---fails to reach dynamical Heisenberg scaling. In contrast, critical probes that approach the static optimality condition, such as the XXZ model as $\gamma_z \to 1^-$, successfully combine near-optimal static sensitivity with optimal preparation time to approach the ultimate bound in Eq.~\eqref{boundGSHeis} arbitrarily closely.

With these results, our work clarifies that reaching precision beyond shot-noise scaling (and even Heisenberg scaling) is possible with short-range local interacting systems operating at criticality, even when the preparation time is taken into account. This is compatible with the shot-noise bound of Ref.~\cite{Shi2024SNL}, which applies at fixed protocol duration, see also Ref.~\cite{Gietka2021adiabaticcritical}: Heisenberg scaling emerges only along protocols of duration $\tau\sim1/\Delta$ growing with $N$.
More generally, our results provide guiding principles for the design of (short-range) many-body ground-state sensors that are optimal both in terms of particle number and sensing time.

Looking ahead, there are several other promising directions for future research. Assessing the robustness to noise and imperfections of our results is crucial. In this respect, the metrological advantage of ground states is expected to be more resistant to certain imperfections, such as particle loss, than that of delicate entangled states. Furthermore, ground-state metrology could be used to characterize non-linear properties of quantum signals, an approach that has been recently leveraged to recover properties of time-dependent signals~\cite{puig2026background,alderete2026thresholded}.
An interesting future question is how to improve the dynamical sensitivity by using more sophisticated ground-state preparation methods beyond adiabatic dynamics~\cite{lin2020near,PlanckianTime,puig2026warm,puig2025variational,gluza2024double,ding2025end}. 
Likewise, extending our framework to multiparameter Hamiltonian estimation is a natural direction, since it is crucial for describing complex systems \cite{Albarelli2020prespective,Mihailescu2024,10.21468/SciPostPhys.13.4.077,Previdi2025OHmolecule,Ullah2026Configuration}. A first step in this direction has recently been made in \cite{pedroni2026dynamicalthermalgroundstatemultiparameter}. 

\textit{Note added}.---Upon completion of this work, we became aware of a recent preprint \cite{chen2026quantum} that also investigates quantum ground-state metrology in short-range interacting systems, including the 1D Ising and XXZ models. Ref. \cite{chen2026quantum} focuses on optimal measurements and the impact of imperfections, making the two studies complementary.

\section*{Acknowledgements}
The authors thank Paolo Abiuso, Mohammad Mehboudi and Javier Lalueza-Puértolas for fruitful discussions. This research was funded in part by the Austrian Science Fund (FWF) ["NOQUS", Grant DOI: 10.55776/PAT1969224]. V. I. and M.P.-L. acknowledge funding 
from  the ATRAE Program (Grant ATR2024-154621) funded by MICIU/AEI/10.13039/501100011033. E.L.A. and V.I. acknowledge the support of the Catalonia Quantum Academy through its Student Mobility Fellowships program. J.C. acknowledges
support from ICREA Academia. Additional support was provided by grant PID2022-141283NB-I00 (MICIU/AEI/10.13039/501100011033). R.P. acknowledges the support of the SNF Quantum Flagship Replacement Scheme (Grant No. 215933). The authors acknowledge TU Wien Bibliothek for financial support through its Open Access Funding Programme.
\bibliography{extrabib} 
\onecolumngrid

\newpage 

\appendix
\begin{center}
    \textbf{\large{Supplementary Material for \\
    ``Reaching the Limits of Ground-State Metrology with Many-Body Probes''
}}
\end{center}

\section{Basic principles of universal phenomena in quantum critical systems}\label{Appx_critical_theo}
In this Appendix, we provide a brief summary of the universal finite-size scaling relations used in the main text for many-body systems close to a second-order quantum phase transition. Let us consider a $d$-dimensional $N$-particle Hamiltonian, $H(g)$, with linear size $L=N^{1/d}$ depending on a single control parameter, $g$. Then, a continuous quantum phase transition is a qualitative change in the ground state of $H(g)$ when the control parameter is tuned to the critical point, $g=g_c$, at zero temperature. Near this point, universal collective phenomena emerge and microscopic details become irrelevant at sufficiently large length scales. An example of this behavior is the fact that the characteristic length scale of quantum fluctuations, i.e., the correlation length $\xi$, diverges as follows:
\begin{equation}
    \xi\sim|g-g_c|^{-\nu}
\end{equation}
in the thermodynamic limit ($L\to\infty$) and sufficiently close to $g_c$. This non-analyticity is characterized by the correlation-length critical exponent $\nu$. Another relevant quantity in second-order quantum phase transitions is the characteristic energy scale $\Delta=E_1-E_0$, i.e., the spectral gap of $H(g)$. In the thermodynamic limit, this quantity closes as
\begin{equation}
    \Delta(g, N\to\infty)\sim\xi^{-z}\sim|g-g_c|^{\nu z},
\end{equation}
where $z$ is the so-called dynamical critical exponent. Naturally, by renormalization group arguments, this universal power-law behavior deviates for $\abs{g-g_c} \sim \mathcal{O}(1)$, i.e., deep within either phase \cite{Privman1990, Cardy1996}.

From the definition of $\xi$, it follows that in finite systems, finite-size effects emerge when $L \lesssim \xi$. That is,
\begin{equation}\label{crit_window_def}
    |g-g_c|\lesssim N^{-1/(d\nu)}.
\end{equation}
We will refer to this region as the critical window. The behavior of $\Delta$ for finite systems within and away from the critical window but sufficiently close to $g_c$ can be combined into a single expression. This is the well-known finite-size scaling ansatz~\cite{Fisher1972, Brzin1985, Binder2010}: 
\begin{equation}\label{FSS_gap}
    \Delta(g,N)=N^{-z/d}f\left((g-g_c)N^{1/(d\nu)}\right),
\end{equation}
with the universal dimensionless scaling function $f(x)$ such that
\begin{equation}
    f(x)\begin{cases}
        =\mathrm{const.}, \hspace{1cm}x\to0\\
        \sim |x|^{\nu z} ,\hspace{1.2cm}|x|\gg 1
    \end{cases},
\end{equation}
where $x\coloneq (g-g_c)N^{1/(d\nu)}$ corresponds to the scaling variable that controls the finite-size effects. Hence, within the critical region, the spectral gap scales as $\Delta\sim N^{-z/d}$, since $f(x)$ does not scale with system size there. As a last remark, we note that for sufficiently large system size, $\Delta$ acquires a dependence on $N$ only within the critical window; away from this region $\Delta\sim\mathcal{O}(1)$ for any $g$.

The last quantity of interest is the ground-state QFI at criticality, which grows with system size as follows \cite{CamposVenuti2007, GU2010FidelityQPT}:
\begin{equation}\label{QFIcritical}
    \mathcal{F}_\theta\sim N^{2/(d\nu)}.
\end{equation}
\section{Derivation of the universal scaling of the variance near a second-order quantum phase transition}\label{Appx_var_crit}
In this Appendix, we derive a condition on the critical exponents under which the ground-state variance of the signal Hamiltonian at criticality achieves quadratic scaling with the number of particles. Following Refs.~\cite{Gu2009, GU2010FidelityQPT}, we consider a $d$-dimensional system of linear size $L=N^{1/d}$, with $H_\theta=\theta H_\mathrm{S}+H_\mathrm{C}$ and $H_\mathrm{S}, o^{(r)}$ given in Eq.~\eqref{localtermsHS}. Therefore, we can write the variance of $H_\mathrm{S}$ with respect to the ground state $\ket{0}$ of $H_\mathrm{C}$ as follows:
\begin{equation}\label{variance_correl}
\begin{aligned}
\mathrm{Var}_{\ket{0}}[H_\mathrm{S}]
&=
\sum_{r r'}\langle o^{(r)}o^{(r')}\rangle_{\ket{0}}-\left(\sum_r\langle o^{(r)}\rangle_{\ket{0}}\right)^2\\
&=\sum_{lr'}
\left(
\langle o^{(l)}o^{(0)}\rangle_{\ket{0}}
-
\langle o^{(l)}\rangle_{\ket{0}}
\langle o^{(0)}\rangle_{\ket{0}}
\right) \\
&=
\sum_{lr'} C(l)
=
L^d\sum_{l} C(l).
\end{aligned}
\end{equation}
where in the second equality we used the translational invariance of the ground state and defined $l \coloneq r-r'$. In the last equality, we defined the correlator
\begin{equation}
    C(r) \coloneq \langle o^{(r)}o^{(0)}\rangle_{\ket{0}}
-
\langle o^{(r)}\rangle_{\ket{0}}
\langle o^{(0)}\rangle_{\ket{0}}.
\end{equation}
Near the critical point, the following scaling transformations hold:
\begin{equation}
    r \mapsto s r, \hspace{2cm} o^{(r)}\mapsto s^{-\Delta_o}o^{(r)},
\end{equation}
where $s>1$ and $\Delta_o$ is the scaling dimension of $o^{(r)}$. From these scaling transformations, it follows that $C(r)\mapsto s^{-2\Delta_o}C(r)$. As $\theta$ is the parameter that controls the phase transition, we will refer to $H_\mathrm{S}$ as the tuning Hamiltonian.

Therefore, going to the continuum in Eq.~\eqref{variance_correl}, we obtain:
\begin{equation}\label{var_scaling_crit}
    \mathrm{Var}_{\ket{0}}[H_\mathrm{S}]\sim \begin{cases}
        L^{(2/\nu-2z)}, \hspace{1cm} 2\Delta_o<d\\
        L^d\log L,\hspace{1.2cm} 2\Delta_o=d\\
        L^d,\hspace{2cm} 2\Delta_o>d
    \end{cases},
\end{equation}
where we used that $\Delta_o=d+z-1/\nu$, as $H_\mathrm{S}$ is the tuning Hamiltonian.

Hence, $\mathrm{Var}_{\ket{0}}[H_\mathrm{S}]\sim N^2$ is equivalent to the \hyperref[condition:ground-state-optimality-critical]{\textit{Ground-State Optimality Condition for Critical Models}} whenever $2\Delta_o<d$.

\section{Local adiabatic matrix elements in translationally invariant critical models}\label{App:matrels_crit_models}
As we present in Sec.~\ref{Sec:tprep_crit}, we divide the preparation path of $H(g)$ into three regions, which are defined in terms of $s \coloneq |g-g_c|$. We further assume that $g$ is the single relevant tuning parameter of the phase transition. In this Appendix, we show how the matrix elements in the local adiabatic condition, $\abs{\matrixel{1 (g)}{\partial_g H}{0 (g)}}$, scale with $N$ in each of these regions.

\subsection{Region (I): \texorpdfstring{$s\sim\mathcal{O}(1)$}{s ~ O(1)}}
Consider that $\partial_g H=\sum_iv^{(i)}$ is a translationally invariant operator acting on a $d$-dimensional lattice of $N$ sites. Since the ground state $\ket{0(g)}$ lives in the zero total momentum sector ($k_\mathrm{total}=0$), the state $\partial_gH\ket{0(g)}$ also has zero total momentum, and it will only couple to excited states that also have $k_\mathrm{total}=0$. Furthermore, we assume a quasiparticle description in which the driving operator preserves a symmetry that forbids single-quasiparticle transitions, and the lowest coupled excitation is a pair with opposite momenta. Then, the lowest excited state yielding a non-zero coupling will be the two-quasiparticle state
\begin{equation}
    \ket{1}=\ket{\mathbf{k}_\mathrm{min},-\mathbf{k}_\mathrm{min}},
\end{equation}
where the smallest coupled momentum satisfies $k_{\min}\propto N^{-1/d}$. We assume periodic boundaries for the momentum-space argument; extending the resulting scaling to open boundaries requires that boundary effects do not alter the leading size dependence.

From the previous argument, we compute
\begin{equation}
    \matrixel{1}{\partial_g H}{0}=\sum_{i=1}^N\matrixel{\mathbf{k}_\mathrm{min},-\mathbf{k}_\mathrm{min}}{v^{(i)}}{0}=N\matrixel{\mathbf{k}_\mathrm{min},-\mathbf{k}_\mathrm{min}}{v^{(0)}}{0}\eqqcolon \mathcal{F}(\mathbf{k}_\mathrm{min}),
\end{equation}
where in the second equality, we used the translational invariance of $\partial_g H$, i.e., $\matrixel{\mathbf{k}_\mathrm{min},-\mathbf{k}_\mathrm{min}}{v^{(i)}}{0}=e^{-i\mathbf{k}_\mathrm{total}r_i}\matrixel{\mathbf{k}_\mathrm{min},-\mathbf{k}_\mathrm{min}}{v^{(0)}}{0}$. Furthermore, since $k_\mathrm{total}=0$, the sum $\sum_{i=1}^{N}$ produces the overall factor $N$. In the last step, we have defined the form factor, i.e., the intensive matrix element of the local driving operator between the ground state and a quasiparticle pair of momenta $\pm \mathbf{k}$, as
\begin{equation}
    \mathcal{F}(\mathbf{k}) \coloneq N\,\matrixel{\mathbf{k},-\mathbf{k}}{v^{(0)}}{0}.
\end{equation}
Within the assumed description in terms of extended quasiparticles, the factor $N$ from the site sum cancels the two \(N^{-1/2}\) normalization factors of the quasiparticle modes. Thus, at fixed $g$ in the gapped region, we take $\mathcal F(k)$ to have no leading explicit dependence on $N$. The size dependence of the lowest-transition matrix element then enters through $k_{\min}$.

For sufficiently large $N$, $k_\mathrm{min} \to 0$. Assuming that the form factor is analytic near $k=0$ in this gapped region, its leading non-zero term gives
\begin{equation}\label{eq:Fk-taylor}
    \mathcal{F}(\mathbf{k}_\mathrm{min})\sim k_\mathrm{min}^p\propto \left(\frac{\pi}{N^{1/d}}\right)^p.
\end{equation}
Here, $p\ge0$ is the order of the first non-zero term in the small-momentum form factor. A non-zero limit gives $p=0$, whereas a linear zero gives $p=1$, as in the TFIM. Its value depends on the driving operator and the excitation channel, rather than on the presence of an internal symmetry alone.

In summary, the local adiabatic matrix element in region (I) (where $s\sim\mathcal{O}(1)$) follows the scaling
\begin{equation}
    \abs{\matrixel{1}{\partial_g H}{0}}\sim N^{-p/d},
\end{equation}
with $p\ge0$ determined by the small-momentum form factor.

\subsection{Region (II): \texorpdfstring{$1\gg s>N^{-1/(d\nu)}$}{1 >> s > N\textasciicircum(-1/(d nu))}}
In this region, the correlation length $\xi \sim s^{-\nu}$ emerges as an additional characteristic scale in the vicinity of the critical region. Consequently, the physical behaviour of the system is fundamentally governed by the dimensionless parameter $k\xi$.

Consider the following definition of the fidelity susceptibility:
\begin{equation}\label{fid_sus_matrelsg}
    \chi_F=\sum_{n>0}\frac{\abs{\matrixel{n}{\partial_gH}{0}}^2}{(E_n-E_0)^2},
\end{equation}
 where we have taken $\partial_g H$ as the perturbation. In this region, if $\partial_gH$ is the relevant tuning operator of the phase transition \cite{Gu2009, CamposVenuti2007},
\begin{equation}\label{eq_crit_exp_fid}
    \frac{\chi_F}{N}\sim s^{d\nu-2},
\end{equation}
where we have used that $d \nu < 2$. We assume that the leading singular contribution to the fidelity susceptibility is captured by the two-quasiparticle channel introduced above. Taking the continuum limit of the corresponding momentum sum then gives
\begin{equation}\label{eq_continuum_matrels_R2}
    \frac{\chi_F}{N}\sim\int\frac{\dd^dk}{(2\pi)^d}\frac{\abs{\mathcal{M}(\mathbf{k},s)}^2}{E(\mathbf{k},s)^2},
\end{equation}
where we have introduced the continuum intensive matrix element density, $\mathcal{M}(\mathbf{k},s)$ and rescaled the ground-state energy to zero. Following the arguments in Refs.~\cite{DeGrandi2010, CamposVenuti2007}, we consider the following universal scale-invariant forms of $\mathcal{M}(\mathbf{k},s)$ and $E(\mathbf{k},s)$:
\begin{equation}\label{universal_eq_matrls_energy}
    \begin{split}
        \mathcal{M}(\mathbf{k},s)&=s^\beta\Phi(\mathbf{k}\xi)\\
        E(\mathbf{k},s)&=s^{\nu z}\Omega(\mathbf{k}\xi),
    \end{split}
\end{equation}
where $\beta$ is to be found and we note that, for $k_\mathrm{min}\propto N^{-1/d}\xrightarrow[]{N\xrightarrow[]{}\infty}0$, we recover $E(0,s)\sim\Delta\sim s^{\nu z}$.

Substituting the quantities of Eq.~\eqref{universal_eq_matrls_energy} into Eq.~\eqref{eq_continuum_matrels_R2} and performing a change of variables such that $\mathbf{q}=\mathbf{k}\xi$, we find
\begin{equation}\label{int_regionII_matrls}
   \frac{\chi_F}{N}\sim s^{d\nu+2\beta-2\nu z}\int\dd^d q\frac{\Phi^2(\mathbf{q})}{\Omega^2(\mathbf{q})}.
\end{equation}
For large $q$, universality\footnote{This can be seen from the fact that $q$ large amounts to $s\xrightarrow[]{}0$, hence we expect a power-law behavior of the universal functions in this case. For instance, $\Omega(q)\sim q^\zeta$ for some $\zeta$ when $q\gg1$. Then, by Eq.~\eqref{universal_eq_matrls_energy} we find $E(k,s)=s^{\nu (z-\zeta)}k^\zeta$, and by imposing that this quantity is $s$-independent we find $\zeta=z$, i.e., $\Omega(q)\sim q^z$.} fixes $\Phi(q)\sim q^{z-1/\nu}$ and $\Omega(q)\sim q^z$. Therefore, the integrand of Eq.~\eqref{int_regionII_matrls} converges for $d\nu<2$ and $q\gg1$, while being regular for other values of $q$. Then, we conclude that this integral is a universal constant dominated by $q\sim\mathcal{O}(1)$. Equating Eq.~\eqref{int_regionII_matrls} to Eq.~\eqref{eq_crit_exp_fid} we obtain $\beta=\nu z-1$, i.e.,
\begin{equation}\label{density_matrels_II}
    \mathcal{M}(\mathbf{k},s)=s^{\nu z-1}\Phi(\mathbf{k}\xi).
\end{equation}
Lastly, we compute the actual scaling of $\abs{\matrixel{1}{\partial_g H}{0}}$ in this region. To do so, we evaluate Eq.~\eqref{density_matrels_II} at $k_\mathrm{min}\sim N^{-1/d}$:
\begin{equation}
    \abs{\matrixel{1}{\partial_g H}{0}}\sim\mathcal{M}(k_\mathrm{min},s)\sim s^{\nu z-1}\Phi\left(\frac{\xi}{N^{1/d}}\right).
\end{equation}
Since $N^{1/d}\gg\xi$ in this region, $x:=\xi/N^{1/d}\ll1$. We assume that the small-momentum form factor retains the behavior $\Phi(x)\sim x^p$, with the same exponent $p$ as in region (I). It follows that
\begin{equation}
    \abs{\matrixel{1}{\partial_g H}{0}}\sim N^{-p/d}s^{\nu(z-p)-1}.
\end{equation}

\subsection{Region (III) \texorpdfstring{$N^{-1/(d\nu)}\geq s\geq 0$}{N\textasciicircum(-1/(d nu)) >= s >= 0}}
In region (III), finite-size effects cut off the critical length scale at $L=N^{1/d}$, and the modes with momenta of order $L^{-1}$ control the leading singular contribution to the fidelity susceptibility. We assume that the lowest coupled transition carries a non-zero, size-independent fraction of this contribution. At the level of scaling,
\begin{equation}
    \chi_F\sim\frac{\abs{\matrixel{1}{\partial_g H}{0}}^2}{\Delta_c^2},
\end{equation}
where $\Delta_c\sim N^{-z/d}$ and $g$ is the relevant tuning parameter of the transition. Inside the critical window, $\chi_F\sim N^{2/(d\nu)}$ (see Eq.~\eqref{QFIcritical}). Hence, from the previous equations, it readily follows that
\begin{equation}\label{eq:C16}
    \abs{\matrixel{1}{\partial_g H}{0}}\sim N^{(1/\nu-z)/d}
\end{equation}
Inserting these matrix-element scalings into Eq.~\eqref{preptimeintegral} yields $\tau_{\mathrm{prep}}\sim N^{z/d}\sim1/\Delta_c$, within the lowest-transition approximation.
\section{A succinct introduction to Brillouin-Wigner perturbation theory}\label{App_BWPT}
As we will resort to the machinery of Brillouin-Wigner Perturbation Theory (BWPT) in the next section, we briefly present here the basics of this framework; the interested reader may consult the more exhaustive treatments in, e.g., \cite{Hubac2009-ta, mila}. BWPT is chosen as it straightforwardly extends to higher orders in perturbation theory while avoiding the need for separate treatment of the non-degenerate and degenerate cases. As usual, we are interested in solving the following eigenproblem:
\begin{equation}\label{PT_eigenprob}
(H^{(0)}+\lambda V)\ket{\psi_n}=E_n\ket{\psi_n},    
\end{equation}
where $\lambda V$ corresponds to the perturbative term acting on the exactly solvable Hamiltonian $H^{(0)}$, i.e., $H^{(0)}\ket{\psi_n^{(0)}}=E^{(0)}_n\ket{\psi_n^{(0)}}$. We define the following projection operators for a given set of quantum numbers, $\aleph$, and its complement, $\bar{\aleph}$, as follows:
\begin{equation}
    \begin{split}
        P&=\sum_{n\in\aleph}\ketbra{\psi_n^{(0)}}{\psi_n^{(0)}},\\
        Q&=\mathbb{I}-P=\sum_{n\in\bar{\aleph}}\ketbra{\psi_n^{(0)}}{\psi_n^{(0)}}.
    \end{split}
\end{equation}
From the previous definitions, one can convert Eq.~\eqref{PT_eigenprob} into an equation on $P\ket{\psi_n}$:
\begin{equation}\label{eigeneq_BW}
    PH_\mathrm{eff}P\ket{\psi_n}=E_nP\ket{\psi_n},
\end{equation}
where we have defined $H_\mathrm{eff} \coloneq H+\lambda^2VQ(E_n-QHQ)^{-1}QV$, which corresponds to the effective Hamiltonian that lives in the Hilbert subspace spanned by the states associated with $P$. Performing a Taylor series with respect to the perturbative parameter embedded in $V$, we can write $H_\mathrm{eff}$ as an infinite series. As we will be interested in the first non-zero contribution of this effective Hamiltonian, we will truncate the series at the $N$-th order\footnote{In Brillouin–Wigner degenerate perturbation theory, higher-order expressions depend explicitly on the exact eigenvalue of the full Hamiltonian $H=H^{(0)}+\lambda V$. A standard way to remove this dependence is to expand around the unperturbed ground-state energy, $E_0^{(0)}$ \cite{Kato1995, Takahashi1977}. In our leading-order calculations, we will adopt this expansion.}:
\begin{equation}\label{DPT_WB}
H^{(N)}_\mathrm{eff}=H^{(0)}+\lambda V+\lambda^2\sum_{l\not \in \aleph}\frac{V\ket{\psi_l^{(0)}}\bra{\psi_l^{(0)}}V}{E_0^{(0)}-E_{l}^{(0)}}+\cdots+\lambda^N\sum_{l,\dots,l^{(N-1)}\not \in\aleph} \frac{V\ket{\psi_l^{(0)}}\bra{\psi_l^{(0)}}V\ket{\psi_{l'}^{(0)}}\cdots\bra{\psi_{l^{(N-1)}}^{(0)}}V}{\Pi_{k=1}^{N-1}E_0^{(0)}-E_{l^{(k)}}^{(0)}}.
\end{equation}
We now consider the particular case where we constrain our effective Hilbert space to be the subspace spanned by $\left\{\ket{\psi_n^{(0)}},\ket{\psi_{n'}^{(0)}}\right\}$, implying $P=\ketbra{\psi_n^{(0)}}{\psi_n^{(0)}}+\ketbra{\psi_{n'}^{(0)}}{\psi_{n'}^{(0)}}$. Hence, Eq.~\eqref{eigeneq_BW} translates to the following matrix equation:
\begin{equation}
    \begin{pmatrix}
        H_{\mathrm{eff},n,n}^{(N)} & H_{\mathrm{eff},n,n'}^{(N)}\\\\
        H_{\mathrm{eff},n',n}^{(N)} & H_{\mathrm{eff},n',n'}^{(N)}
    \end{pmatrix}
    \begin{pmatrix}
        \braket{\psi_n^{(0)}}{\psi_m}\\\\
        \braket{\psi_{n'}^{(0)}}{\psi_m}
    \end{pmatrix}
    =E_m
    \begin{pmatrix}
        \braket{\psi_n^{(0)}}{\psi_m}\\\\
        \braket{\psi_{n'}^{(0)}}{\psi_m}
    \end{pmatrix},
\end{equation}
where $H_{\mathrm{eff},n,n'}^{(N)}=\bra{\psi_{n}^{(0)}}H_\mathrm{eff}^{(N)}\ket{\psi_{n'}^{(0)}}$ and $H_{\mathrm{eff}}^{(N)}$ is given by Eq.~\eqref{DPT_WB}. Therefore, by solving this eigenproblem we obtain the two lowest eigenstates and their associated eigenvalues in the effective subspace.

Lastly, we show a way to bound the error we are making in truncating the expansion of $H_\mathrm{eff}$ around $\lambda=0$. We first note that we can express the expansion of $H_\mathrm{eff}$ in terms of $\lambda$ as follows:
\begin{equation}\label{Heff_expansion}
    H_\mathrm{eff}=H^{(0)}+\sum_{m\geq 1}\lambda^mT^{(m)},
\end{equation}
where the operators $T^{(m)}$ are of the form
\begin{equation}
    T^{(m)}=V\frac{\ketbra{\psi_l^{(0)}}{\psi_l^{(0)}}}{E_0^{(0)}-E_l^{(0)}}V\dots\frac{\ketbra{\psi_l^{(0)}}{\psi_l^{(0)}}}{E_0^{(0)}-E_l^{(0)}}V,
\end{equation}
with $m$ factors of $V$ and $m-1$ denominators at the $m$-th order. We now define the remainder of the expansion of $H_\mathrm{eff}$ as
\begin{equation}\label{boundRemainderPT}
    R(\lambda)=\norm{H_\mathrm{eff}-H_\mathrm{eff}^{(N)}}_\mathrm{op} =\norm{\sum_{m\geq N+1}\lambda^mT^{(m)}}_\mathrm{op} \leq \sum_{m\geq N+1}\lambda^m\norm{T^{(m)}}_\mathrm{op},
\end{equation}
with $\norm{\cdot}_\mathrm{op}$ the usual operator norm. Then, we can upper bound $\norm{T^{(m)}}_\mathrm{op}$ (and, hence, $R(\lambda)$) by bounding the following quantity:
\begin{equation}
    \norm{V\frac{\ketbra{\psi_l^{(0)}}{\psi_l^{(0)}}}{E_0^{(0)}-E_l^{(0)}}}_\mathrm{op}.
\end{equation}
\section{Analytical derivations of the spectral gap and QFI in the two-mode Bose-Hubbard model}\label{App_BH_gs}
Let us now consider the two-mode Bose-Hubbard Hamiltonian in Eq.~\eqref{H_BH}. Using Schwinger’s oscillator model of angular momentum \cite{Sakurai_Napolitano_2020}, the previous model can be mapped to a spin-squeezing type. To this end, we first define the following operators:
\begin{equation}
    \begin{split}
        J_+&\coloneq a^\dagger_La_R\\
        J_-&\coloneq a^\dagger_Ra_L\\
        J_z&\coloneq\frac{1}{2}(n_L-n_R)
    \end{split}
\end{equation}
It is easy to show that the previous operators satisfy the algebra of angular momentum operators, under the identification $N_L\mapsto j+m_z$ and $N_R\mapsto j-m_z$. Besides, the eigenvalue $N_L$ ($N_R$) can be identified with the number of spin-up (spin-down) particles in the ensemble. Within this correspondence, bosonic states belonging to the two-mode Fock space can be mapped to the eigenstates of the operators $\{J_z,\mathbf{J}^2\}$, e.g., $\ket{N,0}\mapsto\ket{j=N/2,m_z=N/2}$ and $\ket{0,N}\mapsto\ket{j=N/2,m_z=-N/2}$. As we did in the spin-squeezing model, we assume from now on that $j=j_\mathrm{max}=N/2$ with the number of bosons, $N>1$, fixed. Now, from the definitions $n=n_R+n_L$ and $2J_z=n_L-n_R$ it follows that
\begin{equation}
    \begin{split}
        n_L&=n/2+J_z\\
        n_R&=n/2-J_z.
    \end{split}
\end{equation}
Furthermore, using $\mathbf{J}^2=\frac{n}{2}(\frac{n}{2}+1)$ we compute $n_L(n_L-1)+n_R(n_R-1)=2J^2+2J_z^2-2n=2J^2+2J_z^2-4j_\mathrm{max}\mathbb{I}$, where in the last equality, we used that the number operator is diagonal in the subspace spanned by $\{\ket{j,m_z}\}$ for a fixed number of bosons and $j=j_\mathrm{max}$. Therefore, the Hamiltonian in Eq.~\eqref{H_BH} can be rewritten in terms of the collective spin-$N$/2 operators as follows:
\begin{equation}\label{BH_spin_model}
    H_\theta=-2KJ_x+U(J_z^2+J^2)+2\theta J_z.
\end{equation}
Note that we have dropped the constant energy shift $-2Uj_\mathrm{max}\mathbb{I}$.

Now, we compute the QFI and the spectral gap for this new Hamiltonian. First, we note that when $K=0$ and $U<0$, the ground state is doubly degenerate with energy $E_0=U [(\pm N/2)^2+N/2(N/2+1)]$ associated with the states $\ket{j=N/2,m_z=\pm N/2}\eqcolon\ket{\pm}$. Hence, we shall use the machinery of Brillouin-Wigner perturbation theory, presented in Appendix \ref{App_BWPT}, to diagonalize the control part of the Hamiltonian in Eq.~\eqref{BH_spin_model} in the regime of $U<0$ and $\Lambda \ll 1$. The control part of this Hamiltonian can be written as 
\begin{equation}
H_\mathrm{C}=\Lambda\underbrace{(-2|U|J_x)}_{V}+\underbrace{U(J_z^2+J^2)}_{H^{(0)}},
\end{equation}
so that our expansion will be a power series in $\Lambda$ and $V$ is the perturbative Hamiltonian. We restrict ourselves to the degenerate subspace spanned by $\{\ket{+},\ket{-}\}$, so that $P=\ketbra{+}{+}+\ketbra{-}{-}$. This is in order, since the degenerate subspace is well isolated with respect to the next excited level, which has an energy $E_1^{(0)}=U[(N/2-1)^2+N/2(N/2+1)]$. Hence, $E_1^{(0)}-E_0^{(0)}=U(1-N)$. 

We now proceed to compute the matrix elements of the effective Hamiltonian up to the $N$-th order in perturbation theory---see Eq.~\eqref{DPT_WB}---in this subspace. We stress the fact that the first non-zero contribution in the expansion of $H_\mathrm{eff}$ occurs at the order dictated by the fixed number of particles in the system, $N$. First, we recall that $V=-2|U|J_x=-|U|(J_++J_-)$ and that $J_-\ket{j,m_z}=\sqrt{j(j+1)-m_z(m_z-1)}\ket{j,m_z-1}$. From this we see that the first non-zero contributions in the matrix elements $\bra{+}H_\mathrm{eff}^{(N)}\ket{-}$ and $\bra{-}H_\mathrm{eff}^{(N)}\ket{+}$ occur at order $N=2j_\mathrm{max}$. At this order, there exists a set of $N$ quantum numbers, $\{m_z^{(k)}\}$, that allows the connection of $\ket{+}$ with $\ket{-}$, and vice versa, by the action of climbing or descending the ladder $2j_\mathrm{max}-1$ times. Thus, the elements of the set $\{m_z^{(k)}\}$, corresponding to $\aleph$ in Eq.~\eqref{DPT_WB}, are fixed to take the single values $m_z'=j_\mathrm{max}-1,\hspace{0.1cm}m_z''=j_\mathrm{max}-2,\dots,\hspace{0.1cm}m_z^{(N-1)}=j_\mathrm{max}-N+1$. Hence, the summation only consists of one single term, and we can treat separately the computations for the numerator and denominator of $H_\mathrm{eff}^{(N)}$. Therefore, the numerator of $\bra{-}H_\mathrm{eff}^{(N)}\ket{+}$ reduces to 
\begin{equation}\label{BH_nume}
(-|U|)^N\bra{-}J_-^{N}\ket{+}=(-|U|)^NN!. 
\end{equation}
The equality is obtained by first considering the action of the ladder operator $J_-$ and then computing
\begin{equation}
    \begin{split}
        J_-^N\ket{N/2,N/2}&=\prod_{l=0}^{N-1}\sqrt{N/2(N/2+1)-(N/2-l-1)(j-l)}\ket{N/2,-N/2}\\
        &=\prod_{l=0}^{N-1}\sqrt{(l+1)(N-l)}\ket{N/2,-N/2}=\sqrt{\prod_{l=0}^{N-1}(l+1)}\sqrt{\prod_{l=0}^{N-1}(N-l)}\ket{N/2,-N/2}\\
        &=\sqrt{N!}\sqrt{\frac{\Gamma(N+1)}{\Gamma(1)}}\ket{N/2,-N/2}=N!\ket{N/2,-N/2},
    \end{split}
\end{equation}
where in the last equation we have used that $\Gamma(n)=(n-1)!$ valid $\forall n\in\mathbb{Z}^+$, where $\Gamma (n)$ is the Gamma function. Likewise, the numerator of $\bra{+}H_\mathrm{eff}^{(N)}\ket{-}$ becomes $(-|U|)^NN!$.

To compute the denominator of these matrix elements, we note that $E_0^{(0)}=U(N^2/4+N/2(N/2+1))$, since $U<0$ and $E_{m_z^{(k)}}^{(0)}=U[(N/2-k)^2+N/2(N/2+1)]$. Hence:
\begin{equation}\label{BH_deno}
\begin{split}
    \prod^{N-1}_{k=1}E_0^{(0)}-E_{m_z^{(k)}}^{(0)}&=\prod^{N-1}_{k=1}Uk(N-k)=U^{N-1}\left(\prod^{N-1}_{k=1} k\right)\left( \prod^{N-1}_{k=1} (N-k)\right)=U^{N-1}[(N-1)!]^2.
\end{split}
\end{equation}
Therefore, from Eqs.~\eqref{BH_nume} and~\eqref{BH_deno}, the off-diagonal matrix elements read:
\begin{equation}\label{def_q}
    \bra{+}H_\mathrm{eff}^{(N)}\ket{-}=\bra{-}H_\mathrm{eff}^{(N)}\ket{+}=\frac{(-|U|)^NN!}{U^{N-1}[(N-1)!]^2}\coloneq q.
\end{equation}
The diagonal terms are easily computed to be:
\begin{equation}
    \bra{+}H_\mathrm{eff}^{(N)}\ket{+}=\bra{-}H_\mathrm{eff}^{(N)}\ket{-}=E_0^{(0)}=U[N^2/4+N/2(N/2+1)].
\end{equation}
These matrix elements yield the effective Hamiltonian $2 \times 2$ matrix in the degenerate space spanned by $\{\ket{N/2,N/2},\ket{N/2,-N/2}\}$:
\begin{equation}
    H_\mathrm{eff}^{(N)}=\begin{pmatrix}
        E_0^{(0)} & q\Lambda^N\\
        q\Lambda^N & E_0^{(0)}
    \end{pmatrix}=E_0^{(0)}\mathbb{I}+q\Lambda^N\sigma_x,
\end{equation}
with $q$ defined in Eq.~\eqref{def_q}. Hence, the ground and first excited states are given by
\begin{equation}
 \begin{split}
     \ket{0}&=\frac{1}{\sqrt{2}}(\ket{N/2,N/2}+\ket{N/2,-N/2}),\\
    \ket{1}&=\frac{1}{\sqrt{2}}(\ket{N/2,N/2}-\ket{N/2,-N/2}),     
 \end{split}   
\end{equation}
with energies $E_{\mp}=E_0^{(0)}\mp q\Lambda^N$, respectively. From this, the spectral gap of $H_\mathrm{eff}$ is readily computed to be
\begin{align}\label{Gap_BH_appx}
    \Delta&=2|q|\Lambda^N+\mathcal{O}(\Lambda^{N+2})=2|U|\dfrac{N}{(N-1)!}\Lambda^{N}+\mathcal{O}(\Lambda^{N+2}).
\end{align}
Furthermore, one can see that the correction to the first non-zero contribution occurs at order $N+2$, as there is no path that connects the states $\ket{+}$ and $\ket{-}$ in $N+1$ ladder steps. We note that in $\mathcal{O}(\Lambda^{N+2})$ a constant involving the fixed number of particles, $N$, is hidden. This constant decays faster than the first contribution.

The ground-state QFI follows from Eq.~\eqref{gs_QFI} by considering only the coupling of $H_\mathrm{S}$ between the ground and first excited states:
\begin{equation}\label{QFI_BH_app}
\mathcal{F}_\theta
=
\frac{4}{\Delta^2}
\left|\matrixel{1}{2J_z}{0}\right|^2
+
O(\Lambda^{-2N+2})
=
\frac{4N^2}{\Delta^2}
+
O(\Lambda^{-2N+2})
=
\frac{\norm{H_\mathrm{S}}^2}{\Delta^2}
+
O(\Lambda^{-2N+2}).
\end{equation}
Hence, the upper bound on the QFI---see Eq.~\eqref{boundGS}---is asymptotically attained in the limit $\Lambda\to0$ for a fixed $N$, as shown in the main text. 

We now keep $K>0$ and $U<0$ fixed and let $N\to\infty$, so that $\Lambda=K/|U|$ remains fixed. Starting from either endpoint Fock state, $\ket{N,0}$ or $\ket{0,N}$, transferring one boson to the empty mode increases the interaction energy by $|U|(N-1)$. The corresponding matrix element of the hopping term $H_{\mathrm{hop}}=-K(a_L^\dagger a_R+a_R^\dagger a_L)$ has magnitude $K\sqrt{N}$. Their ratio therefore vanishes as $N\to\infty$:
\begin{equation}
    \frac{K\sqrt{N}}{|U|(N-1)} =\frac{\Lambda\sqrt{N}}{N-1}\sim\frac{\Lambda}{\sqrt{N}}\longrightarrow0.
\end{equation}
The remaining Fock configurations are separated from the endpoint states by interaction-energy costs of at least $|U|(N-1)$. At fixed $\Lambda$, hopping within this remaining subspace preserves an energy separation of order $|U|N$ for sufficiently large $N$. Consequently, the total probability outside the span of $\{\ket{N,0},\ket{0,N}\}$ is $\mathcal{O}(\Lambda^2/N)$ in either of the two lowest eigenstates. Exchange symmetry therefore implies that these eigenstates approach the symmetric and antisymmetric cat states. Hence $\abs{\matrixel{1}{H_\mathrm{S}}{0}}/N\to1$, and considering the dominant first excited contribution,
\begin{equation}
    1\leftarrow\frac{\abs{\matrixel{1}{H_\mathrm{S}}{0}}^2}{N^2}\leq\frac{\Delta^2\mathcal{F}_\theta}{4N^2}\leq 1,
\end{equation}
that is $\Delta^2\mathcal{F}_\theta=4N^2$ for $\Lambda$ fixed and $N\to\infty$.

The spectral gap can be obtained from \cite{Garg1998DiscreteWKB} in the limit of fixed $\Lambda$ and $N\to\infty$,
\begin{equation}
    \Delta=\sqrt{\frac{2}{\pi}}\abs{U}N^{3/2}\left(\frac{e\Lambda}{N}\right)^N,
\end{equation}
which coincides with the leading term of Eq.~(\ref{Gap_BH_appx}) using Stirling's formula in the latter. Therefore, the leading contribution in the expansion of the spectral gap for a fixed $N$ in the limit $\Lambda\to0$ coincides with the large-$N$ asymptotic gap at fixed $\Lambda$.

The QFI therefore takes the following asymptotic form in this limit:
\begin{equation}
    \mathcal{F}_\theta=\frac{2\pi}{\abs{U}^2N}\left(\frac{N}{e\Lambda}\right)^{2N},
\end{equation}
which, again, coincides with the leading term of Eq.~\eqref{QFI_BH_app} using Stirling's formula in the latter.

\section{Analytical derivations of the spectral gap and QFI in the transverse-field 1D Ising model}\label{App_Ising_gs}
Consider the 1D transverse-field Ising model with periodic boundary
conditions, where we fix the
exchange interaction parameter to unity ($J=1$):
\begin{equation}\label{H_ising}
    H_\theta=-\sum_{i=1}^{N}\sigma_x^{(i)}\sigma_x^{(i+1)}-\theta\sum_{i=1}^N\sigma_z^{(i)}.
\end{equation}
We will consider an even number of particles, $N$, in order to have a non-degenerate ground state. The ground state of this Hamiltonian undergoes a second-order quantum phase transition at the critical point $\theta=J=1$, with critical exponents $z=1$ and $\nu=1$. As we always assume that we estimate the unknown parameter around the central value $\theta=0$, we shift its value such that $\theta\mapsto\theta+1$ so that the Hamiltonian is now critical at $\theta=0$. Hence,
\begin{equation}\label{H_ising_displaced_app}
    H_\theta=-\sum_{i=1}^{N}\sigma_x^{(i)}\sigma_x^{(i+1)}-(\theta+1)\sum_{i=1}^N\sigma_z^{(i)},
\end{equation}
so that the interaction term, $H_\mathrm{C}$, is always located at the critical point. In order to analytically find the spectrum of this term one can perform a Jordan-Wigner transformation, yielding \cite{PFEUTY197079}
\begin{equation}\label{spectrum_isign1d}
H_\mathrm{C}=\sum_{k>0}\Lambda_k(\eta_k^\dagger\eta_k-1),
\end{equation}
where $\Lambda_k=2\sqrt{\varepsilon_k^2+\zeta_k^2}$, with $\varepsilon_k\coloneq J\cos k+\theta$ and $\zeta_k\coloneq J\sin k$. Likewise, $\eta_k$ refers to the fermionic annihilation operator. Given periodic boundary conditions and even $N$, the momenta follow the relation $k=\frac{(2n+1)\pi}{N}$, with $n=0,\dots,N/2-1$.

From Eq.~\eqref{spectrum_isign1d} and rescaling the zero-point energy such that the ground-state energy is zero, the spectral gap $\Delta$ at the critical point can be computed as
\begin{equation}
    \Delta=\Lambda_{k_\mathrm{min}}=2\sqrt{2-2\cos \pi/N}=4\sin \frac{\pi}{2N}\approx2\pi/N,
\end{equation}
where in the second equality we have used that $k_\mathrm{min}=\pi-\pi/N$, while in the approximation we took $N\gg1$. As computed in \cite{Damski2013}, the zero-temperature QFI of the Hamiltonian in Eq.~\eqref{H_ising}, which here we will denote as $\mathcal{G}_\theta$, is given by
\begin{equation}
    \mathcal{G}_\theta=\frac{N^2}{4\theta^2}\frac{\theta^N}{(\theta^N+1)^2}+\frac{N}{4\theta^2}\frac{\theta^N-\theta^2}{(\theta^N+1)(\theta^2-1)},
\end{equation}
which, at the critical point $\theta_c=1$, becomes
\begin{equation}
    \mathcal{G}_{\theta_c=1}=\frac{N^2}{8}-\frac{N}{8}\approx\frac{N^2}{8},
\end{equation}
where again the approximation holds for $N\gg1$. To compute the zero-temperature QFI associated with the Hamiltonian in Eq.~\eqref{H_ising_displaced_app}, we use the fidelity susceptibility in Eq.~\eqref{fid_sus_def}. This quantity is invariant under constant shifts of the parameter, $\theta\mapsto\theta+b$, with $b\in\mathbb{R}$. Such a shift only adds a constant to the energy spectrum and leaves the eigenstates and the perturbation $\partial_\theta H_\theta$ unchanged. Therefore,
\begin{equation}
    \Delta^2\mathcal{F}_{\theta_c=0}=\Delta^2\mathcal{G}_{\theta_c=1}\approx \pi^2/2< 4N^2 =\norm{H_\mathrm{S}}^2,
\end{equation}
as shown in the main text.

For completeness, let us now compute $\mathrm{Var}_{\ket{0}}[H_\mathrm{S}]$. We recall from the Jordan-Wigner transformation that $\sigma_z^{(i)}=1-2\eta_i^\dagger\eta_i$. Hence,
\begin{equation}
    H_\mathrm{S}=\sum_{i=1}^N\sigma_z^{(i)}=\sum_{i=1}^N1-2n_i=N-2\mathcal{N},
\end{equation}
where $n_i\coloneq\eta^\dagger_i\eta_i$ and $\mathcal{N}\coloneq\sum_{i=1}^Nn_i$. It is well known that, for even $N$ and with periodic boundary conditions, the ground state of $H_{\rm C}$ takes the following form \cite{Sachdev2011}:
\begin{equation}\label{GS_TFIM}
    \ket{0}=\bigotimes_{k>0}(u_k+v_k\eta_k^\dagger\eta_{-k}^\dagger)\ket{\mathrm{vac}},
\end{equation}
where $\ket{\mathrm{vac}}$ denotes the zero-excitation state, $u_k=\frac{\Lambda_k+2(\theta+J\cos{k})}{\sqrt{2\Lambda_k(\Lambda_k+2(\theta+J\cos{k}))}}$, $v_k=\frac{2iJ\sin{k}}{\sqrt{2\Lambda_k(\Lambda_k+2(\theta+J\cos{k}))}}$ and $|u_k|^2+|v_k|^2=1$. We further define $\mathcal{N}_k\coloneq n_k+n_{-k}$, where $n_k$ is the Fourier-transformed number operator at momentum $k$. From the previous equations, it can be shown that $\expval{\mathcal{N}_k}_{\ket{0}}=2|v_k|^2$ and $\expval{\mathcal{N}_k^2}_{\ket{0}}=4|v_k|^2$. Therefore,
\begin{equation}
    \mathrm{Var}_{\ket{0}}[\mathcal{N}_k]=4|v_k|^2(1-|v_k|^2).
\end{equation}
From here, it follows that
\begin{equation}
    \begin{split}
        \mathrm{Var}_{\ket{0}}[H_\mathrm{S}]=4\mathrm{Var}_{\ket{0}}[\mathcal{N}]&=4\mathrm{Var}_{\ket{0}}\left[\sum_{k>0}\mathcal{N}_k\right]=16\sum_{k>0}|v_k|^2(1-|v_k|^2),
    \end{split}
\end{equation}
where we have used the fact that $\mathrm{Cov}_{\ket{0}}[\mathcal{N}_l,\mathcal{N}_q]=0$ for $l\neq q$, since the ground state in Eq.~\eqref{GS_TFIM} factorizes into independent ($k$, $-k$) sectors. Then, using that $|v_k|^2=\frac{1-\cos{k}}{4(1+\cos{k/2})}$ at the critical point, we obtain
\begin{equation}
    \mathrm{Var}_{\ket{0}}[H_\mathrm{S}]=2\sum_{k>0}(1-\cos{k}).
\end{equation}
Recalling the quantization condition, $k=\frac{(2l+1)\pi}{N}$ for $l=0,\dots,N/2-1$, we arrive at the final expression
\begin{equation}
        \mathrm{Var}_{\ket{0}}[H_\mathrm{S}]=N-2\sum_{l=0}^{N/2-1}\cos{\frac{(2l+1)\pi}{N}} = N,
\end{equation}
which manifests the sub-optimality of the transverse-field 1D Ising model for ground-state metrology.

\section{Numerical analysis of the static external-field XXZ model}\label{Appx:XXZ_numeric}

In this Appendix, we present the numerical data obtained for the external-field XXZ model discussed in Sec.~\ref{Sec_XXZ_GS}. We consider the open XXZ chain,
\begin{equation}
    H_{\theta} =
    -\sum_{i=1}^{N-1}
    \left[
        \sigma_x^{(i)}\sigma_x^{(i+1)}
        +\sigma_y^{(i)}\sigma_y^{(i+1)}
        +\gamma_z\sigma_z^{(i)}\sigma_z^{(i+1)}
    \right]
    +\theta\sum_{i=1}^{N}\sigma_x^{(i)},
\end{equation}
All numerical calculations are performed at $\theta=0$, using $N=96,128,160,192$ and $\gamma_z=0.99,0.995,0.999,0.9999,0.99999,0.999999$. The numerical sweep below evaluates the spectral gap $\Delta$ and the signal variance $\operatorname{Var}_{|0\rangle}[H_{\mathrm S}]$. As shown in Sec.~\ref{Sec_XXZ_GS}, $4\operatorname{Var}_{|0\rangle}[H_{\mathrm S}]\approx\Delta^2\mathcal{F}_\theta$ in the regime considered.

\subsection{Fits of the signal variance}
\begin{table*}[t]
    \centering
    \small
    \caption{
        Fits of
        $4\operatorname{Var}_{|0\rangle}[H_{\mathrm S}]
        =c_{\mathrm{fit}}N^{p_{\mathrm{fit}}}
        +a_{\mathrm{fit}}N$
        using $N=96,128,160,192$.
        All three parameters are free.
        The critical prediction
        $p_{\mathrm{crit}}=2-\arccos(\gamma_z)/\pi$
        is included for comparison.
    }
    \label{tab:xxz_variance_fits}
    \begin{tabular}{rrrrr}
        \toprule
        $1-\gamma_z$
        & $c_{\mathrm{fit}}$
        & $p_{\mathrm{fit}}$
        & $a_{\mathrm{fit}}$
        & $p_{\mathrm{crit}}$ \\
        \midrule
        $10^{-2}$ & 2.372417 & 1.9555223 & 2.109473 & 1.9549466 \\
        $5 \times 10^{-3}$ & 2.277653 & 1.9687650 & 2.210137 & 1.9681557 \\
        $10^{-3}$ & 2.132049 & 1.9868748 & 2.560756 & 1.9857636 \\
        $10^{-4}$ & 2.018146 & 1.9983435 & 3.701369 & 1.9954984 \\
        $10^{-5}$ & 2.000377 & 1.9999660 & 3.993350 & 1.9985765 \\
        $10^{-6}$ & 2.000004 & 1.9999996 & 3.999927 & 1.9995498 \\
        \bottomrule
    \end{tabular}
\end{table*}
For each anisotropy, we fit
\begin{equation}
    4\operatorname{Var}_{|0\rangle}[H_{\mathrm S}]
    =
    c_{\mathrm{fit}}N^{p_{\mathrm{fit}}}
    +a_{\mathrm{fit}}N,
    \label{eq:xxz_variance_fit}
\end{equation}
with all three parameters free, minimizing squared relative
residuals. Here, $c_{\mathrm{fit}}$ is the coefficient of the
leading term in this finite-size fit, while
$a_{\mathrm{fit}}$ accounts for a linear correction.
The notation connects with the coefficient $c$ used in the
main text, where the leading critical form is written as
$4\operatorname{Var}_{|0\rangle}[H_{\mathrm S}]
\sim cN^{2-\delta}$. The results are reported in Table~\ref{tab:xxz_variance_fits}, together with the critical prediction
\begin{equation}
    p_{\mathrm{crit}}\coloneq2-\delta,
    \qquad
    \delta\coloneq\frac{\arccos(\gamma_z)}{\pi}.
\end{equation}
For $\gamma_z=0.99,0.995,0.999$, the fitted exponents are
close to the critical prediction. Closer to the endpoint,
$p_{\mathrm{fit}}$ approaches $2$ more rapidly than
$p_{\mathrm{crit}}$ over the accessible sizes, while
$c_{\mathrm{fit}}$ approaches $2$ and $a_{\mathrm{fit}}$
approaches $4$.
This behavior is consistent with the collective limit of
the finite chain,
\begin{equation}
    4\operatorname{Var}_{|0\rangle}[H_{\mathrm S}]
    \simeq 2N^2+4N.
\end{equation}

\subsection{Fits of the spectral gap}
For each anisotropy parameter, we fit
\begin{equation}
    \Delta(N)=\frac{b_{\mathrm{fit}}}{N},
    \label{eq:xxz_gap_fixed_fit}
\end{equation}
with $b_{\mathrm{fit}}$ as the only free parameter,
minimizing squared logarithmic residuals. The exponent is fixed at $-1$. The notation connects with the coefficient $b$ used in the main text, where the leading gap scaling is written as $\Delta\sim b/N$. The results are reported in Table~\ref{tab:xxz_gap_fits}, together with a comparison with the collective prediction $b\simeq4(1-\gamma_z)$.
\begin{table*}[t]
    \centering
    \small
    \caption{
        Fits of $\Delta=b_{\mathrm{fit}}/N$ using
        $N=96,128,160,192$.
        Only the coefficient $b_{\mathrm{fit}}$ is fitted;
        the exponent is fixed at $-1$.
        The last column compares the fitted coefficient
        with the collective prediction $4(1-\gamma_z)$.
    }
    \label{tab:xxz_gap_fits}
    \begin{tabular}{rrr}
        \toprule
        $1-\gamma_z$
        & $b_{\mathrm{fit}}$
        & $\displaystyle
           \frac{b_{\mathrm{fit}}}{4(1-\gamma_z)}$ \\
        \midrule
        $10^{-2}$ & $0.041661725$              & $1.04154313$ \\
        $5 \times 10^{-3}$ & $0.020565655$              & $1.02828273$ \\
        $10^{-3}$ & $0.0040433726$             & $1.01084316$ \\
        $10^{-4}$ & $0.00040079014$            & $1.00197536$ \\
        $10^{-5}$ & $4.0009311\times10^{-5}$   & $1.00023277$ \\
        $10^{-6}$ & $4.0000951\times10^{-6}$   & $1.00002378$ \\
        \bottomrule
    \end{tabular}
\end{table*}

At $\gamma_z=0.99$, the fitted coefficient differs from
$4(1-\gamma_z)$ by approximately $4.15\%$.
As $\gamma_z$ approaches $1$, the ratio
$b_{\mathrm{fit}}/[4(1-\gamma_z)]$ approaches unity
over the studied size window.
This behavior is consistent with the collective expression
\begin{equation}
    \Delta\simeq\frac{4(1-\gamma_z)}{N}.
\end{equation}\subsection{Endpoint comparison and convergence checks}

At $\gamma_z=1-10^{-6}$, we compare the raw data directly
with the collective expressions
\begin{equation}
    \operatorname{Var}_{|0\rangle}[H_{\mathrm S}]
        \simeq \frac{N^2}{2}+N,
    \qquad
    \Delta \simeq \frac{4(1-\gamma_z)}{N}.
\end{equation}
We quantify the agreement through
\begin{align}
    d_V
    &=
    \left|
        \frac{\operatorname{Var}_{|0\rangle}[H_{\mathrm S}]}
             {N^2/2+N}-1
    \right|,\\
    d_\Delta
    &=
    \left|
        \frac{N\Delta}{4(1-\gamma_z)}-1
    \right|.
\end{align}
The results in Table~\ref{tab:xxz_endpoint_comparison}
are obtained directly from the raw data and do not
involve a fit in $N$.

\begin{table}[t]
    \centering
    \caption{
        Direct comparison with the collective endpoint
        expressions at $\gamma_z=1-10^{-6}$.
        The discrepancies are computed from the raw data.
        They measure disagreement with these expressions,
        not independently certified numerical uncertainties.
    }
    \label{tab:xxz_endpoint_comparison}
    \begin{tabular}{rrr}
        \toprule
        $N$ & $d_V$ & $d_\Delta$ \\
        \midrule
        96
        & $9.41\times10^{-9}$
        & $1.58\times10^{-5}$ \\
        128
        & $2.25\times10^{-8}$
        & $2.11\times10^{-5}$ \\
        160
        & $4.42\times10^{-8}$
        & $2.64\times10^{-5}$ \\
        192
        & $7.65\times10^{-8}$
        & $3.17\times10^{-5}$ \\
        \bottomrule
    \end{tabular}
\end{table}
All 24 gap points passed the configured convergence checks,
with a relative acceptance tolerance of $10^{-3}$ and
bond-dimension caps $\chi=64,128,256$.
The calculation compared the lowest states in magnetization
sectors $m=0,1,2$, where $m$ is the eigenvalue of
$J_z=\frac12\sum_i\sigma_z^{(i)}$, together with an
orthogonal excited state within $m=0$.
The lowest checked excitation was always in the $m=1$
sector.

At $\gamma_z=1-10^{-6}$, relative changes in the gap upon
increasing the bond dimension were below
$1.26\times10^{-8}$.
The sum of the absolute changes in the two relevant
shifted energies, divided by the gap, was below
$2.24\times10^{-8}$.
These are practical stability diagnostics rather than
rigorous error bounds.
The code's conservative floating-point resolution
diagnostic, divided by the gap, reaches
$1.31\times10^{-4}$ at $N=192$.
Agreement to four significant digits is therefore a
good description of the endpoint comparison.

Finally, combining the leading forms
$4\operatorname{Var}_{|0\rangle}[H_{\mathrm S}]
\sim cN^{2-\delta}$ and $\Delta\sim b/N$
with the approximate relation
$\Delta^2\mathcal{F}_\theta
\simeq4\operatorname{Var}_{|0\rangle}[H_{\mathrm S}]$
established numerically in Sec.~\ref{Sec_XXZ_GS},
we obtain
\begin{equation}
    \mathcal{F}_\theta
    \simeq \frac{c}{b^2}N^{4-\delta},
\end{equation}
which gives the ground-state QFI expression reported
in the main text.

\section{Local adiabatic matrix element of the transverse-field 1D Ising model}\label{Appx:matrels_TFIM}
In this Appendix, we derive the analytical expression of the local adiabatic matrix element, $\abs{\matrixel{1}{\partial_g H}{0}}$, for the TFIM. Here, $\ket{1}$ denotes the first excited state in the same symmetry sector as the ground state. 

We start by considering that, after a Jordan-Wigner transformation, the Hamiltonian in Eq.~\eqref{H_Ising_prep} has a two-level structure in each momentum sector \cite{Sachdev2011}. That is,
\begin{equation}\label{H_sectors_TFIM}
    H_k(g)=\frac{1}{2}\Vec{d}_k(g)\cdot\Vec{\tau},
\end{equation}
with $\Vec{\tau}$ being the pseudo-spin Pauli matrix vector in this effective two-dimensional space, and $\Vec{d}_k(g)\coloneq \left( 2g\sin{k},0,4-2g(1+\cos{k})\right)^T$. Next, let us consider the ground and first excited states
\begin{equation}\label{eq_dyn_TFIM_GS_1}
    \begin{split}
        \ket{0}&=\bigotimes_{k>0}\ket{0_k}\\
        \ket{1}&=\ket{1_{k_\mathrm{min}}}\bigotimes_{k>0,k\neq k_\mathrm{min}}\ket{0_k},
    \end{split}
\end{equation}
where $\ket{0_k}=\cos{\zeta_k}\ket{vac}+i\sin\zeta_k\ket{k,-k}$, $\ket{1_k}=i\sin{\zeta_k}\ket{vac}+\cos{\zeta_k}\ket{k,-k}$, being $\ket{vac}$ the zero-excitation state, with $\tan{2\zeta_k}=g\sin{k}/(2-g-g\cos{k})$ and $k_\mathrm{min}=\pi/N$ corresponding to the lowest quantized momentum. 

Therefore, using Eq.~\eqref{H_sectors_TFIM} and Eq.~\eqref{eq_dyn_TFIM_GS_1}, one obtains (after some algebra)
\begin{equation}
    \abs{\matrixel{1}{\partial_g H}{0}}=\frac{4\sin{(\pi/N)}}{\Delta(g)}\approx\frac{4\pi}{N\Delta(g)},
\end{equation}
where $|\Vec{d}_{k_\mathrm{min}}(g)|=\Delta(g)$ corresponds to Eq.~\eqref{Gap_ising_dyn}, and the approximation is valid for $N\gg 1$.
\section{Dynamical QFI of the spin-squeezing Hamiltonian}\label{SSappendix}
In this Appendix, we derive the dynamical QFI for the spin-squeezing Hamiltonian in Eq.~\eqref{H_spin_squeez_time} within the GSAM approach. To do so, we first derive the profile of $g(t)$, with $0\leq t\leq \tau$, such that the local adiabatic condition in Eq.~\eqref{localadiabaticcondition} is fulfilled. Using the arguments presented in Sec.~\ref{sec:LRpreparation}, and given that $\Delta (g (t)) = g(t)$ for this model, we obtain
\begin{equation}\label{integratedgap}
g (t) = \dfrac{1}{\frac{1}{g(0)} + \epsilon t} = \dfrac{g(0)}{1 + \epsilon g(0) t},
\end{equation}
where $g(0)$ is the initial value of the schedule. Now, let us compute the dynamical QFI by applying Eq.~\eqref{dynamicalqfieq}. If $H_{\theta=0} (t) \equiv H(t)$ admits an eigendecomposition $H (t)~= \sum_k E_k (t) \Pi_k$, the QFI is given by $\mathcal{F}_\theta (\tau) = 4 \Var [H_\mathrm{eff} (\tau)]_{\ket{0 (0)}}$, with $H_\mathrm{eff} (\tau) = \int_0^{\tau} \dd{t} U_0^\dagger (t,0) H_\mathrm{S} U_0 (t,0)$ and $U_0 (t,0) = \mathcal{T} \exp \left( -i \int_0^t \dd{t'} H_\mathrm{C} (t') \right)$. Additionally, if $[H_{\rm C}(t),H_{\rm C}(t')]=0$ for all $t,t'$, the time-ordering symbol can be dropped to write
\begin{equation}
H_\mathrm{eff} (\tau) = \tau \sum_k \Pi_k H_\mathrm{S} \Pi_k + \sum_{k \neq l} \Pi_k H_\mathrm{S} \Pi_l \int_0^{\tau} \dd{t} e^{-i \int_0^t \dd{s} \Delta_{lk} (s)},
\end{equation}
with $\Delta_{lk} (s) \coloneqq E_l (s) - E_k (s)$, which is the generalization of Result 1 in \cite{PRXQuantum.6.030309} for time-dependent $H_\mathrm{C}$. In this case, within the fixed-$j=N/2$ sector, the relevant gaps are
proportional to the schedule,
\begin{equation}
\Delta_{lk}(t)=g(t)\Delta_{lk},
\end{equation}
where $\Delta_{lk}$ is time independent. Thus, let us first compute the action of the effective Hamiltonian in the ground state of $H_{\rm C}$:
\begin{align}
H_\text{eff} \ket{\dfrac{N}{2}, 0} &= \Pi_1 H_{\rm S} \Pi_0 \int_0^{\tau} \dd{t} e^{-i \int_0^t \dd{s} g(s) (E_0 - E_1)} \ket{\dfrac{N}{2}, 0} \\
&= \sqrt{\dfrac{N}{2} \left( \dfrac{N}{2} + 1 \right)} \int_0^{\tau} \dd{t} e^{i \int_0^t \dd{s} g(s)} \left( \ket{\dfrac{N}{2}, +1} + \ket{\dfrac{N}{2}, -1} \right) \\ 
&= \sqrt{\dfrac{N}{2} \left( \dfrac{N}{2} + 1 \right)} \int_0^{\tau} \dd{t} e^{i \int_0^t \dd{s} \frac{g(0)}{1 + \epsilon g(0) s}} \left( \ket{\dfrac{N}{2}, +1} + \ket{\dfrac{N}{2}, -1} \right) \\
&= \sqrt{\dfrac{N}{2} \left( \dfrac{N}{2} + 1 \right)} \int_0^{\tau} \dd{t} e^{\frac{i}{\epsilon} \log \left( 1 + \epsilon g(0) t \right)} \left( \ket{\dfrac{N}{2}, +1} + \ket{\dfrac{N}{2}, -1} \right) \\
&= \sqrt{\dfrac{N}{2} \left( \dfrac{N}{2} + 1 \right)} \dfrac{(1 + \epsilon g(0) \tau)^{1 + \frac{i}{\epsilon}}-1}{g(0) (i + \epsilon)} \left( \ket{\dfrac{N}{2}, +1} + \ket{\dfrac{N}{2}, -1} \right).
\end{align}
Now, we just need to use that $\mathcal{F}_\theta (\tau) = 4 \Var \left[ H_\mathrm{eff} (\tau) \right]_{\ket{0 (0)}}$ to obtain 
\begin{equation}
    \mathcal{F}_\theta (\tau) = \dfrac{4N}{g(0)^2 (1 + \epsilon^2)} \left( \dfrac{N}{2} + 1 \right) \abs{(1 + \epsilon g(0) \tau)^{1 + \frac{i}{\epsilon}} - 1}^2
\end{equation}
which reduces to $\mathcal{F}_\theta (\tau) = \frac{2 N^2 \tau^2}{\epsilon^{-2} + 1} + \mathcal{O} (\tau^2 N) + \mathcal{O} (\tau N^2)$ in the large-$\tau$ limit, as we wanted to show.
\end{document}